\documentclass[opre]{informs3_no_remark}
\usepackage[numbers]{natbib} 
\RequirePackage{tgtermes}
\RequirePackage{newtxtext}
\RequirePackage{newtxmath}
\RequirePackage{bm}
\RequirePackage{endnotes}

\OneAndAHalfSpacedXI

\usepackage{algorithm}
\usepackage{algpseudocode}
\usepackage{tikz}

\usepackage{natbib}
 \bibpunct[, ]{(}{)}{,}{a}{}{,}%
 \def\bibfont{\small}%
 \def\bibsep{\smallskipamount}%
 \def\newblock{\ }%
\usepackage{comment}   
\usepackage{verbatim}

\usepackage{dsfont}

 \usepackage{amsmath}
\usepackage{amsfonts}
\usepackage{bbm}
\usepackage{amsmath, amssymb, amsfonts}

\usepackage{graphicx}
\usepackage{subfigure}
\usepackage{color}

  \usepackage{hyperref}
\usepackage{xspace}
\usepackage{framed}
\usepackage{mathtools}
\usepackage{caption}

\EquationsNumberedThrough    

\TheoremsNumberedThrough     
\ECRepeatTheorems  %

\newcommand{\Exp}[1]{\mathbb{E}\!\left[#1\right]}
\newcommand{\abs}[1]{\left\lvert #1 \right\rvert}

\newcommand{\lrset}[1]{\left\{{#1}\right\}}
\newcommand{\lrp}[1]{\left({#1}\right)}

\DeclareOldFontCommand{\bf}{\normalfont\bfseries}{\mathbf}

\MANUSCRIPTNO{MOOR-0001-2024.00}

\begin{document}



\RUNTITLE{ABS: Simulating larger models using smaller ones}

\TITLE{Agent Based Simulators for Epidemic Modelling: Simulating Larger Models Using Smaller Ones}

\ARTICLEAUTHORS{%
\AUTHOR{Daksh Mittal}
\AFF{Columbia University, Graduate School of Business,  \EMAIL{dm3766@gsb.columbia.edu}}

\AUTHOR{Sandeep Juneja}
\AFF{Ashoka University, Safexpress Centre for Data, Learning and Decision Sciences, \EMAIL{sandeep.juneja@ashoka.edu.in}}
} 

\ABSTRACT{%
Agent-based simulators (ABS) are a popular epidemiological modelling tool to study the impact of various non-pharmaceutical interventions in managing an epidemic in a city (or a region). They provide the flexibility to accurately model a heterogeneous population with time and location varying, person-specific interactions as well as detailed governmental mobility restrictions. Typically, for accuracy, each person is modelled separately. This however may make computational time prohibitive when the city population and the simulated time is large.  In this paper, we dig deeper into the underlying probabilistic structure of a generic, locally detailed ABS for epidemiology to arrive at modifications that allow smaller models (models with less number of agents) to give accurate statistics for larger ones, thus substantially speeding up the simulation. We observe that simply considering a smaller aggregate model and scaling up the output leads to inaccuracies. We exploit the observation that in the initial disease spread phase, the starting infections create a family tree of infected individuals more-or-less independent of the other trees and are modelled well as a multi-type super-critical branching process. Further, although this branching process grows exponentially, the relative proportions amongst the population types stabilise quickly. Once enough people have been infected, the future evolution of the epidemic is closely approximated by its mean field limit with a random starting state. We build upon these insights to develop a shifted, scaled and restart-based algorithm that accurately evaluates the ABS's performance using a much smaller model while carefully reducing the bias that may otherwise arise. We apply our algorithm for Covid-19 epidemic and theoretically support the proposed algorithm through an asymptotic analysis where the population size increases to infinity. We develop nuanced coupling based
arguments  to show that the epidemic process
is close to the branching process early on in the simulation. }






\maketitle


\section{Introduction}
\label{sec:Introduction}

Agent-based simulators (ABS) are a popular tool in epidemiology  \citep{hunter2017taxonomy, ferguson2020report}. As is well known, in an ABS for epidemic simulation of a city, a  synthetic replica of the city is constructed on a computer, capturing the population's interaction spaces and the details of disease spread over time. Typically, each individual in the city is modelled as an agent,  making the total number of agents equal to the city's population.  The constructed individuals reside in homes, children may go  to schools, adults may go to work. Individuals also engage with each other in community spaces to capture interactions in marketplaces, restaurants, public transport, and other public places. Homes, workplaces, schools sizes and locations and  individuals associated with them,   their gender and age,  are created to match the city census  data and are distributed to match its geography. Government policies, such as partial, location specific  lockdowns for small periods of time, case isolation of the infected and home  quarantine of their close contacts, closure of schools and colleges, partial openings of workplaces, etc. lead to mobility reductions and are easily modelled. Similarly, variable compliance behaviour in different segments of the population that further changes with time, is easily captured in an ABS.  Further, it is easy  to introduce new variants as they emerge,  the individual vaccination status, as well as the protection offered by the vaccines against different variants as a function of the evolving state of the epidemic and individual characteristics, such as age, density of individual's interactions, etc.  \textit{Thus, this microscopic level  modelling flexibility allows ABS to become an effective strategic and operational tool to manage and control the disease spread. }

\vspace{3mm}

\noindent {\bf Key drawback:} 
However, when a reasonably large population is simulated, especially over a long time horizon, an ABS can take huge computational time—its key drawback.  This becomes particularly
prohibitive when multiple runs are needed using different parameters. 
For instance, any predictive analysis involves simulating a large number of scenarios
to provide a comprehensive view of potential future sample paths. 
In model calibration, 
the key parameters such as transmission rates, infectiousness of new variants are fit to
the observed infection data over relevant time horizons, requiring many 
computationally demanding simulations.

\vspace{3mm}

\noindent {\bf Key contributions:} To address the challenge of excessive computational time, we developed the Shift-Scale-Restart (SSR) algorithm. We empirically validate its effectiveness using an agent-based simulation of Covid-19 epidemic in a city. Additionally, we provide theoretical support through asymptotic analysis as the population size $N$ approaches infinity. These contributions are detailed below.
\noindent
\begin{enumerate}
    \item  \textbf{Shift-scale-restart (SSR) algorithm:} SSR algorithm carefully adjusts the output from the smaller model to accurately match those from
 the larger one, thereby significantly reducing simulation time. We demonstrate that the naive approach of using smaller models with fewer agents and directly scaling their outputs does not work. Instead, our SSR algorithm overcomes this limitation by strategically exploiting the closeness
of the underlying infection process—which tracks the number of infected individuals of each type at each time point—first to a multi-type supercritical branching process, and subsequently, when appropriately normalized, to the mean field limit of the infection process. This approach ensures that outputs from the smaller model precisely match those from the larger model.

Essentially, both the smaller and the larger model, with identical initial conditions of small number of infections
evolve similarly in the early days of the infection growth.  
Interestingly, in a super-critical branching process, while the number of infections grows exponentially, the proportions of the different types infected quickly stabilises, and this allows us to shift a scaled
path from a smaller model to a later time with negligible change in the underlying distribution.
Therefore, once there are enough infections in the system, output from the smaller model, when scaled, matches that from the larger model at a later `shifted' time.
This shifting and scaling of the paths from the smaller model does a good job of representing output from the larger model when there are no interventions
to the system. However, realistically, the government intervenes and population mobility behaviour changes with increasing infections. 
To get the timings of these interventions right, we restart the smaller model and synchronise the timings of interventions in the shifted and scaled path to the actual timings in the original path of the larger model.

\item \textbf{Empirical evaluation:} 
We evaluate the SSR algorithm's effectiveness using an agent-based simulator (ABS) designed to model a Covid-19 epidemic in a city of 12.8 million people. While Covid-19 serves as a case study, the algorithm is broadly applicable to other epidemics. The ABS realistically captures interactions in settings such as homes, schools, workplaces, and communities, incorporating interventions like lockdowns, home quarantine, case isolation, school closures, and restricted workplace attendance.

    Using the SSR algorithm, we replicate the dynamics of the 12.8 million population model with a smaller model of 1 million agents (see Figures \ref{fig:shift_scale_100_infections} and \ref{fig:shift_scale_restart_real_intervention}), achieving an approximately 12.8-fold reduction in computational time. Similar results are observed with a smaller 0.5 million population model, although further reductions increase errors due to the limited duration of the branching process phase and the breakdown of mean-field approximations (see Section \ref{sec:limits_smaller_city}). To ensure robustness in less dense populations, we adjust interaction parameters and confirm the algorithm's effectiveness in sparser cities (see Section \ref{sec:sparse_cities}). Additionally, we demonstrate the algorithm's robustness under parameter uncertainty, accounting for imprecise epidemic parameters in the initial phase (see Section \ref{sec:parameter_uncertainty}). 
The underlying principles apply generally to any epidemic in a large city with an initial exponential growth phase that slows as the infected population grows (see \citet{GABS_3, FriasMartinez2011SocialCom}), supported by additional experiments.

\item \textbf{Theoretical insights:}
We provide theoretical support for the SSR algorithm through an asymptotic analysis as the population size $N$ approaches infinity. We show that, early on till time $\frac{\log (N / (I_0 \log N))}{\log (\rho)}$, where $\rho$ is the exponential growth rate in the early epidemic phase and $I_0$ is the initial number of exposed individuals, the epidemic process closely resembles a multi-type branching process. This is established through a precise coupling between the epidemic and the branching process during this period. Beyond time $\frac{\log (\epsilon N)}{\log (\rho)}$ for any $\epsilon > 0$ and large $N$, the epidemic process aligns with its mean-field limit, described by a high-dimensional discretized ordinary differential equation (ODE). Although our simulation model allows for an uncountable types of individuals, and the proposed algorithm is numerically seen to be effective in this setting, our theoretical analysis assumes a finite number of types. This simplifies the analysis considerably while still providing clear insights into the SSR algorithm's effectiveness.

Our theoretical framework encompasses general compartmental models. In Appendix \ref{sec:compartmental_models}, we analyze the behaviour of simple compartmental models, such as the susceptible-infected-recovered (SIR) model and its generalizations, during the early branching process phase before transitioning to ODE-governed dynamics.

Finally, we show why a naive rescaling of a smaller model fails, providing insights into limitations of such heuristic approaches.

\end{enumerate}

\vspace{3mm}

 \noindent\textbf{Structure of the remaining paper:}    We review related literature in Section \ref{sec:lit_review}. 
 Section \ref{sec:ABS_brief} provides a concise overview of our agent-based simulator. 
 To demonstrate the core concepts, we present the implementation of our algorithm in a realistic scenario in Section \ref{sec:speeding_abs}. 
 The Shift-Scale-Restart (SSR) algorithm is detailed in Section \ref{sec:ssr_algo}.
 In Section \ref{sec:Theoretical_results}, we provide a theoretical asymptotic analysis supporting the algorithm’s effectiveness, with detailed proofs provided in Appendix \ref{sec:proofs}. Section \ref{sec:further_experiments} evaluates the algorithm’s performance under different scenarios such as sparser cities, parameter uncertainty etc. Finally, we conclude in Section \ref{sec:conclusion}.

\section{Related Work}
\label{sec:lit_review}

Agent-based simulators (ABS) are widely used in epidemiological modelling to evaluate the effectiveness of various non-pharmaceutical interventions (NPIs) in managing epidemics within urban populations \citep{ferguson2005strategies, ferguson2006strategies, hunter2017taxonomy}. Foundational large-scale simulations include Eubank et al.'s city-wide contact network model \citep{Eubank2004Nature}, while open, census-based platforms such as FRED \citep{Grefenstette2013FRED} have enabled reproducible studies of NPIs. More recently, Covid-19-specific simulators—such as Covasim and OpenABM–Covid-19—have explored diverse contact patterns and interventions \citep{Kerr2021Covasim, Hinch2021OpenABM, ferguson2020report}. Our proposed shift–scale–restart (SSR) approach is complementary to these efforts: it specifically targets the computational cost of running ABS models, enabling faster parameter sweeps and counterfactual scenario analysis while preserving accuracy.

A related body of literature formalizes early epidemic dynamics using Galton–Watson branching processes \citep{harris1963theory, AthreyaNey1972}, which estimate extinction probabilities, growth rates, and the composition of infectives prior to significant susceptible depletion \citep{BallMollisonScaliaTomba1997, BallSirlTrapman2009, BallSirlTrapman2012}. Extensions incorporating household and meso-scale structures introduce two-level mixing and yield principled approximations for the stochastic early phase of epidemics \citep{Britton_Pardoux_2019, epidemic_notes}. The connection between exponential growth and reproduction numbers offers an empirical link between observed case growth and disease transmissibility \citep{WallingaLipsitch2007, Cori2013EpiEstim}, while next-generation matrix methods and $R_0$ analyses form the basis for both network-aware and compartmental models \citep{DHR2010NGM, VDW2002NGM}. Metapopulation models incorporating commuting and air traffic patterns explain spatial transmission and timescale separation across regions \citep{Balcan2009PNAS}, and real-world mobility signals have been used to calibrate contact intensities and assess NPIs \citep{FriasMartinez2011SocialCom}.
Beyond the initial stochastic phase, classical density-dependent Markov chain limits justify deterministic (mean-field) approximations and associated diffusion corrections for large populations \citep{Kurtz1970ODE, Kurtz1971Limits, Kurtz1981Approximation, EthierKurtz1986, DarlingNorris2008, BenaimLeBoudec2008}. While the literature has previously explored how epidemic processes resemble branching processes in their early stages and transition to mean-field dynamics later, a key contribution of our work lies in how we integrate these perspectives within an algorithmic framework—yielding computationally efficient estimates from ABS models.

Another line of research accelerates simulations through parallelization, distributing propagation computations over large graphs \citep{Bisset2009SC, Barrett2013EpiSimdemics}. Our approach is complementary: rather than optimizing system-level performance through hardware or software architecture, we propose a principled algorithmic method that estimates the behaviour of large-scale models using smaller-scale simulations.

Our work is most closely related to national-scale ABS models involving tens of millions of agents, which typically require supercomputing infrastructure. To make simulations feasible, many models down-sample the synthetic population by a factor of 10–100, adjusting contact frequencies and stochastic seeding to preserve macroscopic outputs \citep{Scale_Matters_2023, NYC_Scaled_ABS_2020}. These approaches often assume that metrics such as peak timing and the relative efficacy of interventions remain robust under down-scaling, without explicitly modelling the scale of infections. Covasim, for instance, operationalizes this idea through dynamic rescaling, wherein multiple real individuals are merged into a single agent as prevalence grows—allowing runtime to remain approximately constant while maintaining realistic transmission dynamics \citep{Kerr2021Covasim}. However, these strategies are typically implemented in an ad hoc manner. In contrast, our work introduces a principled approach to down-scaling, and we demonstrate that naïve scaling can lead to significant errors—particularly when the initial number of infections is small.

\section{Agent Based Simulator} 
\label{sec:ABS_brief}
We provide a brief overview of the key drivers governing the dynamics of our agent-based simulation (ABS) model. A comprehensive discussion is presented in Appendix \ref{sec:detailed_abs_model}.

The model begins by generating a synthetic representation of the city, comprising individual agents and various interaction spaces including \textbf{households, schools, workplaces, and community spaces.} Infected individuals interact with susceptible individuals within these interaction spaces. The demographic and structural characteristics of the model—including household size distributions, age demographics, employment and educational status, as well as the size and composition of schools and workplaces—are calibrated to match available city-specific data.

The simulation proceeds in discrete time steps of constant duration $\Delta t$. At initialization (time zero), a small number of individuals are assigned to exposed, or symptomatic states to establish the initial infection seed. At each time step $t$, an infection rate $\lambda_j(t)$ is calculated for each susceptible individual $j$, based on their interactions with infected individuals across different interaction spaces (households, schools, workplaces, and community settings).
During the subsequent time interval $\Delta t$, each susceptible individual transitions to the exposed state with probability $1 - \exp\{ - \lambda_j(t) \cdot \Delta t\}$, with transitions occurring independently across all individuals. Concurrently, disease progression proceeds independently for the already-infected population during the same interval $\Delta t$. The simulation time is then advanced to $t + \Delta t$, and individual states are updated to reflect new exposures, changes in infectiousness, hospitalizations, recoveries, quarantine measures, and other relevant state transitions occurring during the period from $t$ to $t + \Delta t$. This iterative process continues until the predetermined simulation endpoint is reached.

{Let $X_n$ denote the disease state of each agent in the city at time $n \Delta t$. The state at the subsequent time step, $X_{n+1}$ (at time $(n+1) \Delta t$), is determined by both new exposures due to transmission and the progression of disease among already-infected individuals. For the ease of notation we will set $\Delta t =1$ in the rest of the paper.}

\subsection{Computing infection rate $\lambda_j(t)$} \label{sec:computing_rates}
A susceptible individual $j$ at time $t$ receives a total infection rate $\lambda_j(t)$, which is the sum of infection rates from distinct interaction spaces: home ($\lambda_j^{h}(t)$), school ($\lambda_j^{s}(t)$), workplace ($\lambda_j^{w}(t)$) 
 and  community ($\lambda_j^{c}(t)$), contributed by infected individuals in these respective spaces. 
This  is expressed as: 
\[\lambda_j(t)= \lambda_j^{h}(t)+\lambda_j^{s}(t)+\lambda_j^{w}(t)+\lambda_j^{c}(t).\]

The transmission rate $ \beta $ of the virus by an infected individual in each interaction space represents the expected number of effective (infection-spreading) contacts with all individuals in that space. This rate accounts for the combined effect of both the frequency of interactions and the probability of transmission per interaction. An infected individual can transmit the virus in the infective (pre-symptomatic or asymptomatic stage) or in the symptomatic stage. To enhance model's realism, additional factors may be introduced to capture the heterogeneity between the individuals. For instance, in our simulation model we consider two individual-specific parameters: a severity variable, which influences attendance at school or work based on disease severity, and a relative infectiousness variable, which linearly affects virus transmission.

Formally, let ${\mathbf 1}_{\mathrm{inf},\tilde{j}}(t) = 1$ when individual $\tilde{j}$ can transmit the virus at time $t$, ${\mathbf 1}_{\mathrm{inf},\tilde{j}}(t) = 0$ otherwise.  Let $\beta^h_{\tilde{j}}$, $\beta^s_{\tilde{j}}$, $\beta^w_{\tilde{j}}$, and $\beta^c_{\tilde{j}}$  denote the transmission coefficients  for individual $\tilde{j}$ in home, school, workplace and  community spaces, respectively.  A susceptible individual $j$ belonging to home $h(j)$, school $s(j)$, workplace $w(j)$, and community space $c(j)$ experiences the following infection rates at time $t$:
 
\begin{enumerate}
    \item \textbf{Home, School and Workplace transmission:} The home infection rate $\lambda_j^{h}(t)$ represents the average transmission rate incoming from each infected individual within the same household and is expressed as:
     \[\lambda_j^{h}(t)= \sum_{\tilde{j} : h(\tilde{j}) = h(j)}     \frac{\beta^h_{\tilde{j}} }{n_{\tilde{j}}^h} ~{\mathbf 1}_{\mathrm{inf},\tilde{j}}(t),  \]

where $n_{\tilde{j}}^h$ is the number of individuals in the  home of individual $\tilde{j}$.
The infection rates $\lambda_j^{s}(t)$ and $\lambda_j^{w}(t)$ are computed analogously based on infected individuals in the respective interaction spaces. This formulation follows the modeling approach used in \cite{ferguson2006strategies, ferguson2020report}. To maintain clarity in the main text, we defer the detailed description of additional heterogeneity factors including individual-specific infectiousness, severity parameters, and community density factors to Appendix \ref{sec:detailed_abs_model}.

     \item \textbf{Community transmission:}   Community infection rates depends on interactions with all the infected individuals within the city.   
The community infection rate experienced by individual $j$ is expressed as:

 \begin{equation*} 
 \lambda_{j}^{c}(t)=    \sum_{\tilde{j}} \psi_{\tilde{j},j} ~  \frac{\beta_{\tilde{j}}^c }{N} ~ {\mathbf 1}_{\mathrm{inf},\tilde{j}}(t),
  \end{equation*}

where $N$ is the total number of people in the city and the parameter $\psi_{\tilde{j},j}$ models various spatial and social factors, including locality-based community structure, individual mobility patterns within communities, proximity to community centers, and population density variations (with $\sum_{{j}} \psi_{\tilde{j},j} =1$). The exact formulation of $\psi_{\tilde{j},j}$ is provided in Appendix \ref{sec:detailed_abs_model}. 
This rate structure also parallels the community mixing models used in \cite{ferguson2006strategies, ferguson2020report}.

\end{enumerate}

\subsection{Disease progression} 
\label{sec:disease_progression}

Disease progression of an infected individual varies considerably across different diseases. While our algorithm is adaptable to any disease progression model, for concreteness we base our numerical experiments on the Covid-19 progression model described in \citep{verity2020estimates} and \citep{ferguson2020report}. 
We emphasize that our proposed method has broad applicability to epidemic modelling of diverse infectious diseases in urban populations.

In this model, individuals can exist in one of eight distinct states: susceptible, exposed, infective (either pre-symptomatic or asymptomatic), recovered, symptomatic, hospitalised, critical, or deceased. The disease progression follows a structured pathway: upon exposure to the virus, individuals experience an incubation period that follows a Gamma distribution. Subsequently, they remain infectious for a duration that is exponentially distributed, encompassing both pre-symptomatic and potential asymptomatic transmission phases.
 We assume that a fraction of infected patients recover without developing symptoms (asymptomatic cases), while the remaining patients progress to symptomatic disease. Following symptom onset, individuals either recover or require hospitalization after an exponentially distributed time interval (detailed model parameters are provided in Appendix \ref{sec:Numerical_Parameters_Section}). The probability of recovery at each stage depends on the individual's age.
Hospitalised individuals may remain infectious but are assumed to be sufficiently isolated to prevent further disease transmission. The subsequent progression from hospitalization to critical care, and potentially to fatality, is also age-dependent.

\subsection{Public health safety measures (PHSMs)}
\label{sec:PHSM}
We introduce methodologies to model various PHSMs (Public Health and Social Measures) commonly employed to control epidemic spread, including lockdowns, home quarantine, case isolation, social distancing for elderly populations, mobility restrictions, and mask mandates. The fundamental principle underlying these interventions is the reduction of mobility and interpersonal interactions, thereby decreasing the effective transmission rate of the disease. Table \ref{tab:interventions} in the Appendix summarizes the key interventions implemented in our experimental framework.

The aforementioned PHSMs impose restrictions on individual mobility when implemented. However, empirical evidence demonstrates that when multiple restrictions are simultaneously enacted, only a subset of the population adheres to these measures. Consequently, our simulator allows for a feature where we can restrict movement exclusively for the individuals within the compliant fraction of households in the urban area under study.

\section{Speeding up ABS: The big picture and three key phenomena}
\label{sec:speeding_abs}
\begin{figure*}
  \centering
  \begin{minipage}[b]{0.31\textwidth}   \includegraphics[width=\textwidth]{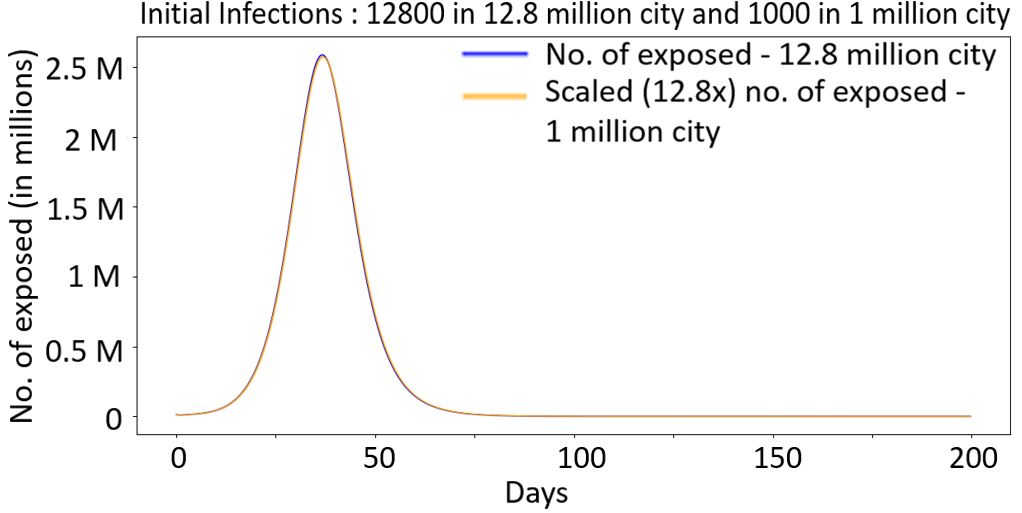}
\caption{Scaled number exposed in the smaller model match the larger model when we start with large, 12800 infections.
  \label{fig:12800_infections}} 
  \end{minipage}
  \hfill
  \begin{minipage}[b]{0.31\textwidth}   \includegraphics[width=\textwidth]{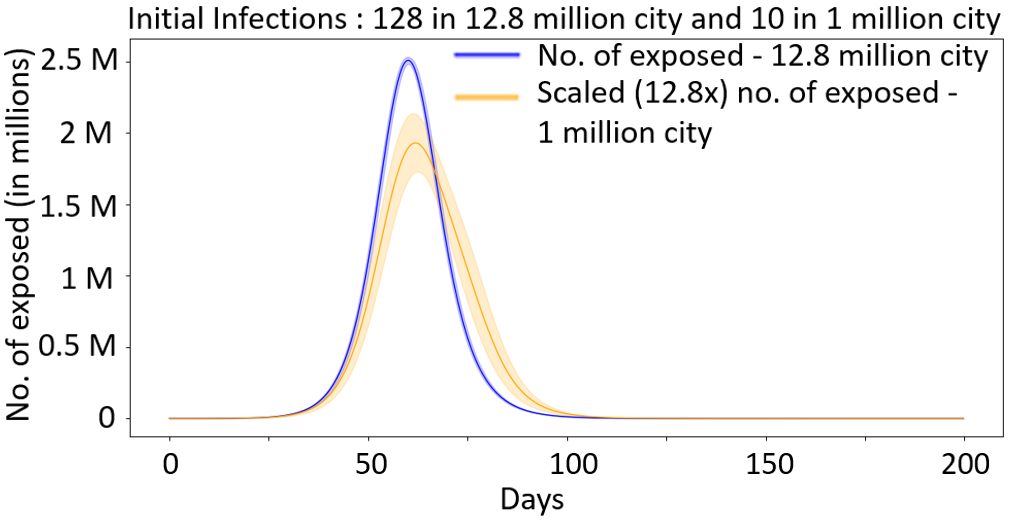}   
    \caption{Scaled number exposed in smaller model do not match the larger model when we start with a few, 128 infections. 
    \label{fig:128_infections}}
  \end{minipage}
  \hfill
  \begin{minipage}[b]{0.31\textwidth}   \includegraphics[width=\textwidth]{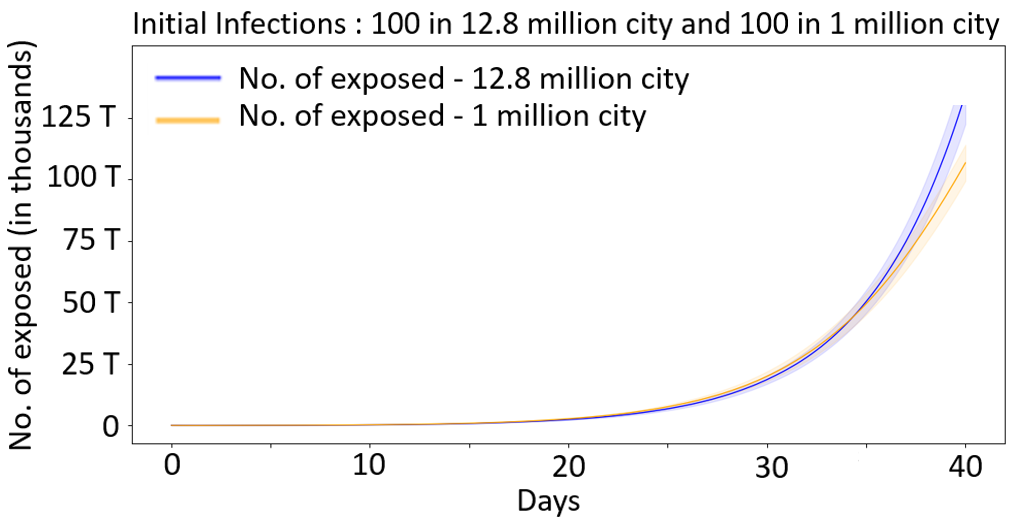}   
    \caption{Smaller and larger model are essentially identical initially when we start with same no. of few, 100 infections.     \label{fig:100_infections_upto_day_35}}
    \end{minipage}
\end{figure*}

\noindent \textbf{Phenomenon 1:} A naive approach to speed up the ABS may be to use a representative smaller population model and scale up the results. For instance, while a realistic model for a city may have 12.8 million agents, 
we may construct a smaller model having, say, a million agents that preserves the essential characteristics of the larger model, and seed the smaller model with a proportionally lesser number of initial infections. Specifically, both models should maintain equivalent infection dynamics such that each infectious individual probabilistically contributes the same total infection rate to all susceptible individuals at each time step. The output from the smaller model can then be scaled by a factor of 12.8 to estimate the corresponding output from the larger model.
We observe a somewhat remarkable finding: this naive scaling approach is actually accurate when the initial seed infections in both models are substantial—in the order of thousands (proportionally scaled)—and are identically distributed across both models. Figure \ref{fig:12800_infections} illustrates this phenomenon by plotting the number of exposed individuals under a counterfactual no-intervention scenario. This comparative accuracy extends equally to other epidemiological statistics, including the number of infected, hospitalised, ICU-admitted, and deceased individuals.

The underlying rationale for this accuracy lies in the sufficient infection prevalence within both models. When infection numbers are large enough, the proportion of infected individuals in both the smaller and larger models closely approximates their identical mean-field limits, thereby ensuring consistent epidemiological dynamics across different population scales.

\vspace{3mm} 
\noindent \textbf{Phenomenon 2:} Modelling initial randomness in disease spread is crucial for several reasons, including: ascertaining the distribution of when and where an outbreak may be initiated, the probability  that some of the initial infection clusters might die out, getting an accurate  distribution of geographical spread of infection over time, 
capturing the  intensity of sample paths (the random variable $W$ in the associated branching process, described in 
Theorem~\ref{theorem_extending_branching_process}), etc (see \citep{sazonov2011two, hartfield2013introducing, hasegawa2016outbreaks, czuppon2021stochastic, kenah2007second}). 
These considerations are typically addressed by initializing the model with a small number of infections, say, around a hundred, at a carefully chosen starting time. However, in such settings, we observe that the scaled output from smaller models (with proportionally fewer initial infections) becomes noisy and biased, rendering simple scaling approaches ineffective (see Figure \ref{fig:128_infections}).
In Section \ref{sec:bias_naive_scaling}, we explain why scaled smaller models exhibit bias and systematically underreport infection numbers compared to larger models. We present an informal theorem below, with the formal version provided in Theorem \ref{thm:bias_justification}.

\noindent \textit{Informal theorem:
Naively scaling a smaller model—starting with a proportionally equivalent number of initial infections—underestimates the total number of individuals who will eventually become infected in a larger city. Furthermore, this underestimation bias decreases as the number of initial infections increases.}

\vspace{3mm}  
\noindent \textbf{Phenomenon 3:} We observe that during the early stages of the epidemic, smaller and larger models initialized with the \textbf{same number of initial infections} and similar distribution behave more or less identically (see Figure \ref{fig:100_infections_upto_day_35}). Consequently, the smaller model  with the \textit{unscaled} number of infections 
provides an accurate approximation to the larger one initially.  During this early phase, each infectious individual in both models contributes approximately the same total infection rate to susceptible populations across home, workplace, school and community. Probabilistically, this is true because both models closely approximate an identical multi-type branching process in the early stages. The shift-scale-restart algorithm outlined in Section \ref{sec:ssr_algo} exploits these observations to accelerate simulator performance, as described below.

\begin{figure}
  \centering
   \begin{minipage}[b]{0.48\textwidth}
    \includegraphics[width=\linewidth, height=4.5cm]{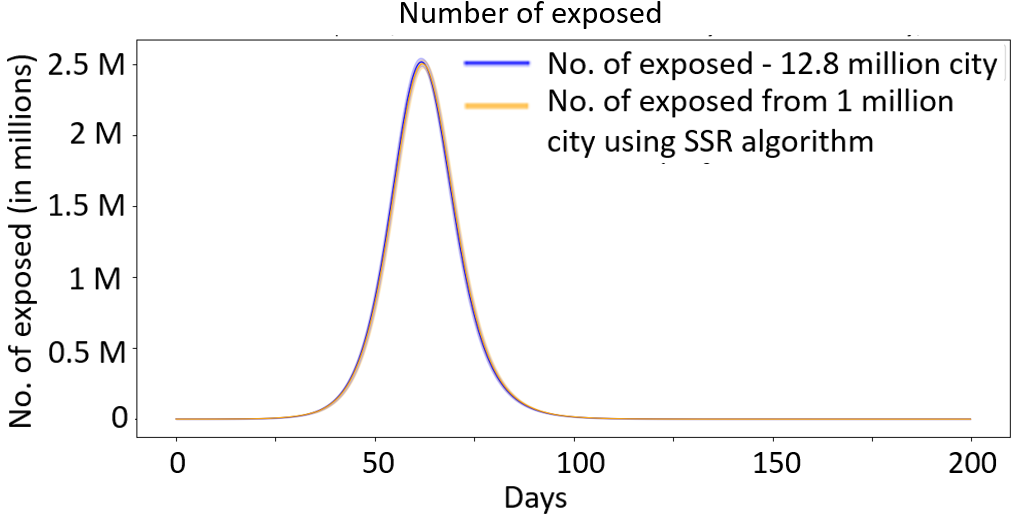}
  \caption{Shift and scale smaller model (no. of exposed) matches the larger model under no intervention scenario. 
\label{fig:shift_scale_100_infections}}
\end{minipage}
\hfill
\begin{minipage}[b]{0.48\textwidth}
\centering
    \includegraphics[width=\linewidth, height=4.5cm]{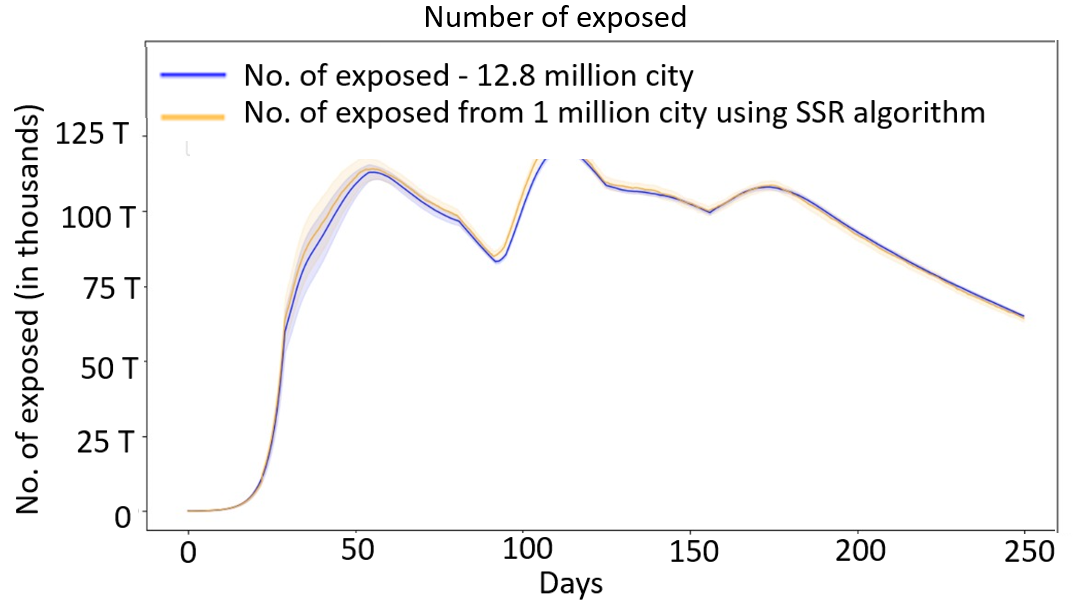}   
   \caption{Shift-scale-restart smaller model matches the larger one under real world intervention scenario over 250 days. 
    \label{fig:shift_scale_restart_real_intervention}}
    \end{minipage}
\end{figure}

\vspace{3mm} 
{\noindent}\textbf{Fixing ideas:} Consider a city with 
an  estimated population of $12.8$ million, and suppose we are interested in the statistics such as expected exposed, infected, hospitalizations and fatalities over time. Rather than running  a $12.8$ million agent model seeded 
with $100$ randomly distributed infections on day zero, we propose an efficient alternative approach. We initiate a smaller 1 million agent model with 100 similarly distributed infections at day zero and generate a complete trajectory for the required duration. To derive the corresponding statistics for the larger population model from this smaller model, we  proceed as follows:

\begin{enumerate}
    \item  \textit{Initial Phase Approximation:} Under the no-intervention scenario, the output of the smaller model closely matches that of the larger model for approximately the first 35 days (Figure \ref{fig:shift_scale_100_infections}). Our theoretical analysis in Section \ref{sec:Theoretical_results} demonstrates that both models closely approximate the associated branching process until time $\log_\rho\left(\frac{N}{I_0\log (N)}\right)$, where $N$ represents the population of the smaller model, $I_0$ denotes the number of initially exposed individuals (100 in this case), and $\rho$ represents the exponential growth rate in early fatalities (epidemic growth rate), estimated from fatality data to be 1.21. This quantity $\log_\rho\left(\frac{N}{I_0\log (N)}\right)$ evaluates to approximately 35 days, confirming that both models remain close to the underlying branching process around day 35.
    \item \textit{Scaling and Time-Shifting Procedure:} Suppose that after this initial 35-day period, the city exhibits an average of $x$ thousand infections. We identify the corresponding day when the city had $\frac{x}{12.8}$ thousand infections, which occurs at day 21.5 in our example. We extract the trajectory from day 21.5 onwards, scale it by a factor of 12.8, and concatenate it to the original path beginning at day 35.
The theoretical foundation for this time shift comes from branching process theory, while a super-critical multi-type branching process exhibits exponential growth with sample path-dependent intensity, the relative proportions among types along each sample path stabilize rapidly and become approximately stationary (see Theorem \ref{theorem_extending_branching_process}), and hence relative proportion amongst infected types at day 21.5 closely resembles those at day 35. This shifted and scaled output after day 35 demonstrates remarkable agreement with the larger model. Figure \ref{fig:shift_scale_100_infections} compares the infection trajectories from the 12.8 million agent model with those from the shifted and scaled 1 million agent model. The choice of day 35 is not critical; similar results can be achieved using earlier transition points, as low as 25 days.
\item \textit{Restart for handling interventions:}  In realistic scenarios, public health interventions typically commence once reported cases begin to increase substantially. Consider the case where intervention occurs on day 40 in our example. In such a scenario, the shifted and scaled path from day 21.5 would need to have the restrictions imposed on day 26.5  so that it approximates day 40 in the larger model. We implement this by first using the initially generated path until day 35, computing the appropriate scaled path time (21.5 days, in this case),
 and then, using common random numbers, restarting an identical path from time zero that has restrictions imposed from day 26.5. This path is scaled from day 21.5 onwards and concatenated to the original path at day 35.
 As noted in Section \ref{sec:Theoretical_results}, the restarted path need not employ common random numbers; independent generation also yields similar results.
\end{enumerate}

Figure \ref{fig:shift_scale_100_infections} compares the number of exposed individuals in the 12.8 million city simulation (no-intervention scenario) with estimates from the shift-scale-restart algorithm applied to the smaller 1 million city model. 
Figure \ref{fig:shift_scale_restart_real_intervention} presents a comparison of the exposed population dynamics between the 12.8 million and 1 million population city models using our algorithm under realistic interventions (lockdown, case isolation, home quarantine, masking, etc.) implemented at appropriate times over a 250-day period.  The results demonstrate that the smaller model faithfully replicates the larger one with negligible error. Similar accuracy is observed for other epidemiological metrics, including hospitalizations and cumulative fatalities, as well as for more complex models incorporating viral variants. Comparisons for additional health statistics and experimental details are presented in Appendix \ref{sec:Numerical_Parameters_Section}.

\section{Shift-scale-restart algorithm} 
\label{sec:ssr_algo}

We now introduce the algorithm more formally. 
Let $\mu_0(N)$ denote the initial distribution of the infected population at time zero in our model with population $N$,
and let the  simulation run for a total of $T$ time units.  E.g., for the city under consideration, at a suitably chosen time 0, we select $I_0=100$ people at random from the
 population and mark them as exposed. 
Algorithm 1 summarizes the simulation dynamics. 
\begin{algorithm}
\caption{Simulation Dynamics}\label{alg:base}
\begin{algorithmic}[1]
 \State  At $t=0$, start the simulation with $I_0$ infections distributed as per $\mu_0(N)$, let $X_0$ denote the disease state of the city at $t=0$. 
 
 \While {$t < T$}
 \State For each susceptible individual
 $j$, calculate $\lambda_j(t)$.  The individual's status then changes to exposed with probability  $1-\exp({-\lambda_j(t)})$.
 \State All individuals in states other than susceptible independently transition to another state according to the disease progression dynamics.
 \State Update disease state $X_t \to X_{t+1}$ of the city as described in Section \ref{sec:ABS_brief}.
 \State Update $t \to t+1$.
 \EndWhile
 \State Get performance measure of the interest/disease statistics affected population (e.g., number exposed, number hospitalised,  number of fatalities)  $[y_0,\ldots,y_T]$ from $[X_0,\ldots,X_T]$. 
 \State  The above simulation is independently repeated many times, and the average of performance measures as a function of time are reported.
\end{algorithmic}
\end{algorithm}

\noindent {\bf Scaling the model:} For $k, N \in \mathbb{N}$ with $k>1$, let $kN$ be the number of individuals in the larger model, and $N$ in the  smaller model. 
Roughly speaking, the larger model has approximately $k$ times more homes, schools and workplaces compared to the smaller model.
The joint distribution of people across homes, schools, and workplaces remains unchanged, and transmission rates $\beta^h$, $\beta^s$, $\beta^w$, and $\beta^c$ are unchanged. (Recall that each transmission rate represents the expected number of infection-spreading contact opportunities between an infected individual and all individuals in that interaction space, accounting for the combined effect of meeting frequency and the probability of infection spread during each meeting.)

When we initiate both the larger and smaller models with the same few well-distributed infections, the disease spreads similarly in homes, schools, and workplaces. To understand disease spread through communities, observe that in each community, a susceptible person experiences approximately $1/k$ times the community infection rate in the larger model compared to the smaller model. However, the larger model has $k$ times more susceptible population. The overall no. infected is roughly Poisson distributed with the same rate in the smaller and larger model. . Therefore, early in the simulation, the total number of people infected through communities is essentially identical between the larger and smaller models, and the infection processes in both models evolve very similarly.

 Let 
$t_{S}$ denote the time until which the two models evolve essentially identically. Theoretically, $t_S  =  {\log_\rho\left(\frac{N}{I_0\log (N)}\right)}$. Empirically, one can determine this by comparing the larger and smaller models under a no-intervention scenario. For notational purposes, we denote this as  $\log_\rho N^*$ for  $N^* = N/I_0\log{N}$. Here  $\log_\rho m = \log {m}/\log \rho$ for any $m\in \mathbb{R}^+$, and $\rho$ denotes the initial infection exponential growth rate.

\begin{figure}[h!]
      \centering
\includegraphics[width=\linewidth]{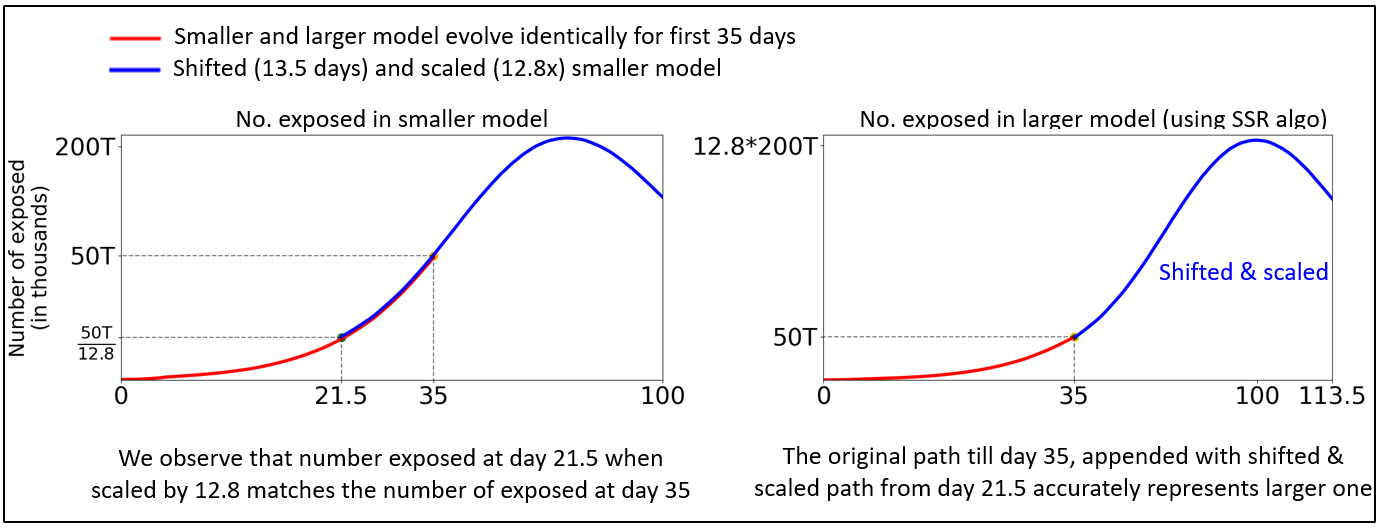}   
    \caption{ Shift-scale under no intervention scenario.} 
    \label{fig:SSR_1_final_1}
  \end{figure}

In this setting, we propose applying a shift and scale algorithm (Algorithm~\ref{alg:cap}),
which builds upon Algorithm \ref{alg:base}, to the smaller city to generate output that resembles the larger city.
 Algorithm~\ref{alg:cap} is graphically illustrated in Figure \ref{fig:SSR_1_final_1}.

\begin{algorithm}
\caption{Shift and scale algorithm in no-intervention setting}\label{alg:cap}
\begin{algorithmic}[1]
 \State   At $t=0$, start the simulation of the smaller city with $I_0$ infections distributed according  $\mu_0(N)$.
Generate the simulation sample path 
 $[X_0,X_1,...,X_{T}]$ 
 where $X_t$ denotes the disease state of population   at time $t$. 

  \State Obtain disease statistics (performance measure of interest) $[y_0,y_1,...,y_{T}]$ from $[X_0,X_1,...,X_{T}]$. 
 
 \State Suppose there are $x$ infections at $t_{S}$ in the smaller city. Determine an earlier time $t_{x/k}$ in the
  simulation when there were approximately $\frac{x}{k}$ infections in the city. 

\State  The statistics of affected population for the larger city are then obtained as:  
\[
 [y_0,y_1,...,y_{t_S}, k \times y_{t_{x/k}+1},...,k \times y_{T-( t_S-t_{x/k})}].
 \]

\State 
As in Step 9, Algorithm 1. 
\end{algorithmic}
\end{algorithm}

\begin{figure}[h!]
   \centering
   \includegraphics[width=\linewidth]{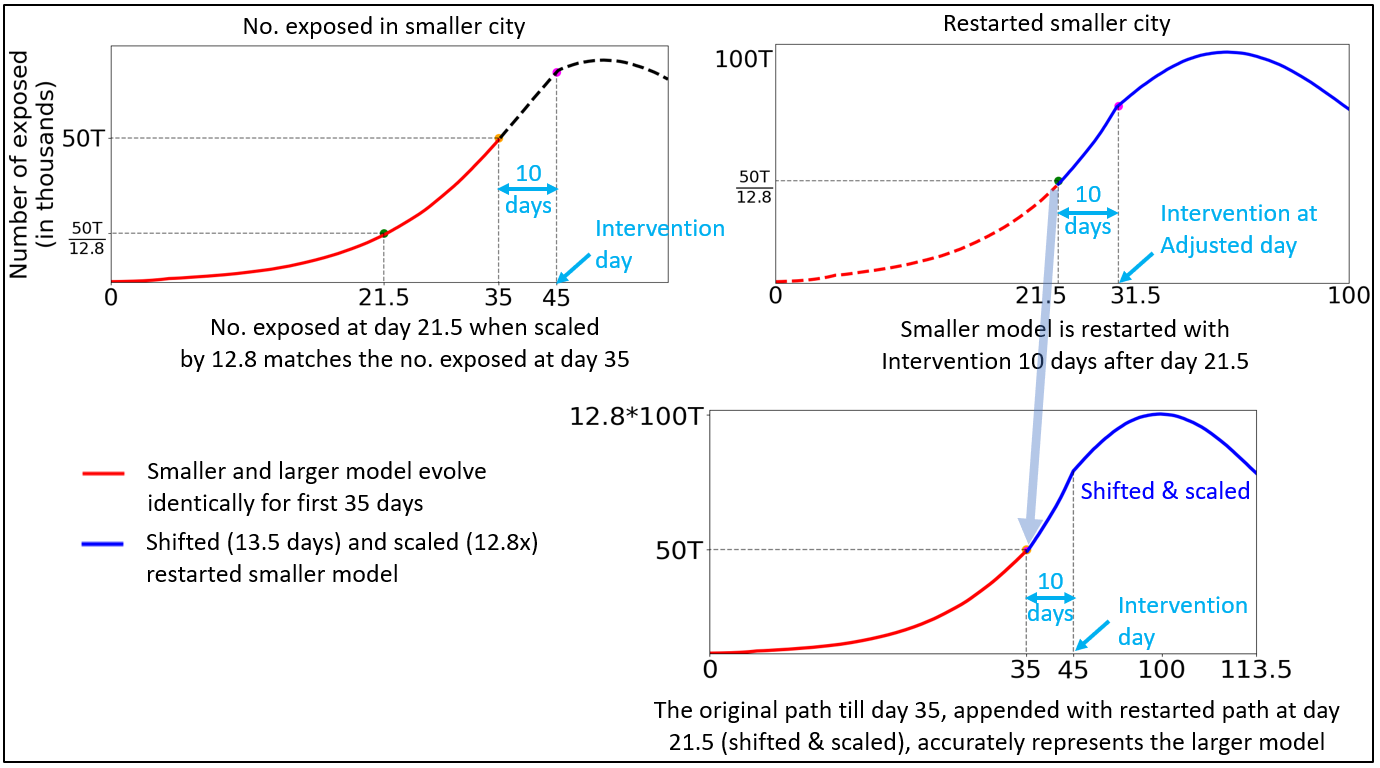}
    \caption{ Shift-scale-restart algorithm under intervention. 
    \label{fig:SSR_2_final_1}}
  \end{figure}

In a realistic scenario, as the infection spreads, the administration will intervene and impose mobility restrictions. Thus, our simulation adjustments to the small city need to account for the timing of these interventions accurately. Let $t_I$ denote the first intervention time (e.g., lockdown; typically occurring after $\beta \log_\rho N$ time for small $\beta \in (0,1)$). Let $t_{min} \approx \min\{t_I, t_S\}$. We need to restart our simulation to ensure that the shifted and scaled path incorporates the intervention at the correct time. Algorithm~\ref{alg:cap2}
achieves this and is graphically illustrated in Figure  \ref{fig:SSR_2_final_1}.

\begin{algorithm}
\caption{Shift, scale and restart algorithm}\label{alg:cap2}
\begin{algorithmic}[1]
\State   At $t=0$, start the simulation with $I_0$ infections in the smaller city distributed according to $\mu_0(N)$.
 Generate the simulation sample path  
 $[X_0,X_1,...,X_{t_{min}}]$, where $X_t$ denotes the disease state of the city at time $t$. 

 \State Suppose there are $x$ infections at $t_{min}$. Determine an earlier time $t_{x/k}$ in the
  simulation when there where approximately $\frac{x}{k}$ infections in the city. 

 \State Restart 
 a new simulation of the city using common random numbers, but with the intervention introduced 
 at time $t_{\frac{x}{k}}+t_{I}-t_{min}$. Simulate it up to time $T-(t_{min}-t_{x/k})$. Denote the simulation path of this restarted simulation as 
\[[Z_1,Z_2,...,Z_{t_{x/k}},...,Z_{T-(t_{min}-t_{x/k})}].\]

 \State Compute the statistics from the simulation path  $[y_0,y_1,...,y_{t_{min}}]$ from  $[X_0,X_1,...,X_{t_{min}}]$.
Denote the time series of statistics obtained from the  restarted simulation
by $[z_0,z_1,...,z_{t_{x/k}},...,z_{T-(t_{min}-t_{x/k})}]$ using  
$[Z_0,Z_1,...,Z_{t_{x/k}},...,Z_{T-(t_{min}-t_{x/k})}]$. 

 \State  The approximate statistics of the affected population for the larger city are then obtained as:  
 \[
 [y_1,y_2,...,y_{t_{min}},k \times z_{t_{x/k}+1},...,k \times z_{T+t_{x/k}-t_{min}}]
 .\]

\State 
As in Step 9, Algorithm 1.

\end{algorithmic}
\end{algorithm}

As we observe empirically, and as suggested by Proposition \ref{initial_proposition}, in Algorithm \ref{alg:cap2}, the evolution of the infection process after time $t_{x/k}$, given the state at that time, is more or less deterministic (even though the process may be close to the associated branching process at this time).

\section{Asymptotic Analysis} 
\label{sec:Theoretical_results}

In the SSR algorithm we need a rigorous theoretical justification for the fact that during early phase of the outbreak epidemic, one can take the infection path from small-city simulation at certain time period, scale it, and stitch it to the path at another appropriately chosen
time period, to accurately generate the trajectories for the larger city. 
We provide this justification through an asymptotic analysis
as the city population increases to infinity. Specifically, we prove  that in the initial disease spread phase,  the epidemic process is close to a multi-type super-critical branching process by coupling the two processes.
Further we prove that, once a sufficiently large number of people have been infected, the subsequent evolution of the epidemic is well approximated by its mean‑field limit with a random starting state.

 In the following analysis, we first formalize and define the epidemic process in Section \ref{sec:epidemic_process}. Further, we define a multi-type super-critical branching process tailored to the epidemic process in Section \ref{sec:associate_bp}. We describe the coupling between the two processes  in Section \ref{sec:defining_coupling}.  We then state the results demonstrating
 the closeness of the epidemic and the branching process in the early disease spread phase  in Section \ref{sec:branching_results}. The results demonstrating the closeness
 of the epidemic process to its mean field limit once the epidemic process has grown are given in   Section \ref{sec:deterministic_results}. Finally, in Section \ref{sec:bias_naive_scaling}, we provide theoretical justification for why naively scaling the smaller model underestimates the infections in larger city.

\subsection{Epidemic process dynamics}
\label{sec:epidemic_process}

Recall that the synthetic city is generated in two stages. First, we instantiate individuals, households, schools, workplaces, and communities from demographic statistics.  
Second, we assign the individuals randomly to these units.  
As in Algorithm~\ref{alg:cap} (Section~\ref{sec:ssr_algo}), the household, school, and workplace size distributions are the same in both the large and the small city models, so the numbers of homes, schools, and workplaces scale linearly with population size.  
The number of communities, however, is kept fixed; hence the number of individuals per community grows proportionally with the population.

This construction yields two different kinds of interaction networks for an individual:
\begin{enumerate}
    \item \emph{Home, school, and workplace contacts.}  
   Because the interactions in these places depend only on the sizes of units, these remain same in both larger and smaller city.
   \item \emph{Community contacts.}  
   In the larger city each individual meets more people in the community, but the per‑pair contact probability is proportionally lower (see Section~\ref{sec:ABS_brief}); the converse holds in the smaller city.
\end{enumerate}

To capture both interaction types, we analyze a \textit{group‑structure} city model: the population is partitioned into $N$ groups, each representing a local network of households whose children may attend the  schools and whose adults work in the workplaces within the group. All home, school, and workplace interactions occur within a group, whereas inter‑group interactions occur only through community contacts. Thus an infectious individual can transmit to another group solely via the community infection. We examine an asymptotic sequence of cities indexed by $N$, the total number of groups. As $N$ grows, the numbers of individuals, homes, schools, and workplaces all grow linearly with $N$.

More concretely, each group  has a fixed number of individuals $g$ in that group. The individuals within a group can have different age, disease progression profile (e.g., some may be more infectious than others), transmission rates (e.g., individuals living in congested areas may be modelled to have higher transmission rates), mobility in the community (e.g., elder population may travel less to the community compared to the working age population), distances from the centre of the community, belong to different interaction spaces (workplaces, schools, homes). Across groups, however, individual $j \in [g]$ is probabilistically identical, that is their attributes and interaction patterns are identically distributed in every group.

\begin{remark}
{  In principle and in our simulations, the population graph in the whole city might be connected to each other through schools, workplaces and homes (apart from being connected through community).  This implies that the state space of the population network might go to $\infty$ as $N\to\infty$. However, by assuming the group structure, we ensure that state of the city takes fixed, finitely many values
as $N\to\infty$.  This finiteness is crucial for tractable analysis while preserving the model’s epidemiological richness.
Further, extending our analysis to cities with multiple communities  and heterogeneous groups, each representing a different network of individuals and their home, school and workplace,  is straightforward only adding to the notational complexity,  so long as the overall state space remains finite. } 
\end{remark}

\subsubsection{Notation}

\begin{itemize}
\item The city comprises of $N$ groups, indexed by $n \in [N]:=\{1,\ldots, N\}$.

\item Each group contains $g$ individuals. 
Individual $j\in[g]:=\{1,\ldots,g\}$ in group $n$ is denoted by $(j,n)$. Individuals with the same first index $j$ are probabilistically identical across groups.

\item
Each individual can be in one of the following disease states: Susceptible (S), exposed (E), infective (I), symptomatic (Sy), hospitalised (H), critical (C), dead (D) or recovered (R).  Individuals are infectious only in infective or symptomatic states. Denote all the disease states  by $\{S,E,I,\ldots\}:=\mathcal D$. For simplicity, we ignore possible reinfections and vaccinations, although incorporating them   would not alter our conclusions.

\item
Each group may be in disease state  $ {\bf s}:=\{s_1,\ldots,s_g\} \in \mathcal D^{g}$, based on the disease states of each individual $j\in[g]$. Let $\mathcal S =  {\mathcal D}^{g}$ be the set of possible group disease states.

\item   Let $u\in\mathcal S$ denote the group state where all the
    individuals  are susceptible, and set $\mathcal U:=\{u\}$. Thus, $\mathcal S \setminus \mathcal U$ denote the set of group states where atleast one individual has already been affected (that is, has been exposed to the disease at some point in the past).
Further, let $\eta=\abs{\mathcal S \setminus \mathcal U}$. 

\item For each group $n$, let $S_n(t)$ denote its state at time $t$. 

\item  For $t\in \mathbb{N}$, let  $X_t^N({\bf s})$  be  the number of groups of type ${\bf s}$ at time $t$,  and set 
${\bf X}_t^N =( X_t^N({\bf s}): {\bf s} \in \mathcal S \setminus \mathcal U) \in {\mathbb{Z^+}}^\eta$. Note that for ${\bf s} \in \mathcal S \setminus \mathcal U$,
\[X_{t}^N({\bf{s}}) = \sum_{n=1}^N {\mathbf 1}\{S_{n}(t)={\bf s}\} \quad \forall \, {\mathbf{s}}\in\mathcal S \setminus \mathcal U\]

\item

Define $A_t^N = \sum_{{\bf s}\in\mathcal S \setminus \mathcal U } X_t^N({\bf s})$ to be the total number of affected groups in the system $N$ at time $t$.

\end{itemize}

\subsubsection{Dynamics}

At time zero, for each $N$,  ${{\bf X}_0^N}$ is initialised by setting a suitably selected small and fixed  number of people 
randomly from some distribution $\mu_0(N)$ and assigning them 
 to the exposed state. All other individuals  are set as susceptible. 
 The distribution $\mu_0(N)$ is assumed to be independent of $N$, so we can set ${{\bf X}_0^N} = {\bf X}_0$  for all $N$.   Given ${{\bf X}_t^N}$, the state ${{\bf X}_{t+\Delta t}^N}$ at time $t+\Delta t$  is  arrived at  through two mechanisms (for the ease of notation we will set $\Delta t =1$):

 \begin{enumerate}
 \vspace{3mm}
     \item \textbf{\emph{Disease Transmission:}}  Infectious individuals at time $t$  make Poisson distributed infectious contacts with 
 the rest of the population (at homes, schools, workplaces within each individual's group and in the community) moving the contacted susceptible population to exposed state. Specifically,  
     \begin{enumerate}
         \item \emph{Community transmission:}  An infectious individual with index $j$ spreads the disease in the community with transmission rate $\beta_{j}^c$. That is, the total number of infectious contacts it makes  with all the individuals (both susceptible and affected)  in one time step is Poisson distributed with rate $\beta_{j}^c$.

      Further, the fraction of contacts hitting individuals with index $\tilde{j}$ is $\psi_{j,\tilde{j}}$ (with $\sum_{\tilde{j}=1}^g\psi_{j,\tilde{j}}=1$), and each such contact is made with an individual $\tilde{j}$ uniformly chosen group among the $N$ groups.      ${\psi_{j,\tilde{j}}}$  helps model biases such as different mobility of individuals across within a community, for e.g., 
          individuals in dense regions living are more likely to infect other.

         \item \emph{Local transmission:} For an infectious individual $j$, the local transmission rates are $\beta_{j}^h$ in homes,   $\beta_{j}^s$ in schools and $\beta_{j}^w$ in workplaces. Thus, the number of infectious contacts an infectious individual $j$ makes with each individual in his home (school, workplace) in one time step is Poisson distributed with rate $\frac{\beta_{j}^h}{n^h_{j}}$ ($\frac{\beta_{j}^s}{n^s_{j}}$, $\frac{\beta_{j}^w}{n^w_{j}}$), where 
 $n^h_{j}$( $n^s_{j}$, $n^w_{j}$) is the number of individuals in the home (school, workplace) 
 of individual $j$. Note that these individuals belong to the same group.

 \item  \emph{New exposures:} A susceptible individual that receives at least one infectious contact becomes exposed at $t+1$; contacts to already affected individuals leave their state unchanged.
     \end{enumerate}
     \vspace{3mm}
 \item \textbf{\emph{Disease Progression:}}        After exposure, the disease progression of each individual is independent of all other individuals. The waiting time in each state (except susceptible, dead and recovered) is assumed to be geometrically distributed, with transition probabilities determined by that individual’s characteristics (disease progression profile).
Let $Q_j(\cdot|s)$ be the one–step transition kernel for an index-$j$ individual
(geometric waiting times yield such a Markov kernel), recovered and dead are absorbing.

 \vspace{3mm}
 
\item \textbf{Formalizing the epidemic dynamics:}
Suppose individual $(j,n)$ gets infected at time $\tau_{j,n}$. For each $\zeta \in \{1,2,\ldots\}$ time step after $\tau_{j,n}$, define the following random variables:
\begin{align*}
    s_{j,n}(\zeta) & := \text{ disease state with } s_{j,n}(\zeta) \sim Q_j(\cdot| s_{j,n}(\zeta-1))
    \\ 
    \{C_{j,n}^{h}(\zeta,\tilde{j}):\tilde j\in[n^h_j]\} & := \text{ no. of infectious contacts $j$ makes with individual $\tilde{j}$ in home ($\zeta$ steps after $\tau_{j,n}$).}
\end{align*}
where $C_{j,n}^{h}(\zeta,\tilde{j}) \sim Poi\left({\beta^h_j }/{n^h_j} \right) $  if individual $(j,n)$ is  
infectious  at $\zeta-1$ time steps after $\tau_{j,n}$,  and $0$ otherwise.
 Analogously define the infectious contacts made at school and workplace: \[ \{C_{j,n}^{s}(\zeta,\tilde{j}):\tilde j\in[n^s_j]\} \quad \text{and} \quad \{C_{j,n}^{w}(\zeta,\tilde{j}):\tilde j\in[n^w_j]\}. \] 
Additionally, the number  of infectious contacts $j$ makes with all individuals $\tilde{j}$ in community ($\zeta$ time steps after exposure) is denoted by 
\begin{align*}
 \{C_{j,n}^{c}(\zeta,\tilde{j}):\tilde j\in[g]\}  
\end{align*}
where $C_{j,n}^{c}(\zeta,\tilde{j}) \sim Poi\left( {\beta^c_j \psi_{j,\tilde{j}}}  \right)$ if individual $(j,n)$ is 
 infectious  at $\zeta-1$ steps after $\tau_{j,n}$,  and $0$ otherwise. Moreover, for each such community contact $k \in [C_{j,n}^{c}(\zeta,\tilde{j})] $, the group selected for that  contact is denoted by   $U_{j,n}^{c}(\zeta,\tilde{j},k)$, which is drawn uniformly from $\{1,\ldots, N\}$.

At time $t$, individual $(\tilde j,\tilde n)$ receives $ C^{\mathrm{all}}_{\tilde{j}, \tilde{n}}(t)$, total number of infectious contacts, 
\begin{align*}
 C^{\mathrm{all}}_{\tilde{j}, \tilde{n}}(t) &=     \sum_{n=1}^N\sum_{j=1}^g \sum_{k=1}^{C_{j,n}^c(t-\tau_{j,n}, \tilde{j})} {\mathbf 1}\{\tau_{j,n}<t\} {\mathbf 1}\left\{U_{j,n}^{c}(t-\tau_{j,n},\tilde{j},k)=\tilde{n}\right\}
    +\sum_{j:h(j)=h(\tilde{j})} {\mathbf 1}\{\tau_{j,\tilde{n}}<t\} C_{j,\tilde{n}}^h(t-\tau_{j,\tilde{n}}, \tilde{j}) \\ & +\sum_{j:w(j)=w(\tilde{j})} {\mathbf 1}\{\tau_{j,\tilde{n}}<t\} C_{j,\tilde{n}}^w(t-\tau_{j,\tilde{n}}, \tilde{j})  
    +\sum_{j:s(j)=s(\tilde{j})} {\mathbf 1}\{\tau_{j,\tilde{n}}<t\} C_{j,\tilde{n}}^s(t-\tau_{j,\tilde{n}}, \tilde{j}) 
\end{align*}

Thus an individual $(\tilde j,\tilde n)$ in state $s$ at time $t$ transitions to $s'$ at time $t+1$ as
\[s' =\begin{cases}
S 
 &  C^{\mathrm{all}}_{\tilde{j}, \tilde{n}}(t) = 0, ~~ s=S \\[1ex]
    E 
 &  C^{\mathrm{all}}_{\tilde{j}, \tilde{n}}(t)\geq 1, ~~ s=S\\[1ex]
s' &  s'\sim Q_j(\cdot|s), ~~ s\in \mathcal D\setminus\{S\}.
\end{cases}\]
For a group $\tilde n$ in state $S_{\tilde n}(t)$ at time $t$, the next state at time $t+1$ is $S_{\tilde n}(t+1)$. Each individual $\tilde j\in[g]$ transitions according to the rules described above. Thus, if
$S_{\tilde n}(t)=\mathbf s:=\{s_1,\ldots,s_g\}$, then
$S_{\tilde n}( t+1)=\mathbf s':=\{s_1',\ldots,s_g'\}$. Hence
\[X_{t+1}^N({\bf{s}}) = \sum_{n=1}^N {\mathbf 1}\{S_{n}(t+1)={\bf s}\} \quad \forall \, {\mathbf{s}}\in\mathcal{D}^g \setminus \mathcal{U}\]

\vspace{3mm}

\item
\noindent \textbf{Transition probability of ${\bf X}^N_t$:} 
Given the global count vector $\{X_t^N({\bf s}):{\bf s}\in\mathcal S\setminus \mathcal{U}\}$, the \emph{community} infection rate to a fixed individual of index $\tilde{j}$ at time $t$ is:
\[ \lambda^{\mathrm{c},N}_{t}(\tilde j)= \sum\limits_{{\bf q}\in \mathcal S \setminus  \mathcal U} \frac{X^N_{t}({\bf q})}{N} \sum_{{j} =1 }^g {\bf{1}}({\bf{q}}({j})= \text{infectious}) \psi_{{j},\tilde{j}} \beta^c_{{j}},\]
where ${\bf{1}}({\bf{q}}({j})= \text{infectious})$ indicates whether  individual ${j}$ in group disease state $\bf{q}$ is infectious or not. 
For an individual $\tilde{j}$ in a group with disease state $\bf s$, home infection rate is (this depends only on the disease state $\bf s$ of the group):
\[\lambda^{{h}}(\tilde j|{\bf s}) = \sum_{j:h(j)=h(\tilde{j})} 1({\bf s}(j)= \text{infectious}) \frac{\beta^h_j}{n^h_j}\]

Similarly we can define the infection rates from school and workplace.
Hence the total one–step infection rate to a susceptible of index $\tilde j$ in a group state $\bf s$ is
\[
\lambda^{N}_{t}(\tilde j, {\bf s})\;:=\;\lambda^{\mathrm{c},N}_{t}(\tilde j)+\lambda^{h}({\tilde j}|{\bf s})+\lambda^{s}({\tilde j}|{\bf s})+\lambda^{w}({\tilde j}|{\bf s}).
\]
By Poisson superposition and thinning, contacts to distinct recipients are conditionally independent given ${\bf X}_t^N$. The probability that  individual $\tilde{j}$ in group state $\bf{s}$ receives atleast one infectious contact in one step is
$p^{N}_{t}(\tilde j, {\bf s})\;:=\;1-e^{-\lambda^{N}_{t}(\tilde j,{\bf s})}$. 
Therefore, the individual one-step transition kernel is
\[ \pi_{t}^N(s'\mid \tilde{j}, {\bf s})\;=\;\begin{cases}
e^{-\lambda_{t}^N(\tilde{j},{\bf s})}\,\mathbf 1\{s'=S\}
+\bigl(1-e^{-\lambda_{t}^N(\tilde{j},{\bf s})}\bigr)\,\mathbf 1\{s'=E\}, & {\bf s}(\tilde{j})=S,\\[1ex]
Q_j(s'|{\bf s}(\tilde{j})), & {\bf s}(\tilde{j})\in \mathcal D\setminus\{S\}.
\end{cases}
\]
This is a row–stochastic kernel on $\mathcal S$, and it depends on $X_t^N$ only through $\lambda_{t}^N(\tilde{j},{\bf s})$.
For a single group in state ${\bf s}=(s_1,\dots,s_g)$ at time $t$, the probability of next state ${\bf s}'=(s_1',\ldots,s_g')$ at $t+1$ is

\[
P_t^N({\bf s'}|{\bf s})\;=\;\prod_{\tilde{j}=1}^g \pi_{t}^N({\bf s'}(\tilde{j})\mid \tilde{j}, {\bf s}),
\]
Hence, 
\[
\Exp{X_{t+1}^N({\bf s'})\mid X_t^N}
\;=\;\sum_{{\bf s}\in\mathcal S} X_t^N({\bf s})\,P_t^N({\bf s'}|{\bf s}).
\]

 \end{enumerate}

\subsection{The associated branching process dynamics}
\label{sec:associate_bp}

Before describing the branching process tailored to epidemic dynamics, we first provide a brief review of the multi-type branching process \citep{branching_process_notes, harris1963theory}.

  \vspace{3mm}
\noindent \textbf{Brief review:} 
A multi-type branching process is a probabilistic model for population growth (or eventual extinction) where the population consists of several different types of individuals. Formally, consider a probability space $(\Omega,\mathcal{F},\mathbb{P})$ and fix an integer $\eta\ge 1$. A multi-type branching process is a stochastic process $\mathbf{B}=(\mathbf{B}_t)_{t\ge 0}$ with state space $\mathbb{N}^{\eta}$, where $\mathbf{B}_t$ is an $\eta$-dimensional vector whose $i$-th component denotes the number of individuals of type $i$ at time $t$. In a multi-type branching process, at the end of each time period, an individual may give birth to children of different types before dying.

The number of children produced by each individual of type $i$ is independent and identically distributed.  
For each type $i\in\{1,\ldots,\eta\}$, let $\boldsymbol{\xi}^{(i)}=(\xi^{(i)}_1,\ldots,\xi^{(i)}_\eta)$ be an $\mathbb{N}^{\eta}$-valued random vector describing the offspring counts of a single parent of type $i$, with law
\[
p^{(i)}_{\mathbf{k}}
= \mathbb{P}\!\big(\boldsymbol{\xi}^{(i)}=\mathbf{k}\big),
\qquad \mathbf{k}\in\mathbb{N}_0^{\eta}.
\]
 If $\mathbf{B}_t = (b_1, \ldots, b_\eta)$, then $\mathbf{B}_{t+1}$ represents the sum of independent offsprings from $b_1$ type-1 parents, $b_2$ type-2 parents, and so forth. Thus, $\mathbf{B}_{t+1}$ is the sum of $b_1 + b_2 + \cdots + b_\eta$ independent random vectors in $\mathbb{N}^{{\eta}}$. Specifically, 
\begin{equation*}
  \mathbf{B}_{t+1}
  \;=\;
  \sum_{i=1}^{\eta}\;\sum_{m=1}^{b_i}\boldsymbol{\xi}^{(i)}_{t,m},
\end{equation*}
where $\boldsymbol{\xi}^{(i)}_{t,m}$ are i.i.d.\ copies of $\boldsymbol{\xi}^{(i)}$ and the entire collection $\{\boldsymbol{\xi}^{(i)}_{t,m}: t\ge 0,\ m\ge 1,\ 1\le i\le \eta\}$ is mutually independent.
Consequently, the multi-type branching process forms a Markov chain $({\bf {B}}_t \in \mathbb{N}^{\eta}:t \geq 0)$. A comprehensive review of super-critical multi-type branching processes and standard convergence results for $B_t$ (appropriately normalized) is provided in Appendix \ref{sec:define_standard_MTBP}.

\vspace{3mm}
\noindent\textbf{Branching process tailored to epidemic process:} For the branching process tailored to epidemic dynamics, our fundamental unit is the group, whose different possible disease states form the different types in the standard branching process framework. As in the epidemic process, each group contains $g$ individuals, and each individual $j \in [g]:=\{1,\ldots,g\}$ is probabilistically identical to its counterpart in the epidemic process.

For each time $t$, let ${\bf B}_t=(B_t({\bf s}))_{{\bf s}\in\mathcal S\setminus\mathcal U}\in\mathbb Z_+^\eta$ represent the number of groups in each state $\mathbf{s}$ at time $t$ in the branching process. Let $A_t^B = \sum_{{\mathbf s}\in\mathcal S \setminus \mathcal U  } B_t({\bf s})$ denote the total number of affected groups up to time $t$.

\subsubsection{Dynamics} At time zero ${{\bf B}_0} = {{\bf X}_0}$. Given ${\bm {B}_t}$, we arrive at ${\bm {B}_{t+1}}$ through the following mechanisms: 

\begin{enumerate}
\vspace{3mm}
    \item \textbf{\emph{Births:}} 
 At time $t$, every infectious individual $j$ gives birth to independent Poisson-distributed offspring groups in state ${\bf e}_{\tilde{j}}$ (representing a group with individual $\tilde{j}$ exposed  and all other individuals susceptible) at time $t+1$  with rate $ (\psi_{j,\tilde{j}}\beta^c_{j})$  for each $\tilde{j}$. The count ${\bm {B}_{t+1}}(\bf s)$ is increased accordingly. 

\vspace{3mm}
\item \textbf{\emph{Within‑group transitions:}}   A group in state $\bf s \in  \mathcal S \setminus \mathcal U$ can undergo transition to state $\bf \tilde{s} \in \mathcal S \setminus \mathcal U$  through  two processes:

\begin{enumerate}
    \item \textit{Infectious contacts within groups:}  An infectious individual $j \in [g]$ makes infectious contacts  with each individual in their home (school, workplace) in one time step. These contacts follow a Poisson distribution with rate $\frac{\beta_{j}^h}{n^h_{j}}$ (or $\frac{\beta_{j}^s}{n^s_{j}}$, $\frac{\beta_{j}^w}{n^w_{j}}$), where 
 $n^h_{j}$ ( $n^s_{j}$, $n^w_{j}$) represents the number of individuals in home (school, workplace) 
 of individual $j$.  Susceptible individuals may become exposed through these infectious contacts.

 \item \textit{Disease progression:}  An already affected individual may transition from one disease state to another, following the same transition probabilities as in the epidemic process.

\end{enumerate}

\vspace{3mm} 
\item \textbf{Formalizing the disease progression:}  As new groups are born in the branching process, we assign each a unique index. The full details of this indexing strategy are provided in Appendix~\ref{sec:defining_branching_process_rv}. For the ease of notation, we denote this group index by $(\cdot)$, and an individual within such a group by  $(j, \cdot)$. If individual $(j,\cdot)$ is infected at time $\tau_{j,\cdot}$, then for each $\zeta\in\{1,2,\ldots\}$ time steps after $\tau_{j,\cdot}$, we define following random variables:
 \[s_{j,\cdot}(\zeta), ~~   \{C_{j,\cdot}^{h}(\zeta,\tilde{j}):\tilde j\in[n^h_j]\}, ~~ \{C_{j,\cdot}^{w}(\zeta,\tilde{j}):\tilde j\in[n^w_j]\} ~ \text{ and } ~ \{C_{j,\cdot}^{s}(\zeta,\tilde{j}):\tilde j\in[n^s_j]\}\] 
 
 which are defined analogously to the epidemic process. For community ``births'', we define:
\begin{align*}
 \{C_{j,\cdot}^{c}(\zeta,\tilde{j}):\tilde j\in[g]\} & := \text{ no.  of births individual $j$ gives to new ${\bf{e}}_{\tilde{j}}$ groups ($\zeta$ time steps after exposure).}
\end{align*}
where $ C_{j,\cdot}^{c}(\zeta,\tilde{j}) \sim Poi\left( {\beta^c_j \psi_{j,\tilde{j}}}  \right) $  if individual $(j,\cdot)$ is 
infectious  at $\zeta-1$ time steps after $\tau_{j,\cdot}$, and $0$ otherwise.
Here,  ${\bf e}_{\tilde j}$ denotes a group state  where individual $\tilde{j}$ is exposed while all other are susceptible.

In Appendix~\ref{sec:defining_branching_process_rv}, we also introduce auxiliary variables $U^{c}_{j,\cdot}(\zeta,\tilde j,k)\sim\mathrm{Unif}\{1,\ldots,N\}$ (for $k=1,\ldots, C^{c}_{j,\cdot}(\zeta,\tilde j)$) to index these newborn groups.  This indexing is not intrinsic to the branching process itself, but serves as a bookkeeping device to facilitate coupling with the epidemic process on a population of size $N$.

At time $t$, an individual $(\tilde j,\cdot)$ receives the following number of \emph{local} infectious contacts (home, school, workplace):
\begin{align*}
 C^{\mathrm{lc}}_{\tilde{j}, \cdot}(t) &=   \sum_{j:h(j)=h(\tilde{j})} {\mathbf 1}\{\tau_{j,\cdot}<t\} C_{j,\cdot}^h(t-\tau_{j,\cdot}, \tilde{j}) +\sum_{j:w(j)=w(\tilde{j})} {\mathbf 1}\{\tau_{j,\cdot}<t\} C_{j,\cdot}^w(t-\tau_{j,\cdot}, \tilde{j})  \\ &  
    +\sum_{j:s(j)=s(\tilde{j})} {\mathbf 1}\{\tau_{j,\cdot}<t\} C_{j,\cdot}^s(t-\tau_{j,\cdot}, \tilde{j}) 
\end{align*}

Thus, an individual in state $s$ at time $t$ transitions to $s'$ at $t+1$ as
\[s' =\begin{cases}
S 
 &  C^{\mathrm{lc}}_{\tilde{j}, \cdot}(t) = 0, ~~ s=S \\[1ex]
    E 
 &  C^{\mathrm{lc}}_{\tilde{j}, \cdot}(t)\geq 1, ~~ s=S,\\[1ex]
s' &  s'\sim Q_j(\cdot|s), ~~ s\in \mathcal D\setminus\{S\}.
\end{cases}\]
For a group in state $S_t(\cdot)={\bf s}=\{s_1,\ldots,s_g\}$, the next state is $S_{t+1}(\cdot)={\bf s}'=\{s_1',\ldots,s_g'\}$ is determined by the disease transition of each individual within that group. Consequently,
\begin{align*}
    \begin{cases}
        B_t({\mathbf{e}}_{\tilde{j}}) = \sum_{(\cdot) \in A_t^B} {\mathbf 1}\{S_{(\cdot)}(t+1)={\mathbf{e}}_{\tilde{j}}\} + \sum_{(\cdot) \in A_t^B} \sum_{j=1}^g {\mathbf{1}(\tau_{j,\cdot}<t) } C_{j,\cdot}^{c}(\zeta,\tilde{j})  & \quad \quad   \,\tilde{j} \in [g] \\
        B_t({\bf{s}}) = \sum_{(\cdot) \in A_t^B} {\mathbf 1}\{S_{(\cdot)}(t+1)={\bf s}\} & \quad {\mathbf{s}}\in\mathcal{S} \setminus \left(\mathcal{U} \cup \{{\mathbf{e}}_{\tilde{j}}:\tilde{j} \in [g]\} \right)
    \end{cases}
\end{align*}

\vspace{3mm}
\item \textbf{Transition probability of ${\bf B}_t$:}  Given ${\bf B}_t$, the \emph{expected} number of newborn groups of type ${\bf e}_{\tilde j}$ over $(t,t+1]$ is:
\[ \sum_{{\bf s}\in\mathcal S} B_t({\bf s})\sum_{j=1}^g {\bf 1}\{{\bf s}(j)=\text{infectious}\} \psi_{j, \tilde{j}}\beta^c_j \quad \text{for all } \tilde{j} \in [g]\]

For an individual $\tilde j$ in a group of state ${\bf s}$, the local (within-group) infection rate is 
\[
\lambda^{\mathrm{lc}}(\tilde j, {\bf s})\;:=\;\lambda^{h}({\tilde j}|{\bf s})+\lambda^{s}({\tilde j}|{\bf s})+\lambda^{w}({\tilde j}|{\bf s}),
\]
and the probability that $\tilde{j}$ individual in group disease state $\bf{s}$ receives atleast one infectious contact is 
$p^{\mathrm{lc}}_{t}(\tilde j, {\bf s})\;:=\;1-e^{-\lambda^{\mathrm{lc}}_{t}(\tilde j,{\bf s})}$. 
Therefore, the individual local transition kernel is
\[ \pi_{t}^{lc}(s'\mid \tilde{j}, {\bf s})\;=\;\begin{cases}
e^{-\lambda_{t}^{\mathrm{lc}}(\tilde{j},{\bf s})}\,\mathbf 1\{s'=S\}
+\bigl(1-e^{-\lambda_{t}^{lc}(\tilde{j},{\bf s})}\bigr)\,\mathbf 1\{s'=E\}, & {\bf s}(\tilde{j})=S,\\[1ex]
Q_j(s'|{\bf s}(\tilde{j})), & {\bf s}(\tilde{j})\in \mathcal D\setminus\{S\}.
\end{cases}
\]
This induces a row-stochastic kernel on $\mathcal S$ that does not depend on ${\bf B}_t$. For a group in state ${\bf s}=(s_1,\ldots,s_g)$ at time $t$, the probability of the next state ${\bf s}'=(s_1',\ldots,s_g')$ is
\[
P_t^{\mathrm{lc}}({\bf s'}|{\bf s})\;=\;\prod_{\tilde{j}=1}^g \pi_{t}^{\mathrm{lc}}({\bf s'}(\tilde{j})\mid \tilde{j}, {\bf s}).
\]
Therefore,  for $\tilde j\in[g]$,
\begin{align*}
    \Exp{B_{t+1}({\bf{e}}_{\tilde{j}})\mid B_t}
\;=\;\sum_{{\bf s}\in\mathcal S} B_t({\bf s})\,P_t^{\mathrm{lc}}({\bf{e}}_{\tilde{j}}|{\bf s}) + \sum_{{\bf s}\in\mathcal S} B_t({\bf s})\sum_{j=1}^g {\bf 1}\{{\bf s}(j)=\text{infectious}\} \psi_{j, \tilde{j}}\beta^c_j,
\end{align*}
and for ${\bf s}'\in(\mathcal S\setminus\mathcal U)\cup\{{\bf e}_{\tilde j}:\tilde j\in[g]\}$,
\begin{align*}
    \Exp{B_{t+1}({\bf s'})\mid B_t}
\;=\;\sum_{{\bf s}\in\mathcal S} B_t({\bf s})\,P_t^{\mathrm{lc}}({\bf s'}|{\bf s}) & \text{for all } {\mathbf{s'}}\in\mathcal{S} \setminus \left(\mathcal{U} \cup \{{\mathbf{e}}_{\tilde{j}}:\tilde{j} \in [g]\}\right).
\end{align*}

 \vspace{3mm}
\item  \textbf{The expected offsprings matrix  $K$ for ${{\bf B}_t}$:} Let each entry  ${K}({\bf s},{\bf q})$ of the matrix represent the expected number of type ${\bf q}$  offsprings of a single type ${\bf s}$ group in one time step. Let  ${\bf{e}}_{j} \in \mathcal S$ denote the group type with individual $j\in [g]$ in exposed state and the remaining  individuals in susceptible state. 

We further define the following key sets:
\begin{itemize}
    \item $\mathcal{H} \subset \mathcal{S}$: group states with at least one individual who is infectious or may become infectious in subsequent time steps (i.e., types with at least one individual in exposed, infective, or symptomatic state),  and $\hat{\eta} = |\mathcal{H}|$.
    \item $\mathcal{\tilde{H}} \subset \mathcal{S}$: all group types with at least one individual in an infectious disease state (i.e., infective or symptomatic state).

   \end{itemize}

As described above, a group in state ${\bf s}\in \mathcal S \setminus \mathcal U$,  may give birth to groups  ${\bf{e}}_{j}$ if it contains an infectious individual, and/or may itself transition to another type in one time step. The matrix $K$ can be written as:

\begin {equation}
\begin{aligned}
K({\bf s},{{\bf e}_{\tilde{j}}}) &= P^{\mathrm{lc}}({{\bf e}_{\tilde{j}}}|{\bf s}) + \sum_{j: j \text{ is infectious in state $\bf s$} } \psi_{j,\tilde{j}}\beta^c_{j} \quad \quad   ~ {\bf s} \in  \mathcal{\tilde{H}}, ~  \tilde{j} \in [g] 
\\
 K({\bf s},{\bf q}) &= P^{\mathrm{lc}}({\bf q}|{\bf s}) \quad \text { otherwise.}  
\end{aligned}
\label{K_definition}
\end{equation}

\end{enumerate}

We observe that,
\begin{equation}
        \Exp{{\bf B}_{t+1}} =  K^T \Exp{ {\bf B}_t } = (K^T)^{t+1} \Exp{ {\bf B}_0 }.
    \label{branching_basic_eq}
\end{equation}
 
Standard convergence results (Appendix \ref{sec:define_standard_MTBP}) for super-critical multi-type branching processes require that the expected offspring matrix $K$ be irreducible. However, $K$ as defined above is not irreducible. 
 Lemma \ref{matrix_structure} below sheds further light on the structure of $K$, and 
  Theorem \ref{theorem_extending_branching_process} demonstrates that standard conclusions continue to hold for the branching process $\{\mathbf{B}_t\}$ associated with epidemic processes $\{\mathbf{X}_t^N\}$.
 
 For any matrix $M$, let $\rho(M)$ denote its spectral radius, that is, the maximum of the absolute values of all its eigenvalues. Henceforth, we assume that $\rho = \rho(K) > 1$ in all subsequent analysis. 

\vspace{3mm}
\begin{lemma} 
Consider  $K$ define in  (\ref{K_definition}).  There exist matrices $K_1 \in {\mathbb {R}^+}^{\hat\eta\times\hat\eta}$,  $C\in{\mathbb {R}^+}^{(\eta-\hat\eta)\times(\eta-\hat\eta)} $ and $M \in {\mathbb {R}^+}^{(\hat{\eta})\times(\eta-\hat\eta)}$ such that 
\[
K = 
\begin{pmatrix}
K_1 & M \\
0   & C
\end{pmatrix},
\]
where $K_1 \in {\mathbb{R}^+}^{\hat\eta\times\hat\eta}$  is  irreducible.
Furthermore, $\rho(C)  \leq \rho(K_1)$. 
\label{matrix_structure}
\end{lemma}

\noindent  Since $K_1$ defined in Lemma \ref{matrix_structure} is irreducible, its Perron-Frobenius eigenvalue equals its spectral radius $\rho(K_1)$.

 \vspace{3mm}
  \begin{theorem}
 The spectral radius of $K$ equals the spectral radius of $K_1$, that is, $\rho(K) =\rho(K_1)$. Furthermore, $\rho= \rho(K)$ is a unique eigenvalue of $K$ with maximum absolute value, and 
\[\lim_{t \to \infty} \frac{K^t}{\rho^t} = {\bf u}{\bf v}^T,\]
where $\mathbf{u}$ and $\mathbf{v}$ are the strictly positive right and left eigenvectors of $K$ corresponding to eigenvalue $\rho = \rho(K)$ such that $\mathbf{u}^T\mathbf{v} = 1$ and $\sum_{i=1}^\eta u(i) = 1$.  Additionally, as  $t \to \infty$,       
       \[\frac{{\bf B}_t}{\rho^t} \to  W{\bf v}, \quad a.s.\]  
      where $W$ is a non-negative random variable such that $P\{W>0\}>0$ iff $B_0({\bf s})\neq0$ for some ${\bf s} \in \mathcal H$ and $\Exp{{W|{\bf B}_0={\bf e_i}}} = u(i)$ for all $i=1,...,\eta$. Let $A=\{\omega:B_t(\omega) \to \infty\}$ as $t \to \infty$. Then, for any $\epsilon > 0$ and for all $\tilde{i}\in[1,\eta]$,
  \begin{equation*}\lim_{t\to \infty} P \{ \omega : \omega \in A, \left|{\frac{{B}_t(\tilde{i})}{\sum_{i=1}^{\eta}B_t(i)}-\frac{{v}(\tilde{i})}{\sum_{i=1}^{\eta}v(i)}}\right|>\epsilon \}=0.
  \end{equation*} 
   \label{theorem_extending_branching_process}  
 \end{theorem}

\subsection{Coupling}
\label{sec:defining_coupling}

 A crucial component in establishing that the epidemic process closely approximates the branching process during its early stages is demonstrating the inequality $A^N_t \leq A^B_t$ a.s. — that is, showing that the number of affected individuals in the epidemic process is dominated by those in the branching process. This inequality is essential for the main results presented in Section \ref{sec:branching_results}, particularly for establishing Lemma \ref{bounding_diff_lemma_final}.

For basic compartmental models such as SIR, the inequality $A^N_t \leq A^B_t$ follows almost immediately.  However, in our framework—where infections can propagate simultaneously across households, schools, workplaces and the broader community—the proof requires considerably more care (see Appendix \ref{sec:defining_coupling_appendix}). To overcome this challenge, we construct a sophisticated coupling between the epidemic and branching processes, with complete details provided in Appendix \ref{sec:defining_coupling_appendix}.

    \subsection{The initial branching process phase}
\label{sec:branching_results}

The following result demonstrates that the epidemic process closely approximates the multi-type branching process until time $ \log_\rho N^*$, where $N^* = \frac{N}{I_0\log (N)}$. We write $X_n \xrightarrow{P} X$ to denote that the sequence of random variables ${X_n}$ converges to $X$ in probability as $n\to\infty$.

\begin{theorem} As $N\rightarrow\infty$, for all ${\bf s}\in \mathcal S \setminus  \mathcal U$,
        \begin{equation}
           \sup\limits_{t \in [0, \log_\rho ({N^*}/{\sqrt{N}})]} \left\lvert{X}_t^N({\bf s})  - {B}_t({\bf s}) \right\rvert \xrightarrow{P} 0.  
           \label{e_b_close_1}
        \end{equation}
        \begin{equation}
        \sup\limits_{t\in \left[0, \log_\rho N^*\right]} \left\lvert{\frac{{X}_t^N({\bf s})}{\rho^t} - \frac{{B}_t({\bf s})}{\rho^t}}\right\rvert \xrightarrow{P} 0.
        \label{e_b_close_2}
        \end{equation}
     \label{e_b_close}   
\end{theorem}
 
 Result (\ref{e_b_close_1}) was previously established for the SIR setting in \citep{epidemic_notes}.  We extend this result to general models and additionally prove result (\ref{e_b_close_2}) in this more general framework.

\begin{lemma} For all ${\bf s}\in \mathcal S \setminus  \mathcal U$ and $t \in \mathbb{N}$ 
\[\Exp{\abs{{{B}}_t ({\bf s}) - {{X}}^N_t({\bf s})}} \leq {e} \frac{\rho^{2t}}{N},\]
for some ${e}>0$.
\label{bounding_diff_lemma_final}
\end{lemma}

Recall from Theorem \ref{theorem_extending_branching_process} that 
$ {{\bf B}_t}/{\rho^t} \to W{\bf v}$ a.s. as  $t \to \infty$,
 where $W$ is a non-negative random variable representing the intensity of the branching process and ${{\bf v}} \in {\mathbb{R}^+}^{\eta}$ is the left eigenvector corresponding to eigenvalue $\rho$ of matrix $K$.
Therefore, initially (until time $\log_{\rho}N^*$), the epidemic process grows exponentially at rate $\rho$, with the sample path dependent intensity determined by $W$. Moreover, both the smaller and larger models closely approximate the same branching process and thus have the same  $W$.

 The following proposition follows directly from Theorems \ref{theorem_extending_branching_process} and  \ref{e_b_close}, and justifies that the proportions across different types stabilize rapidly in the epidemic process. Thus early on—until time  $\log_{\rho} N^*$—paths can be patched from one time period to another with negligible error due to changes in proportions.

\begin{proposition} For $t_N \rightarrow \infty$ as $N \rightarrow \infty$  and  $\limsup_{N \rightarrow \infty}\frac{t_N}{\log_\rho N^*} 
< \infty$, then for all ${\bf s}\in \mathcal S \setminus  \mathcal U$: 
        \[ \left\lvert\frac{{X}_{t_N}^N({\bf s})}{\sum_{{\bf q}\in \mathcal S \setminus  \mathcal U} {X}_{t_N}^N({\bf q})} - \frac{{v}({\bf s})}{\sum_{{\bf q}\in \mathcal S \setminus  \mathcal U} v({\bf q})}\right\rvert \xrightarrow{P} 0. \]
        
\label{initial_proposition}
\end{proposition}

\begin{remark}{ In  Algorithm \ref{alg:cap2}, 
we had suggested that the restarted simulations should use common random numbers as in the original simulation
 paths so as to identically reproduce them. However, in our experiments we observe that even if the restarted paths are generated using independent  samples, that leads to a negligible anomaly.
 To understand this phenomenon, observe that to replicate an original path, we essentially need to replicate $W$ along that path, since after a small initial period, the associated branching process is well-specified once $W$ is known. (Note that this $W$ is implicitly generated in a simulation, it is not explicitly computed). Common random numbers achieve this replication. 
 However, for approximating the statistics of the larger city after $t_{\min}$ (Day 22 in Figure \ref{fig:shift_scale_restart_real_intervention}), 
 independent restarted simulations provide equally valid sample paths as the ones using the common random numbers. 
 The only difference is that the $W$ associated with each independently generated path may not match the $W$ associated with the original path,
 so patching them together at time $t_{\min}$ may result in a mismatch.  Nevertheless, since we report statistics associated with the average of generated paths, the statistics at time $t_{\min}$ depend linearly on the average of the $W$'s associated with the generated sample paths. Thus, if restarted paths are independently generated, while the corresponding average of associated $W$'s may not match the average of $W$'s associated with the original simulations, the difference as observed empirically is negligible as the average of $W$'s appears to have small variance. }
 \end{remark}

\begin{remark}{
Compartmental models are widely used to model epidemics. Typically, in these models, we begin with an infected population that constitutes a positive fraction of the overall population. However, little is known about the dynamics when we start with a small, constant number of infections.  
In Appendix \ref{sec:compartmental_models} we describe the results for some popular compartmental models using Proposition \ref{initial_proposition} and Theorem \ref{e_b_close}.}
 \end{remark}

\subsection{Deterministic phase}
\label{sec:deterministic_results}

As noted in Theorem \ref{e_b_close} and Proposition \ref{initial_proposition}, early in the infection growth—until time  $\log_\rho N^*$ for large $N$—while the number of affected individuals grows exponentially, the proportion of individuals across different types stabilizes. Here, the types corresponding to the susceptible population are not considered because at this stage, the affected individuals constitute a negligible fraction of the total susceptible population. The growth in the affected population during this phase is sample-path dependent and depends on the non-negative random variable $W$ (Theorem \ref{theorem_extending_branching_process}).

However, at time  $\log_{\rho} (\epsilon N)$ for any $\epsilon >0$ and large $N$, this behavior changes as the affected population equals $\Theta(N)$. 
Thereafter, the population growth closely approximates its mean-field limit, whose initial state depends on the random variable $W$ and where the proportions across types may change as time progresses. It is important to recall that in the initial branching process phase, both the smaller and the larger models closely approximate the same branching process and hence share the same random variable $W$. Once $W$ is realized, the evolution of both models becomes deterministic, step by step. As a result, the infections in the larger model can be obtained by scaling up the smaller model in this phase.
Our key result in this setting is Theorem~\ref{deterministic_theorem}. To establish this result, we require Assumption~\ref{deterministic_assumption}.

Let $t_N = \log_{\rho} (\epsilon N)$. Let ${\bf \mu}^N_{t}$ denote the empirical distribution across all types at time $t_N + t$, including the group disease state in which all individuals are susceptible. This corresponds to augmenting the vector ${\bf X}_{t_N+t}^N$ with the number of completely susceptible groups at time $t_N + t$ and scaling the resultant vector by the factor $N^{-1}$.

\begin{assumption} 
There  exists  a random distribution  ${\bf \bar{\mu}}_{0}(W) \in {\mathbb{R}^+}^{(\eta+1)}$ that is independent of $N$ 
such that  ${\bf \mu}^N_{0} \xrightarrow{P} {\bf \bar{\mu}}_{0}(W)$ as $N \rightarrow \infty$.
\label{deterministic_assumption}
\end{assumption}

Observe that ${\bf \bar{\mu}}_{0}(W)$ above is path dependent in that  it depends on  the random variable $W$ (Theorem \ref{theorem_extending_branching_process}). 
The above assumption is observed to hold empirically.  While we do not have a formal proof for this assumption (this appears to be a difficult and open problem), Corollary \ref{e_b_close_epsilon} below supports this assumption. This corollary follows from Lemma \ref{bounding_diff_lemma_final}.

\begin{corollary} For $\epsilon \in (0,1)$, $t=\log_\rho(\epsilon N)$, and for all ${\bf s}\in \mathcal S \setminus  \mathcal U$,
        \begin{equation*}
        \Exp{ \left\lvert{\frac{{X}_t^N({\bf s})}{\rho^t} - \frac{{B}_t({\bf s})}{\rho^t}}\right\rvert } \leq \epsilon {e},
        \end{equation*}
        for some constant ${e}>0$.
     \label{e_b_close_epsilon}   
\end{corollary}

 Further, note that Assumption~\ref{deterministic_assumption} holds along some subsequence ${N_i}$ since ${\bf \mu}^N_{0}$ is a bounded sequence.
\vspace{3mm}

 Recall that, $\lambda^{\mathrm{c},N}_{t}(j)$ 
 denote the total incoming infection rate from the ``community'' as seen by an individual with characteristic ${(j,\cdot)}$ at time $t$ from all  infectious individuals across the city.  This rate is determined by the disease state of all individuals at time  $t$. In our setup, $\lambda^{\mathrm{c},N}_{t}(j)$ equals

  \begin{equation}
   \lambda^{\mathrm{c},N}_{t}(j) =  \sum\limits_{{\bf q}\in \mathcal S \setminus  \mathcal U} \mu^N_{t}({\bf q}) \sum_{\tilde{j} =1 }^g {\bf{1}}({\bf{q}}(\tilde{j})= \text{infectious}) \psi_{\tilde{j},{j}} \beta^c_{\tilde{j}},
    \label{cnt}
    \end{equation}
   where ${\bf{1}}({\bf{q}}(\tilde{j})= \text{infectious})$ denotes the indicator function for whether  individual $\tilde{j}$ in group disease state $ \bf{q}$ is infectious. 
   
  For each group $n$, recall $S_{ n}(t)$ denote its state at time $t$.  Then,  for  ${\bf s}\in \mathcal S$, \[\mathbb{P}\left( S_{ n}(t+1) = {\bf s} | \mathcal \mu_t^N, S_{ n}(t) \right) = {p}(S_{ n}(t), {\bf s},  \lambda^{\mathrm{c},N}_{t}(1),\ldots, \lambda^{\mathrm{c},N}_{t}(g)),\]
 for a continuous function $p$. 
 In particular, the transition probability depends only on the disease state of the individual at the previous time, the disease state to which it is transitioning, and the infection rate incoming to the individuals in the group at that time.

Recall that from Assumption \ref{deterministic_assumption}, we have defined ${\bf \bar{\mu}}_{0}(W) \in {\mathbb{R}^+}^{(\eta+1)}$
such that  ${\bf \mu}^N_{0} \xrightarrow{P} {\bf \bar{\mu}}_{0}(W)$ as $N \rightarrow \infty$. We define  ${\bf \bar{\mu}}_{t}(W) \in {\mathbb{R}^+}^{(\eta+1)}$ such that for all $t\in\mathbb N$,

\begin{equation}
        \label{mu_definition_equation}
     \bar{\mu}_{t}({\bf s},W) := \sum\limits_{{\bf s'} \in \mathcal S} \bar{\mu}_{t-1}({{\bf s'}},W) p({\bf s'}{\bf s}, \bar{\lambda}^{\mathrm{c}}_{t-1}(1,W),\ldots,  \bar{\lambda}^{\mathrm{c}}_{t-1}(g,W) ), \end{equation}
     
     where for all $j\in [g]$, \[\bar{\lambda}^{\mathrm{c}}_{t-1}(j,W) =  \sum\limits_{{\bf q}\in \mathcal S \setminus  \mathcal U} \bar{\mu}_{t-1}({\bf q},W) \sum_{\tilde{j} =1 }^g {\bf{1}}({\bf{q}}(\tilde{j})= \text{infectious}) \psi_{\tilde{j},{j}} \beta^c_{\tilde{j}}.\]

\begin{theorem}  Under Assumption \ref{deterministic_assumption} and for $t\in\mathbb N$, ${\bf \mu}^N_{t} \xrightarrow{P} {\bf \bar{\mu}}_{t}(W)$  as  $N\rightarrow\infty$.

        \label{deterministic_theorem}
        \end{theorem}
\noindent \textit{Proof sketch:} We rewrite $\mu^N_{t+1}(\bm{s}) = \frac{1}{N} \sum\limits_{n=1}^N {\mathbf 1}\left\{S_{ n}(t+1) = \bm{s}\right\}$ as, 

\begin{equation}\label{eq:errorVar} \mu^N_{t+1}(\bm{s})= \sum\limits_{\bm{s'}\in\mathcal S} ~ \left[  {p}(\bm{s'},\bm{s},\lambda^{\mathrm{c},N}_{t}(1), \ldots, \lambda^{\mathrm{c},N}_{t}(g)) \mu^N_{t}(\bm{s'}) \right]  ~+~ \mathcal M^N_{t+1}(\bm{s}),\end{equation}

where $\mathcal M^N_{t+1}(\bm{s}) =
  \frac{1}{N} \sum\limits_{n=1}^N \left[{\mathbf 1}\left\{S_{ n}(t+1) = \bm{s}|S_{ n}(t), \mu^N_t\right\} - {p}(S_{ n}(t), \bm{s}, \lambda^{\mathrm{c},N}_{t}(1), \ldots, \lambda^{\mathrm{c},N}_{t}(g))\right].$

  Observe that $\mathcal M^N_{t+1}(\bm{s}) $ is a sequence of zero-mean random variables whose variance converges to $0$ as $N\rightarrow \infty$. Also,  $\lambda^{\mathrm{c},N}_{t}(j)$ is a continuous and bounded function of $\mu^N_{t}$, and ${p}$ is uniformly continuous in its arguments since it is a continuous function defined on a compact set.  Therefore, ${\bf \mu}^N_{t} \xrightarrow{P} {\bf \bar{\mu}}_{t}(W)$ follows by inducting on $t$. Further details are provided in Appendix \ref{sec:app_proof_DetEv}.

        In particular, if ${\bf \bar{\mu}}_t$ denotes the mean field limit of the normalized process at time $t + \log_{\rho} (\epsilon N)$, then the number of infections observed in a smaller model with population $N$ is approximately $N \cdot {\bf \bar{\mu}}_t$ and that of a larger model is approximately $kN \cdot {\bf \bar{\mu}}_t$. \textbf{Thus, once the system enters the mean-field phase, the larger model infection process can be approximated by the smaller model infection process by scaling it by $k$.}
        
        \begin{remark} {
        While the deterministic equations (conditioned on $W$) in Theorem \ref{deterministic_theorem} can be easily solved when the number of types is small and the initial state is established via simulation, these become much more difficult in a realistic model with all its complexity, where the number of types is extremely large and may be uncountable if non-memoryless probability distributions are involved. \textbf{One may thus view the key role of the simulator as a tool that identifies the random initial state $W$ of these deterministic mean-field equations and then solves them efficiently using stochastic methods.}}
  \end{remark}

\subsection{Why naive scaling does not work}
\label{sec:bias_naive_scaling}

In Section \ref{sec:speeding_abs}, we observed two phenomena. First, when we start with a smaller city with a large number of infections, naively scaling its outputs accurately estimates infections in a larger city that starts with a proportionally larger number of infections (Phenomenon 1). In contrast, when the initial number of infections is small, naively scaling the outputs from the smaller-city model underestimates infections in the larger city (Phenomenon 2).
In this section, we explain why naive scaling fails when we start from smaller number of infections. For clarity, we illustrate the intuition with a stochastic SIR model, though the analysis readily extends to more complex multi-type epidemic models.

Consider a model with population $N$ starting with $I_0$ infections. Let the infectious time period of any individual be exponentially distributed with rate $r$. Further, assume that the number of infectious contacts each individual $j$ makes with every other individual in the city during time $t$ are Poisson distributed with rate $\frac{t \beta}{N}$. Let the total epidemic size in this city be denoted by $R_{N,I_0}$. We have the following result from \citep[Ch. 4]{andersson2012stochastic}.

\begin{theorem} [\citep{andersson2012stochastic}] 
As  ${N \to \infty},$ we have $\frac{R_{N,I_0}}{N} \xrightarrow{\,d\,}   Z_{I_0} $, where $Z_{I_0}$ is distributed as follows:
\begin{align*}
Z_{I_0} =
\begin{cases}
0 & \text{w.p. } p^{I_0} \\
c & \text{w.p. }  1-p^{I_0}
\end{cases}
\end{align*}

where $c \in (0,1)$ is the solution to $1-c = e^{-\frac{\beta}{r}c}$ and $p$ is the solution to  $p = e^{-\frac{\beta}{r}(1-p)}$.
\label{thm:stochastic_SIR_facts}
\end{theorem}
 
This theorem states that with probability $p^{I_0}$, the epidemic dies out in the early branching phase, where $p$ represents the probability that a branching tree starting with one infected individual dies out. When the epidemic takes off, which occurs with probability $1- p^{i_0}$, the total epidemic size is $c$. This is equivalent to saying that if the epidemic takes off, then each person has an equal probability $c$ of becoming infected. 

Since $\frac{R_{N,I_0}}{N}$ is uniformly bounded with $0 \leq \frac{R_{N,I_0}}{N} \leq 1$, we have:
$$\lim_{N \to \infty}\Exp{\frac{R_{N,I_0}}{N}} = c \left(1-p^{I_0}\right)$$

Now let us compare a smaller model (population $N$) with $i_0$ initial infections and a larger model (population $kN$) with proportionally scaled infections $kI_0$. The following theorem is a direct consequence of Theorem \ref{thm:stochastic_SIR_facts}.

\begin{theorem} Consider a stochastic SIR model comparing a smaller model (population $N$) with $I_0$ initial infections and a larger model (population $kN$) with proportionally scaled infections $kI_0$. Then:

    \begin{align*}
    \lim_{N \to \infty} \left(\Exp{ \frac{R_{kN,kI_0}}{kN}} - \Exp{ \frac{R_{N,I_0}}{N}}\right)  
    & = c \left(p^{I_0}-p^{kI_0}\right) 
\end{align*}

\label{thm:bias_justification}
\end{theorem}

This result demonstrates that the scaled smaller model systematically underestimates the epidemic size in the larger city. Furthermore, this discrepancy decreases as the number of initial infections increases. These theoretical findings align with our empirical observations in Figures \ref{fig:12800_infections} and \ref{fig:128_infections}.  The same phenomenon extends readily to multitype epidemic processes; based on \citep[Ch. 6]{andersson2012stochastic}.

\section{Empirical evaluation}
\label{sec:further_experiments}
  
 \subsection{SSR on sparser cities}
\label{sec:sparse_cities}
 
\begin{figure}[h!]
      \centering
      \begin{minipage}[b]{0.49\textwidth}
  \includegraphics[width=\linewidth, height=5cm]{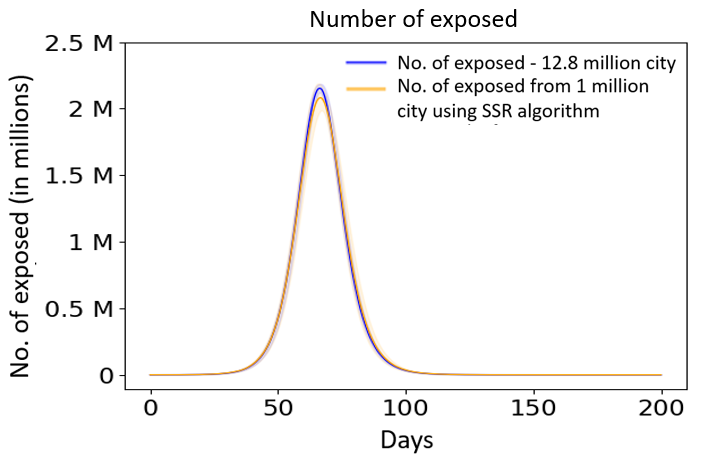}   
    \caption{ \textbf{Sparser city (Scenario 1, No intervention):} Shift and scale smaller model  (no. of exposed) matches the larger model under no intervention ($\beta^c_{new} = \beta^c/10$).} 
    \label{fig:1m-12.8m-beta-c-10-no-inte}
  \end{minipage}
\hfill
\begin{minipage}[b]{0.49\textwidth}
    \includegraphics[width=\linewidth, height=5cm]{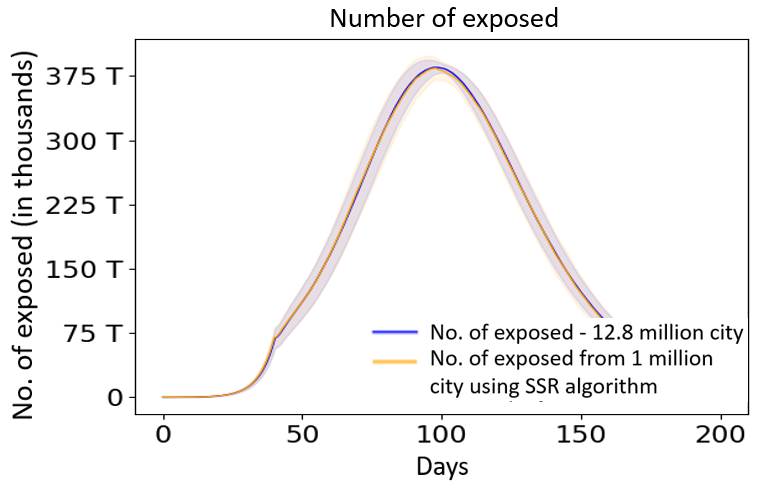}
    \caption{ \textbf{Sparser city (Scenario 1, With intervention):} Shift-scale-restart smaller model  (no. of exposed) matches the larger model under intervention ($\beta^c_{new} = \beta^c/10$).
    \label{fig:1m-12.8m-beta-c-10-inte}}
\end{minipage}
  \end{figure}

Although our earlier experiments focused on a densely populated urban city, it is pertinent to investigate whether these findings extend to less densely populated cities or regions. In this section, we present experimental results for sparser urban settings, achieved by reducing contact rates and, consequently, transmission rates among individuals across various interaction spaces. We examine two representative scenarios:
\begin{itemize}
    \item \textbf{Scenario 1:} The community transmission rate ($\beta^c$) is set to one-tenth the level considered in our earlier experiments described in Section \ref{sec:speeding_abs}. As shown in Figure \ref{fig:1m-12.8m-beta-c-10-no-inte}, the SSR algorithm yields results consistent with those of the larger model in the absence of interventions. Similarly, Figure \ref{fig:1m-12.8m-beta-c-10-inte} illustrates that the SSR algorithm aligns with the larger model when a home quarantine intervention is implemented after 40 days. While extensive testing was not conducted, we anticipate comparable results under more comprehensive intervention strategies.

    \item \textbf{Scenario 2:} The community transmission rate ($\beta^c$) is reduced by a factor of 20, while transmission rates at home ($\beta^h$), school ($\beta^s$), and workplace ($\beta^w$) are reduced by a factor of 4. Figure \ref{fig:1m-12.m-all-beta-no-inte} (Appendix \ref{sec:sparser_city_scenario_2_results}) demonstrates that the SSR algorithm's results align with those of the larger model in the no-intervention scenario. Notably, without interventions, the number of infections remains exceptionally low even after 200 days. When home quarantine is implemented after 40 days, the number of exposed cases in the smaller model is within 20\% of the larger model, though both models report negligible infection numbers, with the peak not exceeding 250 cases (see Figure \ref{fig:12.8m-all-beta-low-intervention} in Appendix \ref{sec:sparser_city_scenario_2_results}).
\end{itemize}

 \subsection{How small can the smaller model be}
\label{sec:limits_smaller_city}
\begin{figure}
      \centering
      \begin{minipage}[b]{0.49\textwidth}
      \includegraphics[width=\linewidth, height=5cm]{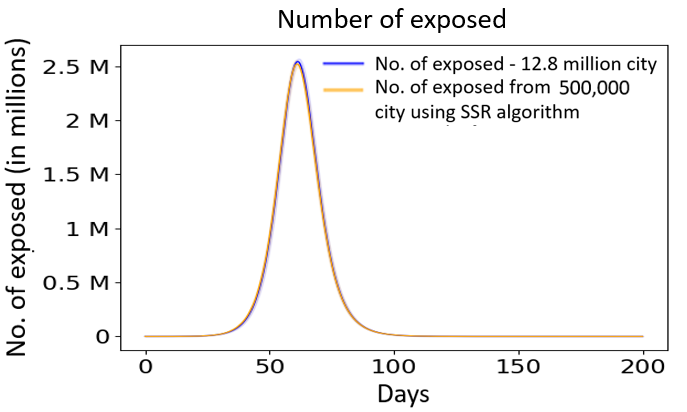} 
    \caption{{\bf 500,000 smaller model (No intervention):} Shift and scale smaller model  (no. of exposed) matches the larger model under no intervention scenario (Larger model - 12.8 million population).
    \label{fig:12.8m-5l-nointe}}
  \end{minipage}
\hfill
      \begin{minipage}[b]{0.49\textwidth}
      \includegraphics[width=\linewidth, height=5cm]{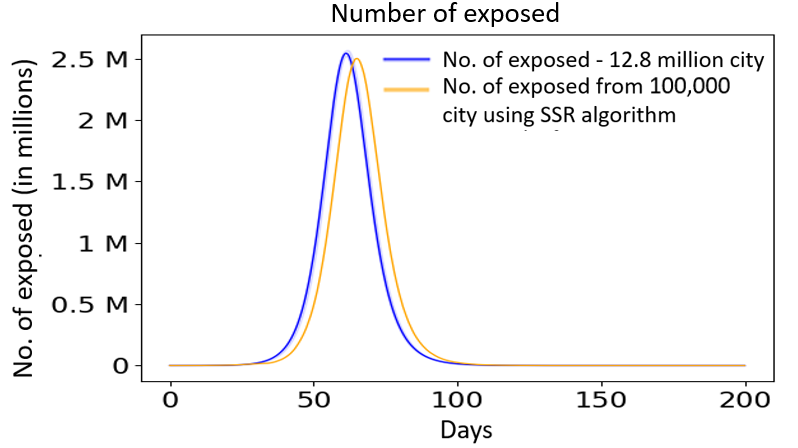} 
    \caption{{\bf 100,000 smaller model (No intervention):} Shift and scale smaller model  (no. of exposed) does not match the larger model (Larger model - 12.8 million population).
    \label{fig:12.8m-1l-no-inte}}
  \end{minipage}
  \end{figure}

In our previous experiments, detailed in Section \ref{sec:speeding_abs}, we compared the outcomes of a 12.8-million-person city model with those of a smaller 1-million-person city model using the SSR algorithm. A pertinent question arises: what is the minimum size of the smaller city for which the SSR algorithm remains accurate? To address this, we evaluated the SSR algorithm's performance on smaller city models with populations of 500,000 and 100,000.
 \begin{itemize}
     \item {\bf 500,000-population smaller city model:} As shown in Figure \ref{fig:12.8m-5l-nointe}, the SSR algorithm's results align closely with those of the larger model in the no-intervention scenario. Similarly, Figure \ref{fig:12.8m-5l-inte} (Appendix \ref{sec:smaller_city_under_intervention_results}) demonstrates that the SSR algorithm accurately matches the larger model's outcomes in a scenario with a single intervention—home quarantine implemented from day 40 onward.
     \item {\bf 100,000-population smaller city model:} Figure \ref{fig:12.8m-1l-no-inte} reveals that the SSR algorithm's results poorly match the larger model's outcomes in the no-intervention scenario. This discrepancy arises because the 12.8-million-person and 100,000-person city models exhibit similar dynamics for only the first 24 days (see Figure \ref{fig:12.8m-1l-initial-same} in Appendix \ref{sec:smaller_city_under_intervention_results}). At day 24, the smaller model reports approximately 8,000 exposed individuals. 
     Scaling down this by a factor of 128 (to align with the larger model) yields 62.5 exposed individuals.
    Consequently, we would need to extract simulation output from the smaller model at the point when the number of exposed individuals equals 62.5 and apply this data to day 24 by scaling it (×128).
     This is infeasible, as the smaller model starts with 100 exposed individuals.
     Ad-hoc adjustments that utilize simulation data when initial exposed individuals number 100 or more result in significantly noisier outputs under the SSR algorithm. Therefore, the SSR algorithm cannot be reliably applied in this configuration, particularly when interventions are introduced.
 \end{itemize}

\subsection{Parameter uncertainty}
\label{sec:parameter_uncertainty}

In many epidemic scenarios, the parameters governing disease transmission and progression are not precisely known and are subject to considerable uncertainty. This section evaluates the robustness of our algorithm under such parametric uncertainty. Specifically, we assume that disease transmission rates are unknown and follow a known probability distribution.

In Algorithm \ref{alg:base}, epidemic performance measures are estimated by conducting multiple independent simulations using identical parameter sets and averaging the results. However, when parameter uncertainty is present, we sample a new parameter set from the known distribution for each simulation run and base the simulation on that sample (see Algorithm \ref{alg:base_parameter_uncertainty}).

This necessitates modifications to Algorithm \ref{alg:cap2}. A crucial step in Algorithm \ref{alg:cap2} involves determining the time $t_S$—the duration during which the larger and smaller models exhibit nearly identical behaviour. When parameters are constant, $t_S$ remains uniform across all simulation runs since it depends on the infection growth rate ($\rho$), which remains constant when parameters are fixed. However, under parameter uncertainty, each run employs different parameters, requiring separate determination of $t_S$ for each run.

To determine $t_S$ for each run individually, we set $t_S = t_{N^{\dagger}}$, representing the time when the smaller city reaches $N^{\dagger} = \frac{N}{\log{N}}$ infections. Consequently, we define $t_{\min} = \min \{t_{N^{\dagger}}, t_I\}$ separately for each parameter set. We then proceed analogously to Algorithm \ref{alg:cap2} for each independent run, as detailed in Algorithm \ref{alg:SSR_parameter_uncertainty}.

\begin{figure}
      \centering
      \begin{minipage}[b]{0.49\textwidth}
      \includegraphics[width=\linewidth, height=5cm]{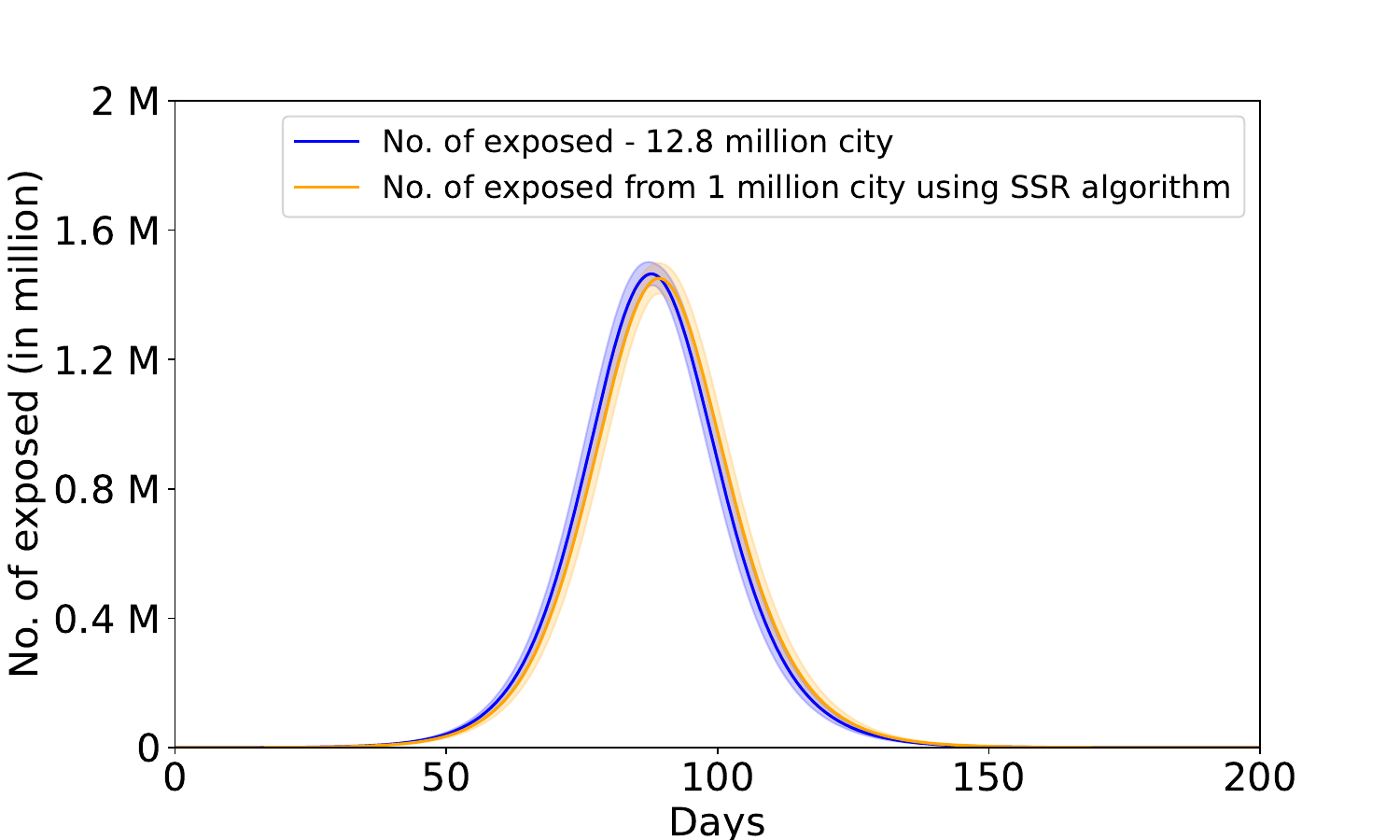} 
    \caption{\textbf{Parameter uncertainty (No intervention):} Shift and scale smaller model  (no. of exposed) matches the larger model under no intervention scenario.
    \label{fig:param-uncertain-no-inte}}
  \end{minipage}
\hfill
\begin{minipage}[b]{0.49\textwidth}
 \includegraphics[width=\linewidth, height=5cm]{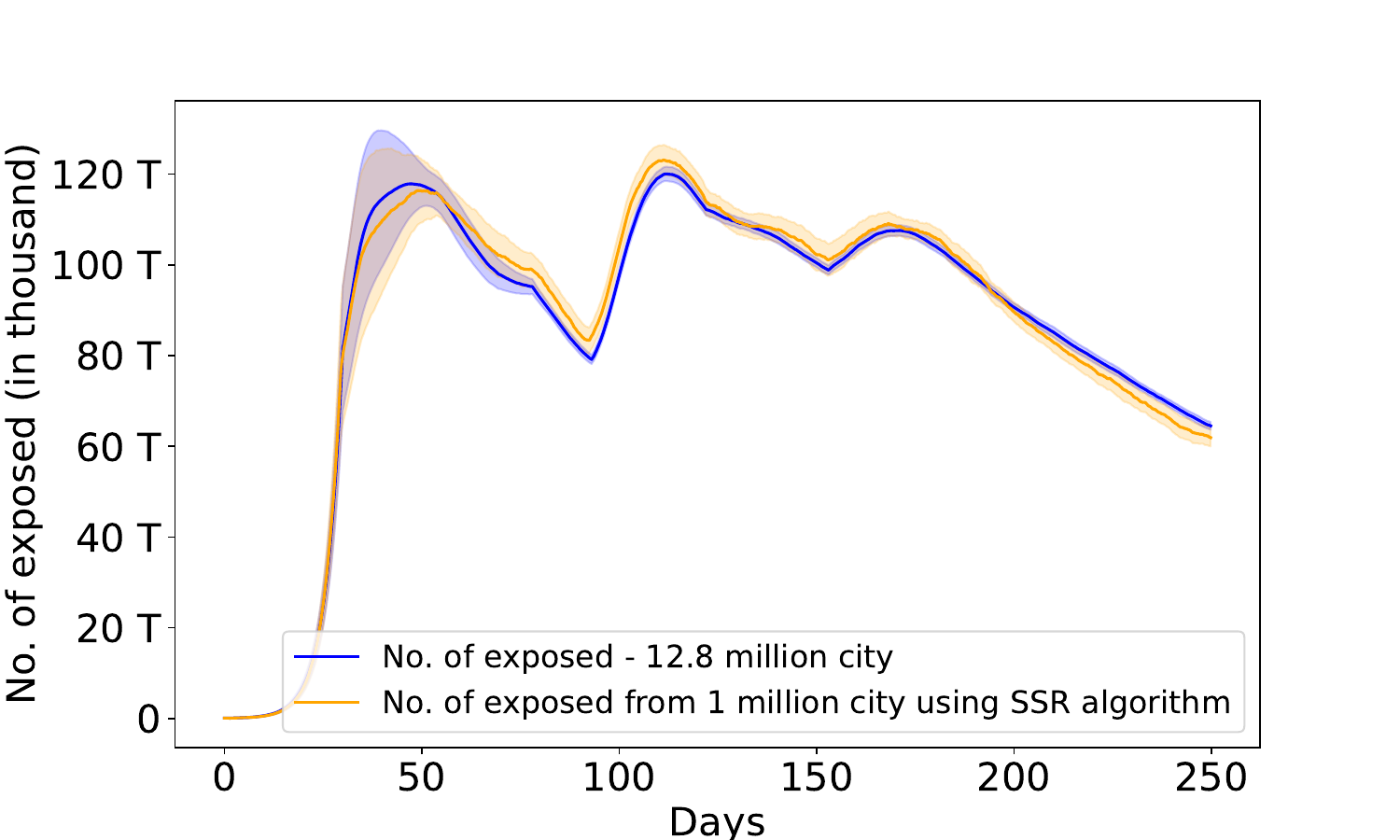}
    \caption{\textbf{Parameter uncertainty (With intervention):} Shift-scale-restart smaller model  (no. of exposed) matches the larger model under intervention.
    \label{fig:param-uncertain-with-inte}}
    \end{minipage}
  \end{figure}

We examine a parameter uncertainty scenario where disease transmission rates ($\beta_h, \beta_w, \beta_c, \beta_s$) are uniformly distributed within a range of ±5\% around the baseline transmission rates used in the experiments described in Section \ref{sec:speeding_abs}.
Figure \ref{fig:param-uncertain-no-inte} demonstrates our algorithm's performance under the no-intervention scenario. The shift-and-scale smaller model (number of exposed individuals) matches the larger model even under parameter uncertainty. Similarly, Figure \ref{fig:param-uncertain-with-inte} shows comparable agreement when interventions are implemented.

\subsection{Multiple strains}
\label{sec:multiple_strains}

Due to space constraints,  we present the evaluation of the SSR methodology under multiple epidemic strains in Appendix \ref{sec:experiments_with_multiple_strains}.



\section{Conclusion}
\label{sec:conclusion}
In this paper we considered large ABS models used to simulate epidemic spread in a city or a region. These models are of great use in capturing 
details of population types, their behaviour, time varying regulatory instructions, etc. A key drawback 
of ABS models is that computational time can quickly become prohibitive as the city size and the simulated time increases. In this paper we proposed a shift-scale-restart algorithm that exploits the underlying probabilistic structure and  allows smaller city simulations to provide extremely  accurate approximations to larger ones using much 
         less computational time. We supported our experiments and the algorithm through asymptotic analysis where we showed
         that initial part of the epidemic process is close to a corresponding multi-type branching process
         and very quickly the evolution of the epidemic process is well approximated
         by its mean field limit.  The fact that in multi-type branching process 
         the distribution of population across types stabilizes quickly helped
         us shift sample paths across time without loss of accuracy. 
         To show the closeness of the epidemic 
         process to a branching process we developed careful coupling arguments.  To show the robustness of the proposed methods, we numerically tested their effectiveness in presence of multiple strains, parameter uncertainty, sparser city populations, and the use of substantially smaller city models to generate output that matches the large city models.

\newpage

\section*{Acknowledgments}
We acknowledge the support of A.T.E. Chandra Foundation and Safexpress Centre for Data, Learning and Decision Sciences at Ashoka University for this research. We further acknowledge the support of the Department of Atomic Energy, Government of India, to TIFR under project no. 12-R\&D-TFR-5.01-0500. We thank Shubhada Agrawal (IISc Bangalore) for insightful initial discussions on this project, and Shane Henderson (Cornell University) for his valuable feedback on our work.

\vspace{3mm}

\bibliographystyle{abbrvnat}

\ifdefined\useorstyle
\setlength{\bibsep}{.0em}
\else
\setlength{\bibsep}{.7em}
\fi

\newpage

\ECSwitch
\ECHead{Appendix}

\noindent Section~\ref{sec:define_standard_MTBP} reviews the standard super‑critical multi‑type branching process.
 Section~\ref{sec:defining_coupling_appendix} introduces the nuanced coupling between the epidemic process and its associated multi‑type branching process.
 In Section~\ref{sec:proofs} we present detailed proofs of the main results stated in Section~\ref{sec:Theoretical_results}.
 Section~\ref{sec:compartmental_models} characterizes the early, branching‑process phase of classical compartmental models (e.g., SIR, SEIR), before the number of infections is large enough for the deterministic ODE description to apply. Section~\ref{sec:detailed_abs_model} provides a detailed account of the ABS model. Finally, Section~\ref{sec:additional_numerics} summarizes the experimental setup, including city‑level statistics, additional numerical results, and supplementary health statistics such as cumulative fatalities.

\section{Supercritical multi-type branching process review}
\label{sec:define_standard_MTBP}

  In this section, we introduce a multi-type branching process and state a fundamental result for the super‑critical case \citep{branching_process_notes, harris1963theory}, together with a set of sufficient conditions under which it holds.

Consider a probability space $(\Omega,\mathcal{F},\mathbb{P})$ and fix an integer $\eta\ge 1$. A multi-type branching process is a stochastic process $\mathbf{B}=(\mathbf{B}_t)_{t\ge 0}$ with state space $\mathbb{N}^{\eta}$, where $\mathbf{B}_t$ is an $\eta$-dimensional vector whose $i$-th component denotes the number of individuals of type $i$ at time $t$. In a multi-type branching process, at the end of each time period, an individual may give birth to children of different types before dying.

The number of children produced by each individual of type $i$ is independent and identically distributed.  
For each type $i\in\{1,\ldots,\eta\}$, let $\boldsymbol{\xi}^{(i)}=(\xi^{(i)}_1,\ldots,\xi^{(i)}_\eta)$ be an $\mathbb{N}^{\eta}$-valued random vector describing the offspring counts of a single parent of type $i$, with law
\[
p^{(i)}_{\mathbf{k}}
= \mathbb{P}\!\big(\boldsymbol{\xi}^{(i)}=\mathbf{k}\big),
\qquad \mathbf{k}\in\mathbb{N}_0^{\eta}.
\]

 If $\mathbf{B}_t = (b_1, \ldots, b_\eta)$, then $\mathbf{B}_{t+1}$ represents the sum of independent offsprings from $b_1$ type-1 parents, $b_2$ type-2 parents, and so forth. Thus, $\mathbf{B}_{t+1}$ is the sum of $b_1 + b_2 + \cdots + b_\eta$ independent random vectors in $\mathbb{N}^{{\eta}}$. Specifically, 
\begin{equation*}
  \mathbf{B}_{t+1}
  \;=\;
  \sum_{i=1}^{\eta}\;\sum_{m=1}^{b_i}\boldsymbol{\xi}^{(i)}_{t,m},
\end{equation*}
where $\boldsymbol{\xi}^{(i)}_{t,m}$ are i.i.d.\ copies of $\boldsymbol{\xi}^{(i)}$ and the entire collection $\{\boldsymbol{\xi}^{(i)}_{t,m}: t\ge 0,\ m\ge 1,\ 1\le i\le \eta\}$ is mutually independent. Consequently, the multi-type branching process forms a Markov chain $({\bf {B}}_t \in \mathbb{N}^{\eta}:t \geq 0)$.  Its one-step transition kernel is the $\mathbb{N}^{\eta}$-valued convolution of $b_i$ copies of $p^{(i)}$ across types:
\begin{equation}\label{eq:kernel}
\mathbb{P}\!\left(\mathbf{B}_{t+1}=\mathbf{y}\,\middle|\,\mathbf{B}_t=\mathbf{b}\right)
= \big(p^{(1)}\big)^{*b_1} * \cdots * \big(p^{(\eta)}\big)^{*b_\eta}(\mathbf{y}),
\qquad \mathbf{y}\in\mathbb{N}_0^{\eta},
\end{equation}
where $*$ denotes convolution on $\mathbb{N}^{\eta}$.

Define the mean off spring matrix ${K}\in {\mathbb{R}^+}^{{\eta} \times {\eta}}$ by 
\[{K}(i,j) = \Exp{{B}_1(j)|{\bf {B}}_0 ={\bf e_i}} =\Exp{\xi^{(i)}_j},\] the expected number of type $j$ children produced by a single type-$i$ individual in one period. Then, 
\begin{equation*}
    \begin{aligned}
        \Exp{{\bf {B}}_{t+1}} &= {K}^T \Exp{ {\bf {B}}_t } = ({K}^T)^{t+1} \Exp{{\bf {B}}_0}. 
    \end{aligned}
\end{equation*}

\noindent 
A non-negative square matrix $A$ is \textbf{irreducible} if there exists an integer $m>0$ such that
all entries of $A^m$ are strictly positive..
Further, Perron Frobenius eigenvalue of a non-negative irreducible matrix is its largest eigenvalue (in absolute value).

The following assumption is standard in multi-type branching process theory  \citep{branching_process_notes, harris1963theory}.

\begin{assumption}
      The matrix  ${K}$ is  irreducible  and has Perron Frobenius eigenvalue ${\rho} >1.$  Moreover,  for all $ 1\leq i,j\leq {\eta}$
   \[
   \Exp{{B}_1(j)\log {B}_1(j)|{\bf {B}}_0={\bf e_i}}<\infty.\] 
   \label{ass:mtbp_std_ass}
\end{assumption}

Under Assumption \ref{ass:mtbp_std_ass}, ${K}$ admits  strictly positive left  and right eigenvectors ${\bf {v}}$ and  ${\bf{u}}$ associated with eigenvalue $\rho$, normalized so that ${\bf {u}}^T{\bf {v}}=1$ and $\sum_{i=1}^{{\eta}}{{u}(i)}  = 1$.  The following classical result (see~\citep{branching_process_notes, harris1963theory}) describes the asymptotic behaviour of the process.
 
\begin{theorem}
Suppose Assumption \ref{ass:mtbp_std_ass} holds. Then,
\[
\lim_{t \to \infty} \frac{{K}^t}{{\rho}^t} = {\bf {u}}{\bf{v}}^T.
\]
Moreover as $ t \to \infty$, 
      \begin{equation*} \label{eq:0001}
       \frac{{\bf {B}}_t}{{\rho}^t} \to W{\bf {v}}, \quad a.s.
       \end{equation*} 
      where $W\geq 0$ is a  random variable with $P\{W>0\}>0$ and for each $i$,
      \[\Exp{W|{\bf {B}}_0={\bf e_i}} ={u}(i).\]  
      Let $A=\left\{\omega: \|{\bf B}_t(\omega)\| \to \infty ~\text{as} ~t \to \infty\right\}$. For any $\epsilon > 0$ and every $j\in\{1, \ldots {\eta}\}$,
  
 \begin{equation*}\lim_{t\to \infty} P \left\{ \omega \in A: \abs{\frac{{{B}}_t(j)}{\sum_{i=1}^{ {\eta}}{B}_t(i)}-\frac{{{v}(j)}}{\sum_{i=1}^{{\eta}}{v}(i)}}>\epsilon \right\}=0.
 \label{proportion_stbp}
 \end{equation*}
  \label{thm:stbp_theorem}
\end{theorem}

\section{Coupling}
\label{sec:defining_coupling_appendix}

In this section, we formalize the nuanced coupling between the epidemic process and its associated branching process. We first introduce the random variables that characterize each process and then construct a coupling of these variables to link the two processes.

\subsection{Random variables associated with the epidemic process}
\label{sec:defining_epidemic_process_rv}

Consider an individual $(j,n)$ who is exposed at time $\tau_{(j,n)}$.  
For each integer $\zeta \ge 0$, let $W_{(j,n)}(\zeta)$ denote the collection of random variables relevant to that individual at time $\tau_{(j,n)}+\zeta$:
\begin{enumerate}
    \item the variables governing the transition of individual $(j,n)$ between disease states at time $\tau_{(j,n)}+\zeta$;
    \item the variables governing the number of infectious contacts that individual $(j,n)$ makes \emph{within} their group i.e.,  with individuals $(\tilde{j},n)$ at home, school, or workplace—at time $\tau_{(j,n)}+\zeta$;
    \item the variables governing the number of infectious contacts that individual $(j,n)$ makes through community i.e.,  with individuals $(\tilde{j},\cdot)$ at time $\tau_{(j,n)}+\zeta$, together with 
    \item the associated uniform random variables used to select, from the $N$ potential recipients $(\tilde{j},\cdot)$, the specific individual contacted.
\end{enumerate}

Define the full collection for individual $(j,n)$ by
$
  C_{(j,n)}:=\{\,C_{(j,n)}(\zeta): \zeta\ge 0\,\}.
$
 For every $j$, the collections $C_{(j,\cdot)}$ are identically distributed.

\subsection{Random variables associated with the branching process }
\label{sec:defining_branching_process_rv}

To describe the branching process rigorously we first introduce a three–component index \((j,\cdot,\cdot)\) that uniquely labels every individual ever generated by the process.  Although the indexing convention refers to the population size \(N\), the dynamics of the branching process itself are independent of \(N\); only the labels depend on $N$.

\vspace{3mm}
\noindent\textbf{Indexing scheme}  

\begin{enumerate}

\item \textbf{First coordinate \((j)\).}  
        This is the \emph{type} of the individual and takes values
        \(j\in[g]\).

    \item  \textbf{Second coordinate.}
\begin{enumerate}
\item Recall that at time zero,  \(\mathbf B_0=\mathbf X_0\).  Each initially–infected individual \((j,n)\) in the epidemic is matched to an individual in the branching process and assigned the label \((j,n,\cdot)\).

\item For every individual \((j,n,\cdot)\), the \(g-1\) other members of its  group receive the labels \((\tilde{j},n,\cdot)\) with \(\tilde{j}\in[g]\).

\item While infectious, an individual of type \(j\) gives birth to Poisson(\(\psi_{j,\tilde{j}}\beta^c_j\)) offspring groups of type \(\mathbf e_{\tilde{j}}\).   In each such group the newly–exposed individual of type \(\tilde{j}\) is assigned a second index \(\tilde n\sim\mathrm{Unif}\{1,\ldots,N\}\); the other members of that group inherit the same second index \(\tilde n\).
\end{enumerate}

\item \textbf{Third coordinate.} 
\begin{enumerate}
\item
All members of a group share the same third index, which records the order in which that group was generated. 

\item
When the \(\ell\)-th offspring group containing an individual with second index \(n\) is produced, its members receive labels \((\tilde{j},n,\ell)\) for \(\tilde{j}\in[g]\).

 \item 

If several such groups with second index $n$ arise at the same time, they are ordered uniformly at random before indexing.

\end{enumerate}
 \end{enumerate}

\vspace{3mm}
\noindent \textbf{Random variables} 

Let \(C_{(j,\cdot,\cdot)}\) be the collection of all random variables
governing the life of an individual with label \((j,\cdot,\cdot)\).  If
that individual is exposed at time \(\tau_{(j,\cdot,\cdot)}\), define
\(C_{(j,\cdot,\cdot)}(\zeta)\) as the set of random variables that are
realised at time \(\tau_{(j,\cdot,\cdot)}+\zeta\). Specifically,

\begin{enumerate}

\item
the variables driving the individual’s disease‑state transition at \(\tau_{(j,\cdot,\cdot)}+\zeta\);
\item
the variables determining the number of infectious contacts the individual makes \emph{within} its own group
\((\tilde{j},\cdot,\cdot)\) via home, school, or workplace;
\item
the variables specifying the number of offspring groups of type \(\mathbf e_{\tilde{j}}\) produced at \(\tau_{(j,\cdot,\cdot)}+\zeta\);
\item
the associated uniform variables used to select the second index \(\tilde n\in\{1,\ldots,N\}\) of the newly‑exposed individual in each such offspring group.
\end{enumerate}

Finally, set
$
   C_{(j,\cdot,\cdot)}:=\{\,C_{(j,\cdot,\cdot)}(i): i\ge 0\,\}.
$ 
The collections \(C_{(j,\cdot,\cdot)}\) are identically distributed over
all individuals of type \(j\), indeed they have the same law as the
corresponding \(C_{(j,\cdot)}\) introduced for the epidemic process.

\subsection{Coupling the two processes}

A crucial step in proving that  the epidemic process is close to the branching process during the early phase is to prove the stochastic domination
\[
A_t^{N}\,\leq\,A_t^{B}\,,
\]
i.e.\ the number of affected individuals in the epidemic process never exceeds that in the branching process.  
This inequality is pivotal for the main results in Section~\ref{sec:branching_results}, in particular for Lemma~\ref{bounding_diff_lemma_final}.

\begin{figure}
  \centering
    \includegraphics[width=0.6\linewidth, height=10cm]{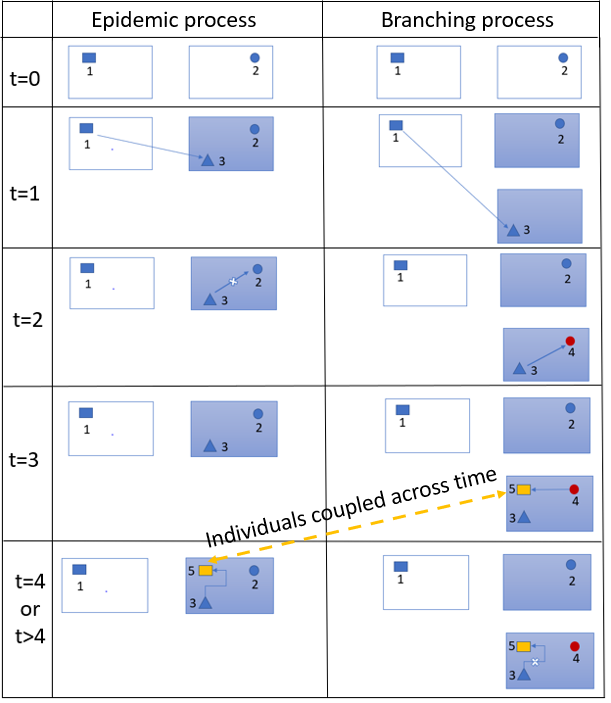}
  \caption{Demonstration for the need of a nuanced coupling. 
\label{coupling-1-alt}}
\end{figure}

Proving \(A_t^{N} \le A_t^{B}\) is not trivial: it requires a carefully designed coupling between the two processes.   Figure~\ref{coupling-1-alt} illustrates how the inequality can fail if the coupling is naively implemented. To follow this figure, assume a single community in which each household contains three types of individuals—rectangle, circle, and triangle.   The disease dynamics in the figure unfold as follows:

\begin{itemize}
    \item \textbf{$\bf t=0:$}  Two infected individuals, labelled 1 and 2, appear in both the epidemic and branching processes, each in a distinct household. Individual~1 in the epidemic process is coupled with its counterpart in the branching process, and likewise for individual~2.
    \item \textbf{$\bf t=1:$}  A rare event occurs-in the epidemic process, individual~1 infects susceptible individual~3 who resides with individual~2. In the branching process, individual~1 instead ``gives birth’’ to a new household whose sole exposed member is individual~3. The two 3’s are coupled.
    \item \textbf{$\bf t=2:$}  In the epidemic process, individual~3 contacts (and would infect) individual~2 inside its household, but individual~2 is already infected, so nothing changes.  By contrast, in the branching process the analogous household contact infects a new susceptible, labelled~4. Individual~4 is therefore \emph{uncoupled}.
    \item \textbf{$\bf t=3:$}  Nothing happens in the epidemic process.   In the branching process, individual~4 infects individual~5 (a rectangle) via another household contact, again producing an uncoupled individual.
    \item \textbf{$\bf t=4:$}  In the epidemic process, individual~3 now infects a rectangle (call this individual~5) within its own household.   In the branching process, however, the corresponding rectangle was already infected at \(t=3\); hence no new infection occurs.  Consequently, the epidemic process acquires an uncoupled individual who can trigger additional community infections, potentially breaking the desired inequality \(A_t^{N}\le A_t^{B}\).
\end{itemize}

To prevent such discrepancies, we permit \emph{cross‑time coupling}: an individual \((j,n)\) in the epidemic process may be coupled with \((j,n,\ell)\) in the branching process even if \(\tau_{(j,n,\ell)} \le \tau_{(j,n)}\). For example, we couple the epidemic rectangle~5 generated at \(t=4\) with the branching rectangle~5 generated at \(t=3\).

\vspace{3mm}
\noindent\textbf{Coupling prescription.}

Every individual \((j,n)\) in the epidemic process is coupled with some \((j,n,\ell)\) in the branching process such that
\[
C_{(j,n)} \;=\; C_{(j,n,\ell)}\,,
\]
i.e., they share \emph{all} random variables that govern disease progression after exposure. Their exposure times may differ, however, with \(\tau_{(j,n,\ell)}\le \tau_{(j,n)}\). 
Once the pair becomes infectious, \(i\geq 0\) time steps post‑exposure, 
\begin{enumerate}
    \item both individuals follow identical disease trajectories, driven by the same underlying randomness;
    \item they produce the same Poisson‑distributed numbers of household, school, and workplace contacts;
    \item the number of community contacts made by \((j,n)\) with individuals of type~\(\tilde{j}\) equals the number of offspring groups \(\mathbf e_{\tilde{j}}\) generated by \((j,n,\ell)\);
    \item the uniform random variable used to pick, from the \(N\) potential recipients \((\tilde{j},\cdot)\), the specific individual contacted by \((j,n)\) is identical to the uniform random variable that selects the second index \(\tilde n\in\{1,\dots,N\}\) of the newly exposed individual in the offspring group of \((j,n,\ell)\).
\end{enumerate}

\vspace{3mm}
\noindent \textbf{Coupling rules --- }\textit{coupling proceeds through the following rules.}
\begin{enumerate}
    \item \textbf{Initial time:}   
      At \(t=0\) as \({\bf B}_0 = {\bf X}_0\), we pair each initially exposed \((j,n)\) in the epidemic process with its counterpart \((j,n,\cdot)\) in the branching process.
    \item \textbf{Community infection:}  
      Suppose \((j,n)\) in the epidemic process infects a susceptible \((\tilde{j},\tilde{n})\) via community contact exactly \(i\) time steps after exposure.  
      The coupled branching individual \((j,n,\ell)\) simultaneously ``gives birth’’ to \((\tilde{j},\tilde{n},\tilde{\ell})\).  
      If \((\tilde{j},\tilde{n})\) is not yet coupled, we now couple it with \((\tilde{j},\tilde{n},\tilde{\ell})\).
    \item \textbf{Within group (household/school/workplace) infection:}  
      Consider a branching process group containing a coupled member \((j,n,\ell)\).  
      In that group, if a contact infects a new individual \((\tilde{j},n,\ell)\) and the epidemic individual \((\tilde{j},n)\) is still uncoupled, we couple the two.
    \item \textbf{Uncoupled lineage in branching process:}  
      All other branching‑process individuals remain uncoupled.  Specifically,
      \begin{enumerate}
          \item If a community contact in the epidemic process hits an already affected individual, the corresponding offspring in the branching process is uncoupled.
          \item Any community offspring produced by an uncoupled branching‑process individual stays uncoupled.
          \item A branching process group without coupled members will remain entirely uncoupled for all time.  Equivalently, if a group does contain a coupled member, its \emph{first} exposed individual must be coupled.
      \end{enumerate}
\end{enumerate}

Rules (1)–(3) guarantee that \emph{every} epidemic individual \((j,n)\) is paired with some \((j,n,\ell)\) in the branching process. Later in Section \ref{sec:proof_lemma_2}, we prove that for all $j$ and $n$,  \(\tau_{(j,n,\ell)}\le \tau_{(j,n)}\), thereby ensuring the domination \(A_t^{N} \le A_t^{B}\).

\section{Detailed proofs of Section \ref{sec:Theoretical_results} results}
\label{sec:proofs}

In this section we give detailed proofs of the asymptotic analysis results in Section \ref{sec:Theoretical_results} of the main paper.

\subsection{Proof of Lemma \ref{matrix_structure}}
\label{sec:proof_lemma_5.1}
\begin{itemize}

\item
Types ${\bf s}\in\mathcal{H}^c\setminus\mathcal U$ represent hospitalised, critical, deceased, or recovered individuals, whereas ${\bf q}\in\mathcal{H}$ correspond to exposed, infectious, or symptomatic states.  
Because an individual in a hospitalised, critical, deceased, or recovered state neither re‑enters an exposed, infectious, or symptomatic state nor produces offspring in those states, it follows that
\[
K({\bf s},{\bf q}) = 0 \quad\text{for all } {\bf s}\in\mathcal{H}^c\setminus\mathcal U,\;{\bf q}\in\mathcal{H}.
\]

\item
Define $K_1\in\mathbb{R}^{+\hat{\eta}\times\hat{\eta}}$ by 
$K_1({\bf s},{\bf q}) = K({\bf s},{\bf q})$ for ${\bf s},{\bf q}\in\mathcal{H}$.  The matrix $K_1$ therefore governs the dynamics of types that are currently infectious or may become infectious in future time steps. Since every infectious individual produces (with positive probability) offspring of each exposed type, $K_1$ is irreducible.

\item
Define $C \in {\mathbb{R}^+}^{(\eta - \hat{\eta}) \times (\eta - \hat{\eta})} $ by
$C({\bf s},{\bf q}) =K({\bf s},{\bf q}) $ for ${\bf s},{\bf q}\in\mathcal{H}^c\setminus\mathcal U$.   The individuals in groups of  type ${\bf s} \in \mathcal{H}^c\setminus\mathcal U$ do not generate new infections; they either transition to the next disease state or remain in their current state with positive probability.   Thus $C$ is upper‑triangular, with diagonal entries equal to the probability that an individual remains in the same state over one time step, each strictly less than~1.  
Consequently, every eigenvalue of $C$ lies in $(0,1]$.

\item
Define $M \in {\mathbb{R}^+}^{\hat{\eta} \times (\eta - \hat{\eta})} $ by 
$M({\bf s},{\bf q}) =K({\bf s},{\bf q}) $ for ${\bf s} \in \mathcal{H}$ and ${\bf q} \in\mathcal{H}^c\setminus\mathcal U$. 
\end{itemize}

Combining these blocks, the next‑generation matrix can be written as

\[
K = 
\begin{pmatrix}
K_1 & M \\
0   & C
\end{pmatrix},
\]

where $K_1$ is irreducible and $\rho(C) < \rho(K_1)$.

 \subsection{Proof of Theorem \ref{theorem_extending_branching_process}}

\label{sec:proof_theorem_5.2}

By Lemma \ref{matrix_structure} the next–generation matrix can be decomposed as  
\[
K=
\begin{pmatrix}
K_1 & M\\
0   & C
\end{pmatrix},
\]
where $K_1\in\mathbb{R}^{+\hat\eta\times\hat\eta}$ is irreducible and $C\in\mathbb{R}^{+(\eta-\hat\eta)\times(\eta-\hat\eta)}$ is upper‐triangular with diagonal entries in $(0,1]$.

\begin{itemize}
\item
Because $K$ is block‑upper‑triangular, its spectrum is the disjoint union of the spectra of $K_1$ and $C$.  
From Lemma \ref{matrix_structure} we have $\rho(C)\le 1$, hence
\[
\rho(K)=\rho(K_1).
\]

\item
Since $K_1$ is irreducible, $\rho(K_1)$ is its Perron–Frobenius eigenvalue; consequently $\rho(K)$ is the unique eigenvalue of $K$ with maximal modulus.

\item
Lemma~8.16 in \citep{matrix_analysis_notes} (p.\,69) then yields
\[
\lim_{t\to\infty}\frac{K^{t}}{\rho^{t}} = {\bf u}{\bf v}^{\!\top},
\]
where ${\bf u},{\bf v}\in\mathbb{R}^{+\eta}$ are the right and left Perron eigenvectors of $K$, respectively.
\end{itemize}

\medskip
The supercritical results for multitype branching processes stated in Theorem \ref{thm:stbp_theorem}  were proved in \citep[Chap.\,V]{branching_process_notes}.  
The key requirement there is precisely the convergence
\[
\lim_{t\to\infty}\frac{K^{t}}{\rho^{t}} = {\bf u}{\bf v}^{\!\top},
\]
now verified for our matrix $K$.

Moreover, for every $1\le i,j\le\eta$ the offspring law satisfies
\[
\mathbb{E}\!\bigl[B_1(j)\,\log B_1(j)\,\big\vert\,{\bf B}_0={\bf e}_i\bigr]<\infty,
\]
because each component of the offspring vector follows a Poisson distribution.  
Therefore all hypotheses of \citep{branching_process_notes} apply, and the conclusions of Theorem \ref{thm:stbp_theorem} hold for the branching process associated with $K$.

\subsection{Proof of Lemma \ref{bounding_diff_lemma_final}}

\label{sec:proof_lemma_2}

Before proceeding to the proof of Lemma~\ref{bounding_diff_lemma_final}, we first establish the domination
\begin{equation}\label{total_affected_ineq_2}
   A_t^{N}\;\le\;A_t^{B}.
\end{equation}

\vspace{3mm}
    
\noindent \textbf{Proof of (\ref{total_affected_ineq_2}):} 
Because every individual $(j,n)$ in  the epidemic process is coupled with a unique individual $(j,n,\ell)$ in the branching process, it suffices to show that the  exposure time of  individual $(j,n)$ is never earlier than that of its partner $(j,n,\ell)$, i.e.\ \(\tau_{(j,n)}\ge \tau_{(j,n,\ell)}\).  
We prove this by induction on~\(t\).

\begin{itemize}
    \item \textbf{Base case:} At time $0$ the two processes coincide, and each individual is coupled by Rule~(1), so the claim is immediate.

    \item \textbf{Induction hypothesis:} Assume that up to (and including) time~\(t\), every exposed individual $(j,n)$ in the epidemic process was exposed \emph{no earlier than} its coupled counterpart $(j,n,\ell)$ in the branching process.

    \item \textbf{Induction step (\(t\to t+1\)):}  
Suppose, for contradiction, that at time \(t+1\) an  individual \((\tilde{j},\tilde{n})\) is exposed in the epidemic process, while no  individual \((\tilde{j},\tilde{n},\cdot)\) with \(\tau_{(\tilde{j},\tilde{n},\cdot)}\le t+1\) exists in the branching process.  
Two cases can give rise to the exposure of individual \((\tilde{j},\tilde{n})\):

\begin{enumerate}
  \item \emph{Community exposure.}  
        Assume \((\tilde{j},\tilde{n})\) is infected by a community contact from \((j,n)\).  
        By the induction hypothesis, the coupled branching individual \((j,n,\ell)\) was exposed no later than \((j,n)\); hence, by Rule~(2), it must have produced an offspring \((\tilde{j},\tilde{n},\cdot)\) by time \(t+1\).  
        This contradicts the assumption that no such branching individual exists.

  \item \emph{Household/school/workplace exposure.}  
        Suppose \((\tilde{j},\tilde{n})\) is infected via a household, school, or workplace contact from \((j,\tilde{n})\).  
        Again, \((j,\tilde{n},\ell)\) was already exposed no later than \((j,\tilde{n})\), so by time \(t+1\) it has contacted \((\tilde{j},\tilde{n},\ell)\).  
        Because \((j,\tilde{n},\ell)\) is coupled, Rule~(4c) ensures that the first exposed individual in its household is also coupled.  
        Rule~(3) therefore couples \((\tilde{j},\tilde{n},\ell)\) as soon as it is exposed, i.e.\ at or before \(t+1\)—again a contradiction.
\end{enumerate}
\end{itemize}

Both cases lead to contradictions, so the induction is complete and inequality~\eqref{total_affected_ineq_2} holds.

\vspace{3mm}
\noindent Next to prove Lemma \ref{bounding_diff_lemma_final}, we analyze the difference between the epidemic and branching processes.

\vspace{3mm}
\noindent\textbf{Analyzing the difference between epidemic and branching processes.} To conduct a rigorous analysis, we define three key group classifications: completely coupled groups, multiple hit groups, and ghost groups.

\begin{enumerate}
    \item 
\textbf{Completely coupled groups:} At time $t$, a group $n$ in the epidemic process is classified as completely coupled if each affected individual $(j,n)$ within group $n$ is paired with a corresponding individual $(j,n,\ell)$ in the branching process (where $\ell$ remains constant across all $j$). Furthermore, each paired individual $(j,n)$ in the epidemic process must be exposed at the same time as its corresponding individual $(j,n,\ell)$ in the branching process. The corresponding group in the branching process is also termed a completely coupled group.
\item 
 \textbf{Multiple hit groups, Ghost groups: } All groups in the epidemic process that are not completely coupled are classified as multiple hit groups. While those in the branching process are designated as ghost groups.

\item
 \textbf{Evolution of the above group classes:} When an infectious individual $(j,n)$ in the epidemic process infects a susceptible individual within a completely coupled group $(\tilde{n})$ at time $t$ through community transmission, the group $(\tilde{n})$ transitions to a multiple hit group. Simultaneously, the corresponding completely coupled group $(\tilde{n}, \tilde{\ell})$ in the branching process becomes a ghost group. Additionally, the coupled individual $(j,n,\ell)$ in the branching process generates another ghost group.

When an infectious individual $(j,n)$ in the epidemic process infects an already affected individual within a completely coupled group $(\tilde{n})$ at time $t$ through community, group $(\tilde{n})$ maintains its completely coupled status with the corresponding group in the branching process. However, the paired individual $(j,n,\ell)$ in the branching process still generates a ghost group.

Note that the number of completely coupled groups at any time $t$ remains identical in both processes. Moreover, a completely coupled group in the epidemic process at time $t$ exhibits the same state as its corresponding group in the branching process, and vice versa. Consequently, only multiple hit groups and ghost groups contribute to the difference between $X_t^N({\bf s})$ and $B_t({\bf s})$.

\item \textbf{Notation:} Let $G_t^N$ denote the number of new ghost groups generated in the branching process at time $t$ due to infectious individuals infecting members of completely coupled groups in the epidemic process. All ghost groups present at time $t$ in the branching process are either newly generated ghost groups born at time $t$ or descendants of ghost groups generated at earlier time points. Let $D_{i,t-i}^N$ represent all descendants of $G_i^N$ after $(t-i)$ time steps, evaluated at time $t$. 
\end{enumerate}

Let $H_t^N$ denote the total number of ghost groups in the branching process at time $t$. Then,
\begin{equation}H_t^N \leq
\sum_{i=1}^t D_{i,t-i}^N.
\label{Ghost_eq_2}
\end{equation}

Since the number of completely coupled groups at time $t$ is identical in both the branching and epidemic processes, and ${A_t^N} \leq {A^B_t}$, it follows that the number of multiple hit groups in the epidemic process at time $t$ is less than the number of ghost groups in the branching process.  Consequently, for all ${\bf s}\in \mathcal{S} \setminus \mathcal{U}$,
 \begin{equation}\abs{{{B}}_t({\bf s}) - {{X}}^N_t({\bf s})}  \le 2 H^N_t. \label{diff_begin_2}\end{equation}

Combining inequalities (\ref{Ghost_eq_2}) and (\ref{diff_begin_2}), we obtain
\begin{equation}
\abs{{{B}}_t({\bf s}) - {{X}}^N_t({\bf s})} \leq 2 \sum_{i=1}^t D_{i,t-i}^N.
\label{diff_1_2}
\end{equation}

Before proceeding with the analysis, we state two key results, inequalities (\ref{diff_2_2}) and (\ref{diff_4_2}), whose proofs will be provided subsequently. 

For some constant $e_1 > 0$,
 \begin{equation}\sum_{i=1}^t \Exp{D_{i,t-i}^N} \le \frac{e_1}{N} {\bf 1}^T \sum_{i=0}^{t} {K}^{t-i} \Exp{(A_{i-1}^B)^2 }{\bf 1}, 
 \label{diff_2_2}
 \end{equation}
 
 where $A_i^B$ represents the total number of affected groups in the branching process at time $i$, and  ${\bf 1} \in {\mathbb{R}^+}^{\eta}$ is a vector with all entries equal to 1.
 
 For some constant $e_4>0$,
 \begin{equation}\Exp{\lrp{A^B_i}^2} \leq e_4 \rho^{2i}. \label{diff_4_2}
 \end{equation}
 
Assuming inequality (\ref{diff_2_2}) to be true, we combine (\ref{diff_1_2}) and (\ref{diff_2_2}) to
 
 \begin{equation}\Exp{\abs{{{B}}_t({\bf s}) - {{X}}^N_t({\bf s})}} \le \frac{2e_1}{N} {\bf 1}^T \sum_{i=0}^{t} {K}^{t-i} \Exp{(A_{i-1}^B)^2 } {\bf 1}.
 \label{diff_3_2}
 \end{equation}

 Since $\lim_{t \to \infty} \frac{K^t}{\rho^t} = {\bf u}{\bf v}^T$ (from Theorem \ref{theorem_extending_branching_process}), there exists a matrix $K_3 \in {\mathbb{R}^+}^{\eta\times\eta}$ such that for all $t \geq 0$,
 \begin{equation}K^t \le {\rho}^t  K_3.\label{matrix_inequality_new_2}
 \end{equation}
 
Assuming equation (\ref{diff_4_2}) holds and combining (\ref{diff_4_2}), (\ref{diff_3_2}), and (\ref{matrix_inequality_new_2}), we derive

\begin{equation*} \label{eq:Bound1_2}
      \begin{aligned}
        \Exp{\abs{{{B}}_t({\bf s}) - {{X}}^N_t({\bf s})}}   &\le \frac{2e_1}{N} {\bf 1}^T \sum_{i=0}^{t} \rho^{t-i} K_3 e_4 \rho^{2i-2} {\bf 1}\\
                                          &\le  \frac{2e_1}{N}  {\bf 1}^T    K_3 e_4 \frac{\rho^{2t-1}}{(\rho -1)}  {\bf 1} \\
                                          &= e\frac{\rho^{2t}}{N} , 
      \end{aligned}
    \end{equation*}
   
    where $e = \frac{2e_1e_4}{ \rho (\rho -1)} {\bf 1}^T    {K_3 }  {\bf 1}$.


\vspace{5mm}
 
\noindent\textbf{Proof of (\ref{diff_2_2}):} Recall that $G_t^N$ denotes  the number of new ghost groups generated in the branching process at time $t$ due to infectious individuals infecting members of completely coupled groups in the epidemic process.  When an infectious individual makes community contact with a susceptible individual in a completely coupled group, two new ghost groups are created. On the other hand, when an infectious individual makes community contact with an already affected individual in a completely coupled group, one new ghost group is generated. In either case, the number of ghost groups created is at most 2 per contact with an affected group.  Moreover, the number of completely coupled groups at time $t-1$ is upper bounded by $A_{t-1}^N$.

Let $Y_t^N$ denote the total number of community contacts in the epidemic process during the time interval $[t-1, t]$. For any given contact, the probability of hitting an individual in an already affected group is at most $g\frac{A^N_{t-1}+Y_t^N}{N}$. Therefore, we have

\begin{equation*}
    \begin{aligned}
    \Exp{{G}^N_t} &= \Exp{\Exp{{G}^N_t | \bf X^N_{t-1}} } \\
    & \leq \Exp{\Exp{2g \frac{(A_{t-1}^N+{Y}_t^N)}{N} {Y}_t^N | \bf X^N_{t-1}}} \\
    & \leq \Exp{\Exp{2g \frac{A_{t-1}^N{Y}_t^N}{N} |\bf X^N_{t-1}}} + \Exp{\Exp{2g \frac{({Y}_t^N)^2}{N} | \bf X^N_{t-1}}}. 
    \end{aligned}
\end{equation*}

To establish an upper bound for  ${Y}_t^N$, we assume that all individuals in affected groups in the epidemic process at time $t-1$ are infectious, and that community contacts by each infectious individual follow a Poisson distribution with parameter  $\beta_{max} = \max_{j\in [g]}\{\beta_{j}^c\}$. Under these assumptions, we have ${Y}_t^N \leq Z_t^N$ where $Z_t^N \sim Poisson(g\beta_{max}A_t^N)$. Then,
\[\Exp{{G}^N_t} \leq 2g^2\Exp{ \beta_{max}\frac{(A_{t-1}^N)^2}{N} } + 2g \Exp{\frac{g\beta_{max} A_{t-1}^N+g^2\beta_{max}^2 (A_{t-1}^N)^2}{N} }. \]

Since $A_{t-1}^N\geq 1$ , and setting $e_1=2(g^3\beta_{max}^2+2g^2\beta_{max})$, we obtain
\[\Exp{{G}^N_t} \leq e_1 \Exp{ \frac{(A_{t-1}^N)^2}{N} }. \]

 To establish an upper bound for $\Exp{D_{i,t-i}^N}$, we assume that each state ${\bf s} \in \mathcal S \setminus \mathcal U$ has $G_i^N$ new ghost groups born at time $i$. Since the new ghost groups $G_i^N$ created at time $i$ produce descendants according to the same branching process dynamics, we have
\begin{equation*} \Exp{D_{i,t-i}^N} \le \frac{e_1}{N} {\bf 1}^T  {K}^{t-i} \Exp{(A_{i-1}^N)^2 }{\bf 1}. 
 \end{equation*}

Summing both sides over all relevant time indices, we obtain
 \begin{equation}\sum_{i=1}^t \Exp{D_{i,t-i}^N} \le \frac{e_1}{N} {\bf 1}^T \sum_{i=1}^{t} {K}^{t-i} \Exp{(A_{i-1}^N)^2 }{\bf 1}. 
 \label{diff_end_2}
 \end{equation}

From inequalities (\ref{total_affected_ineq_2}) and (\ref{diff_end_2}), we conclude that
 \begin{equation*}\sum_{i=1}^t \Exp{D_{i,t-i}^N} \le \frac{e_1}{N} {\bf 1}^T \sum_{i=1}^{t} {K}^{t-i} \Exp{(A_{i-1}^B)^2 }{\bf 1}. 
 \end{equation*}

\noindent \textbf{Proof of (\ref{diff_4_2}):} Recall that $A^B_i=\sum_{{\bf s}\in \mathcal {S} \setminus \mathcal U}B_i({\bf s})$. Then, $A^B_i={\bf 1}^T {\bf B}_i= {\bf {B}}_i^T {\bf 1}$. Hence,
    \begin{equation} 
      \begin{aligned}
        \mathbb{E}\lrp{\lrp{A^B_i}^2} &\leq \Exp{({\bf 1}^T {B}_i)^2 } \\
                            &=  {\bf 1}^T \Exp{{\bf {B}}_i {\bf {B}}_i^T} {\bf 1}.
     \label{affected_square_bound}
      \end{aligned}
    \end{equation}

       Let $\text{Var}({\bf B}_1)\in {\mathbb{R}^+}^{\eta\times\eta}$ denote a matrix whose $({\bf s},{\bf q})$-th entry equals $\Exp{B_1({\bf s})B_1({\bf q})}-\Exp{B_1({\bf s})}\Exp{B_1({\bf q})}$. Let $V_{{\bf s}}$ denote $\text{Var}({\bf B}_1|{\bf B}_0={\bf e}_{{\bf s}})$ and $C_0$ denote the matrix whose $({\bf s},{\bf q})$-th entry is $\Exp{B_0({\bf s})B_0({\bf q})}$. From the branching process literature (Chapter 2, page 37, \citep{harris1963theory}), we have 

\begin{equation}
      \begin{aligned}
      \Exp{{\bf {B}}_{i} {\bf {B}}_{i}^T} & = ({K}^i)^T C_0 {K}^i +  &  \sum_{j=1}^{i} ({K}^{i-j})^T \Big(\sum_{{\bf s}\in \mathcal {S} \setminus \mathcal U}V_{{\bf s}}\Exp{{\bf {B}}_{j-1}({\bf s})}\Big) {K}^{i-j}. 
      \end{aligned}
      \label{bb_eq}
    \end{equation}
 
Furthermore, for our branching process, there exists a constant $e_2>0$ such that $V_{{\bf s}} \le e_2 {K}$ for all $ {\bf s}\in \mathcal {S} \setminus \mathcal U$. Substituting this bound into (\ref{bb_eq}), we obtain

\begin{equation}
      \begin{aligned}
      \Exp{{\bf {B}}_{i} {\bf {B}}^T_{i}} & = ({K}^i)^T C_0 {K}^i +  e_2 \sum_{j=1}^{i} ({K}^{i-j})^T \Big(\sum_{{\bf s}\in \mathcal {S} \setminus \mathcal {U}}\Exp{{B}_{j-1}({\bf s})}\Big) {K}^{i-j+1}.
      \label{matrix_multiply_bound} 
      \end{aligned}
    \end{equation}

Since $\mathbb{E}[\mathbf{B}_{t+1}] = (K^T)^{t+1} \mathbb{E}[\mathbf{B}_0]$, it follows from (\ref{matrix_inequality_new_2}) that  \begin{equation}
  {\bf B}_i \leq \rho^i (K_3)^T {\bf B}_0.
  \label{branching_basic_ineq}
\end{equation}.  

Combining (\ref{matrix_multiply_bound}) and (\ref{branching_basic_ineq}), we derive

\begin{equation*}
    \begin{aligned}
    \Exp{{\bf {B}}_{i} {\bf {B}}^T_{i}} & \le \rho^{2i} (K_3)^T C_0 K_3 + e_2 \tilde{d} \sum_{j=1}^{i} \rho^{2i-j} (K_3)^T  K_3\\
    & \le  \rho^{2i} \bigg((K_3)^T C_0 K_3 + \frac{\rho e_2 \tilde{d}}{\rho-1}  (K_3)^T  K_3 \bigg)\\
    & = \rho^{2i} {K}_4,
    \label{eq:BBT3} 
    \end{aligned}
\end{equation*}
where, $e_3 := \sum_{{\bf s}\in\mathcal S \setminus  \mathcal U} \Big(\Exp{{B}_0^T} K_3\Big)({\bf s})$ and $K_4 := \bigg((K_3)^T C_0 K_3 + \frac{\rho e_2 e_3}{\rho-1}  (K_3)^T  K_3\bigg)$.

Applying this bound to (\ref{affected_square_bound}) and setting $e_4=\mathbf{1}^T K_4 \mathbf{1}$, we conclude that
 
 \begin{equation*}
    \begin{aligned}\Exp{\lrp{A^B_i}^2} \leq e_4 \rho^{2i}. 
    \end{aligned}
\end{equation*}

\subsection{Proof for Theorem \ref{e_b_close}}
\label{sec:proof_theorem_5.3}
  For $\zeta > 0$, we have

    \begin{equation*}
      \begin{aligned} 
      \mathbb{P}\lrp{ \sup\limits_{t\in[0,\log_\rho (N^*/\sqrt{N})]} \abs{{{X}}^N_t({\bf s}) - {{B}}_t({\bf s})} \ge \zeta }  & \le \frac{1}{\zeta} \Exp{\sup\limits_{t\in[0,\log_\rho (N^*/\sqrt{N})]} \abs{{{X}}^N_t({\bf s}) - {{B}}_t({\bf s})}}\\ &
           \le  \frac{1}{\zeta} \sum\limits_{t=0}^{\log_\rho (N^*/\sqrt{N})} ~\Exp{\abs{{{X}}^N_t({\bf s}) - {{B}}_t({\bf s})}}, 
      \end{aligned}
    \end{equation*}
    where the first inequality follows from Markov's inequality, and the second inequality follows from the fact that the maximum of non-negative random variables is bounded by their sum. Similarly, 

     \begin{equation*}
      \begin{aligned}
        \mathbb{P}\lrp{ \sup\limits_{t \in[0, \log_\rho N^*]} \abs{ \frac{{{X}}^N_t({\bf s})}{\rho^t} - \frac{{{B}}_t({\bf s})}{\rho^t} } \ge \zeta } & \le \frac{1}{\zeta} \Exp{\sup\limits_{t\in[0,\log_\rho N^*]} \abs{\frac{{{X}}^N_t({\bf s})}{\rho^t} - \frac{{{B}}_t({\bf s})}{\rho^t}}} \\ &
                \le \frac{1}{\zeta} \sum_{t=0}^{\log_\rho N^*} ~\Exp{\abs{ \frac{{{X}}^N_t({\bf s})}{\rho^t} - \frac{{{B}}_t({\bf s})}{\rho^t}}}.
      \end{aligned}
     \end{equation*} 

    Therefore, it suffices to show that

    \[ \lim\limits_{N\rightarrow\infty} ~ \lrset{  ~ \sum_{t=0}^{\log_\rho (N^*/\sqrt{N})} ~\Exp{\abs{{{X}}^N_t({\bf s}) - {{B}}_t({\bf s})}}}  = 0, \quad \text{and} \quad  \lim\limits_{N\rightarrow\infty} ~ \lrset {  ~ \sum_{t=0}^{\log_\rho N^*} ~\Exp{\abs{ \frac{{{X}}^N_t({\bf s})}{\rho^t} - \frac{{{B}}_t({\bf s})}{\rho^t}}}} = 0. \]

    From Lemma \ref{bounding_diff_lemma_final}, we have

    \begin{equation}
      \begin{aligned}
        \Exp{\abs{{{X}}_t^N({\bf s}) - {{B}}_t({\bf s})}}    &\le {e} \frac{\rho^{2t}}{N} , 
        \label{difference_bound}
      \end{aligned}
    \end{equation}
    
     Taking sum from $t =0$ to  $t=\log_\rho (N^*/\sqrt{N})$ above, we observe that
    
     \[ \sum\limits_{i=0}^{\log_\rho (N^*/\sqrt{N})} ~\Exp{\abs{{{X}}^N_i({\bf s}) - {{B}}_i({\bf s})}} \leq  \frac{\rho^{2 \log_\rho (N^*/\sqrt{N})}}{N}\frac{{e}}{\rho^2-1}.\]

   Recalling that $N^* = N/\log N$, we conclude that
    \[ \lim\limits_{N\rightarrow \infty} ~ \Exp{ \sup\limits_{t\in [0, \log_\rho (N^*/\sqrt{N})]} \abs{{{X}}_t^N({\bf s}) - {{B}}_t({\bf s})} } = 0 . \]

   For the second limit, we use inequality \eqref{difference_bound} to bound $\sum_{i=0}^{t}\frac{1}{\rho^i}\Big(  \mathbb{E}(\abs{{{X}}_{i}^N({\bf s})-{{B}}_{i}({\bf s})})\Big)$.  We have

    \begin{equation*}
    \begin{aligned}
    \sum_{i=0}^{t}\frac{1}{\rho^i}\Big(\  \mathbb{E}(\abs{{X}_{i}^N({\bf s})-{B}_{i}({\bf s})})\Big) &
     \leq  \frac{\rho^{t}}{N}  \frac{{e}}{\rho-1} .
    \label{eq:bound6}
   \end{aligned}
\end{equation*}
    
    Since  $t=\log_\rho N^*$, we obtain
    \[ \sum\limits_{i=0}^{t} ~\Exp{\abs{\frac{{X}^N_i({\bf s})}{\rho^i} - \frac{{B}_i({\bf s})}{\rho^i}}}  \leq \frac{1}{\log N}   \frac{{e}}{\rho-1}.\]
    
   Therefore,
    \[ \lim\limits_{N\rightarrow\infty}  ~ \Exp{\sup\limits_{t\in [0,\log_\rho N^*]} \abs{\frac{{X}^N_t({\bf s})}{\rho^t} - \frac{{B}_t({\bf s})}{\rho^t}}} = 0. \]

  \subsection{Proof of Theorem \ref{deterministic_theorem}} \label{sec:app_proof_DetEv}
  
  We prove Theorem \ref{deterministic_theorem} by induction.  
For brevity we suppress the explicit dependence on $W$ and write ${\bf\bar\mu}_t$ and $\bar{\lambda}^{\mathrm{c}}_{t}(j)$ in place of ${\bf\bar\mu}_t(W)$ and $\bar{\lambda}^{\mathrm{c}}_{t}(j, W)$.
\begin{enumerate}
    \item 
\textbf{Induction base: }  Assumption~\ref{deterministic_assumption} (Section~\ref{sec:deterministic_results}) serves as the base case.

\item  \textbf{Induction hypothesis:}
Assume
\[
{\bf\mu}^{N}_{t}\xrightarrow{\mathrm{P}}{\bf\bar\mu}_{t}.
\]

 \item  \textbf{Induction step:} For every type ${\bf s}\in\mathcal S$ we have

 \[
\mu^{N}_{t+1}({\bf s})=\frac1N\sum_{n=1}^{N}\mathbf 1\!\left(S_{ n}(t+1)={\bf s}\right),
\]

  with conditional transition probabilities
\[
\mathbb P\bigl(S_{ n}(t+1)={\bf s}\mid{\bf\mu}^{N}_{t}, S_{ n}(t)\bigr)
   =p\bigl(S_{n}(t),{\bf s},\lambda^{\mathrm{c},N}_{t}(1),\dots,\lambda^{\mathrm{c},N}_{t}(g)\bigr).
\]

Define the martingale difference
\[
\mathcal M^{N}_{t+1}({\bf s})\;=\;
\frac1N\sum_{n=1}^{N}
\Bigl[\mathbf 1\!\left(S_{n}(t+1)={\bf s}\mid S_{n}(t),{\bf\mu}^{N}_{t}\right)
      -p\bigl(S_{n}(t),{\bf s},\lambda^{\mathrm{c},N}_{t}(1),\dots,\lambda^{\mathrm{c},N}_{t}(g)\bigr)
\Bigr].
\]
Then

\[
\mu^{N}_{t+1}({\bf s})
   =\sum_{{\bf s'}\in\mathcal S} \Big[
      p\bigl({\bf s'},{\bf s},\lambda^{\mathrm{c},N}_{t}(1),\dots,\lambda^{\mathrm{c},N}_{t}(g)\bigr)\,
      \mu^{N}_{t}({\bf s'})
      \Big] +\mathcal M^{N}_{t+1}({\bf s}).   \tag{\theequation}\label{eq:munplus1}
\]

Because $\mathcal M^{N}_{t+1}({\bf s})$ is a zero‑mean term whose variance tends to~$0$ as $N\to\infty$, we have
\[
\mathcal M^{N}_{t+1}({\bf s})\xrightarrow{\mathrm{P}}0.
\]

It therefore suffices to show that, for every ${\bf s},{\bf s'}\in\mathcal S$,
\[
p \bigl({\bf s'},{\bf s},\lambda^{\mathrm{c},N}_{t}(1),\dots,\lambda^{\mathrm{c},N}_{t}(g)\bigr)\,\mu^{N}_{t}({\bf s'})
\;\xrightarrow{\mathrm{P}}\;
p \bigl({\bf s'},{\bf s},\bar \lambda^{\mathrm{c}}_{t}(1),\dots,\bar \lambda^{\mathrm{c}}_{t}(g)\bigr)\,\bar\mu_{t}({\bf s'}).
\tag{\theequation}\label{eq:transition_conv}
\]

\medskip\noindent
Write the left‑hand side of \eqref{eq:transition_conv} as  
\begin{align*}
&\;p\bigl({\bf s'},{\bf s},\bar \lambda^{\mathrm{c}}_{t}(1),\dots,\bar \lambda^{\mathrm{c}}_{t}(g)\bigr)\,\mu^{N}_{t}({\bf s'}) \\
&\quad
+\Bigl[
      p\bigl({\bf s'},{\bf s},\lambda^{\mathrm{c},N}_{t}(1),\dots,\lambda^{\mathrm{c},N}_{t}(g)\bigr)
     -p\bigl({\bf s'},{\bf s},\bar \lambda^{\mathrm{c}}_{t}(1),\dots,\bar \lambda^{\mathrm{c}}_{t}(g)\bigr)
 \Bigr]\mu^{N}_{t}({\bf s'}).
\end{align*}

\begin{itemize}
\item
By the induction hypothesis $\mu^{N}_{t}({\bf s'})\xrightarrow{\mathrm{P}}\bar\mu_{t}({\bf s'})$, so the first term converges to the required limit.

\item
For the second term note that each infection rate
\[
\lambda^{\mathrm{c},N}_{t}(j)=
\sum_{{\bf q}\in\mathcal S\setminus\mathcal U}
   \mu^{N}_{t}({\bf q})
   \sum_{\tilde j=1}^{g}
      \mathbf 1\!\bigl({\bf q}(j)=\text{infectious}\bigr)\,
      \psi_{\tilde j,j}\,\beta^{c}_{\tilde j}
\]
is a continuous function of ${\bf\mu}^{N}_{t}$; hence $\lambda^{\mathrm{c},N}_{t}(j)\xrightarrow{\mathrm{P}}\bar \lambda^{\mathrm{c}}_{t}(j)$.  
Because $p(\cdot)$ is uniformly continuous on its compact domain and both $\lambda^{\mathrm{c},N}_{t}(j)$ and $\bar \lambda^{\mathrm{c}}_{t}(j)$ are bounded, the difference in brackets converges to $0$ in probability.
\end{itemize}

Thus \eqref{eq:transition_conv} holds, and taking limits in \eqref{eq:munplus1} yields
\[
\mu^{N}_{t+1}({\bf s})\xrightarrow{\mathrm{P}}\bar\mu_{t+1}({\bf s})
      :=\sum_{{\bf s'}\in\mathcal S}
          p\bigl({\bf s'},{\bf s},\bar \lambda^{\mathrm{c}}_{t}(1),\dots,\bar \lambda^{\mathrm{c}}_{t}(g)\bigr)\,
          \bar\mu_{t}({\bf s'}).
\]

Therefore the induction step is complete, and Theorem~\ref{deterministic_theorem} follows.

 \end{enumerate}

\section{Behaviour of Simple Compartmental Models in the Early Branching‑Process Phase}
\label{sec:compartmental_models}

\begin{example}[SIR model]\normalfont
The \textit{Susceptible–Infectious–Recovered} (SIR) model is the most basic aggregate compartmental model in epidemiology.  
Every individual in the population is assumed to be identical, i.e., each has the same community–transmission rate and the same disease‑progression parameters.  
Let $S(t), I(t)$, and $R(t)$ denote the numbers of susceptible, infectious, and recovered individuals at time~$t$.  
An infectious individual transmits the disease at rate~$\beta$ and recovers at rate~$r$.  
The dynamics are described by
\[
\frac{dS(t)}{dt} = -\frac{\beta}{N}\,S(t)\,I(t), 
\qquad
\frac{dI(t)}{dt} = \frac{\beta}{N}\,S(t)\,I(t) - r\,I(t), 
\qquad
\frac{dR(t)}{dt} = r\,I(t).
\]

A discrete‑time analogue is more convenient for our setting; we fix the time step to~$1$.  
Again, let $\beta$ be the community–transmission rate, so that the number of contacts an infectious individual makes in one time step is Poisson with mean~$\beta$, and let $r$ be the per‑step recovery rate.  
Up to time $t = \log_{\rho} N^{*}$ the epidemic closely follows a branching process.  
Label infectious individuals as type~1 and recovered individuals as type~2; the offspring (next‑generation) matrix is
\[
K = \begin{pmatrix} 1+\beta-r & r \\[2pt] 0 & 1 \end{pmatrix}.
\]
Hence $K_1 = (1+\beta-r)$, $C = (1)$, $M = (r)$, and $\rho(K)=\rho(K_1)=1+\beta-r$.  
The (right) eigenvector associated with~$\rho$ is
\[
v = \begin{pmatrix} 1 \\[4pt] \dfrac{r}{\beta-r} \end{pmatrix}.
\]
Consequently, for $t \le \log_{\rho} N^{*}$ (and large~$N$) the number of infections grows exponentially at rate $\rho=1+\beta-r$, and the ratio of infectious to recovered individuals quickly approaches $(\beta-r)/r$.
\end{example}

\begin{example}[SEIR model]\normalfont
The \textit{Susceptible–Exposed–Infectious–Recovered} (SEIR) model adds an \emph{exposed} compartment to the SIR framework.  
All individuals are still assumed identical.  
An infectious individual makes Poisson$(\beta)$ contacts per time step, an exposed individual becomes infectious at rate~$p$, and an infectious individual recovers at rate~$r$.

Let type~1 be exposed, type~2 infectious, and type~3 recovered.  
The branching‑process offspring matrix is
\[
K=\begin{pmatrix}
1-p & p & 0 \\ 
\beta & 1-r & r \\ 
0 & 0 & 1
\end{pmatrix}.
\]
Thus $C=(1)$, $M=\begin{pmatrix}0\\ r\end{pmatrix}$, and
\[
K_{1}= \begin{pmatrix} 1-p & p \\ \beta & 1-r \end{pmatrix}, 
\qquad
\rho(K)=\rho(K_{1}) = 1+\frac{1}{2} ~\Big(\sqrt{(p+r)^{2}+4p(\beta-r)}-(p+r)\Big).
\]
A corresponding eigenvector is
\[
v = \begin{pmatrix}
\beta \\[4pt] p+\rho-1 \\[4pt] r+\dfrac{rp}{\rho-1}
\end{pmatrix}.
\]

Hence, up to $t = \log_{\rho} N^{*}$ the epidemic grows like $\rho^t$ with
\[
\rho = 1+\frac{1}{2} ~ \Big(\sqrt{(p+r)^{2}+4p(\beta-r)}-(p+r)\Big),
\]
and the proportions of exposed, infectious, and recovered individuals rapidly converge to the normalized vector $\bigl(\beta,\; p+\rho-1,\; r+rp/(\rho-1)\bigr)$.
\end{example}

\begin{example}[SIR model with two age groups]\normalfont
Different age groups often have distinct disease trajectories.  
We therefore split the population into two age groups, with proportions $\pi_{1}$ and $\pi_{2}=1-\pi_{1}$.  
Each infectious individual makes Poisson$(\beta)$ contacts per time step, distributed at rates $\pi_{1}\beta$ to group~1 and $\pi_{2}\beta$ to group~2.  
Recovery rates are $r_{1}$ and $r_{2}$ for the two groups, respectively.

We track four types: infectious group~1, infectious group~2, recovered group~1, and recovered group~2.  
The offspring matrix is
\[
K=\begin{pmatrix}
\pi_{1}\beta+1-r_{1} & \pi_{2}\beta & r_{1} & 0 \\ 
\pi_{1}\beta & \pi_{2}\beta+1-r_{2} & 0 & r_{2} \\ 
0 & 0 & 1 & 0 \\ 
0 & 0 & 0 & 1
\end{pmatrix}.
\]
With the notation of Lemma~\ref{matrix_structure},
\[
C=\begin{pmatrix} 1 & 0 \\ 0 & 1 \end{pmatrix},
\qquad
M=\begin{pmatrix} r_{1} & 0 \\ 0 & r_{2} \end{pmatrix},
\qquad
K_{1}= \begin{pmatrix} 
\pi_{1}\beta+1-r_{1} & \pi_{2}\beta \\[4pt] 
\pi_{1}\beta & \pi_{2}\beta+1-r_{2} 
\end{pmatrix}.
\]
The dominant eigenvalue is
\[
\rho(K)=1+\frac{1}{2}\Bigl(\sqrt{(\beta-r_{1}-r_{2})^{2}+4\pi_{1}r_{2}(\beta-r_{1})+4\pi_{2}r_{1}(\beta-r_{2})}+(\beta-r_{1}-r_{2})\Bigr).
\]
An associated eigenvector is
\[
v=\begin{pmatrix}
\pi_{1}\beta \\[4pt]
r_{1}+\rho-1-\pi_{1}\beta \\[4pt]
\dfrac{r_{1}}{\rho-1}\,\pi_{1}\beta \\[6pt]
\dfrac{r_{2}}{\rho-1}\,\bigl(r_{1}+\rho-1-\pi_{1}\beta\bigr)
\end{pmatrix}.
\]

Thus, for $t\le\log_{\rho}N^{*}$ (and large~$N$) the epidemic grows at rate
\[
\rho = 1+\frac{1}{2}\Bigl(\sqrt{(\beta-r_{1}-r_{2})^{2}+4\pi_{1}r_{2}(\beta-r_{1})+4\pi_{2}r_{1}(\beta-r_{2})}+(\beta-r_{1}-r_{2})\Bigr),
\]
and the four population types quickly stabilise in the ratio
\[
\Bigl(\pi_{1}\beta,\; r_{1}+\rho-1-\pi_{1}\beta,\; r_{1}\pi_{1}\beta/(\rho-1),\; r_{2}(r_{1}+\rho-1-\pi_{1}\beta)/(\rho-1)\Bigr).
\]
\end{example}

\section{Detailed ABS model}
\label{sec:detailed_abs_model}

\subsection{Synthetic city generation}
The first step in our agent-based model is to model a synthetic city that respects the demographics of the city that we want to study. Our city generator uses the following data as input:
\begin{itemize}
  \item Geo-spatial data that provides information on different localities of a city (components) along with its boundaries.
  \item Population in each locality, with break up on those living in high density and low density areas.
  \item Age distribution in the population.
  \item Household size distribution (in high and low density areas) and some information on the age composition of the houses (e.g., generation gaps, etc.)
  \item The number of employed individuals in the city.
  \item Distribution of the number of students in schools and colleges.
  \item Distribution of the workplace sizes.
  \item Distribution of commute distances.
  \item Origin-destination densities that quantify movement patterns within the city.
\end{itemize}

Taking the above data into account, individuals, households, workplaces, schools, and community spaces are instantiated. Individuals are then assigned to households, workplaces or schools, and community spaces. The algorithms for the assignments do a coarse matching. 
The interaction spaces -- households, workplaces or schools, and community spaces -- reflect different social networks and transmission happens along their edges. There is interaction among these graphs because the nodes are common across the graphs. An individual of school-going age who is exposed to the infection at school may expose others at home. This reflects an interaction between the school graph and the household graph. Similarly other graphs interact.

We now describe how individuals are assigned to interaction spaces.

\begin{enumerate}
    \item 

\textbf{Individuals and households}: $N$ individuals are instantiated and ages are sampled according to the age distribution in the population. Households (based on the $N$ and the mean number per household) are then instantiated and assigned a random number of individuals sampled according to the distribution of household sizes. An assignment of individuals to households is then done to match, to the extent possible, the generational structure in typical households. The households are then assigned to localities so that the total number of individuals in the locality is in proportion to population density in the locality, taken from census data.
The generational gap, household distribution, and age distribution patterns are assumed to be uniform across the localities in the city. Each household in a locality is then assigned a random location in the locality, and all individuals associated with the household are assigned the same geo-location as the household.
Based on the age and the unemployment fraction, each individual is either a student or a worker or neither.

\item
\textbf{Assignment of schools}: Children of school-going ages 5-14 and a certain fraction of the population aged 15-19 are assigned to schools. These are taken to be students. The remaining fraction of the population aged 15-19 and a certain fraction of the population aged 20-59, based on information on the employed fraction,
are all classified as workers and are assigned workplaces. The rest of the population (nonstudent, unemployed) is not assigned to either schools or workplaces.

We assign students to schools on a locality-by-locality basis. In each locality, we have a certain number of students. We pick a school size from a given school size distribution and instantiate a school of this size and place it randomly in that locality. Students who live in that locality are picked randomly and assigned to this school until that school is filled to its capacity. We repeat this procedure until all students in that locality gets assigned to a school, and then we repeat this procedure for all localities. This procedure could lead to at most one school per locality whose capacity is more than its sampled capacity.

\item
\textbf{Assignment of workplaces}: Workplace interactions can enable the spread of an epidemic. 
The assignment of individuals to workplaces is done in two steps. In the first step, for each individual who goes to work, we decide the locality where his/her office is located. This assignment of a ``working locality" is based on an Origin-Destination (OD) matrix. An OD matrix is a square matrix whose number of rows equals the number of localities, and its $(i, j)$th entry tells us the fraction of people who travel from locality $i$ to locality $j$ for work. In the second step, for each locality, we sample a workplace size from a workplace size distribution, create a workplace of this capacity and place it uniformly-at-random in that locality. We then randomly assign individuals who work in that locality to this workplace. Similar to assignment of schools, we continue to create workplaces in this locality until every individual working in that locality gets assigned to a workplace, and we repeat this procedure for all localities. The  OD matrix can be obtained from  the `zone to zone'' travel data of the city. 

\item
\textbf{Community spaces}: Community spaces include day care centres, clinics, hospitals, shops, markets, banks, movie halls, marriage halls, malls, eateries, food messes, dining areas and restaurants, public transit entities like bus stops, metro stops, bus terminals, train stations, airports, etc. Each individual sees a community that is personalised to the individual's location and age based on locality-level common local communities and a distance-kernel based weighting. Details are given later.

\end{enumerate}

The output of all the above is our synthetic city on which infection spreads.

\subsection{Computing infection rate $\lambda_j(t)$}

Recall that a susceptible individual $j$ at time $t$ receives a total infection rate $\lambda_j(t)$, which is the sum of infection rates from distinct interaction spaces, expressed as  
\[\lambda_j(t)= \lambda_j^{h}(t)+\lambda_j^{s}(t)+\lambda_j^{w}(t)+\lambda_j^{c}(t).\]

Further, recall that 
\[\lambda_j^{h}(t)= \sum_{\tilde{j} : h(\tilde{j}) = h(j)}     \frac{\beta^h_{\tilde{j}} }{n_{\tilde{j}}^h} \cdot {\mathbf 1}_{\mathrm{inf},\tilde{j}}(t).  \]

School ($\lambda_j^{s}(t)$), and workplace ($\lambda_j^{w}(t)$) infection rates are defined similarly. Further, the community infection rate was defined as 

 \begin{equation*} 
 \lambda_{j}^{c}(t)=    \sum_{\tilde j} \psi_{\tilde j,j} \cdot \frac{\beta_{\tilde{j}}^c }{n} \cdot {\mathbf 1}_{\mathrm{inf},\tilde{j}}(t).
  \end{equation*}

To introduce heterogeneity into the model, for 
each individual $j$ has three parameters:  relative infectiousness variable $\gamma_j$, infection-stage-related infectiousness $\kappa$  and  severity variable $\alpha_j$.

\begin{enumerate}
    \item  Infectiousness $\gamma_j$ is a random variable that is Gamma distributed with shape 0.25 and scale 4 (so the mean is 1).
    \item   If the individual gets exposed at time $\tau_j$, a relative infection-stage-related infectiousness is taken to be $\kappa(t - \tau_j)$ at time $t$. For the disease progression described in the previous section, this is 1 in the presymptomatic and asymptomatic stages, 1.5 in the symptomatic, hospitalised, and critical stages, and 0 in the other stages.  
    \item  The severity variable captures severity-related absenteeism at school/workplace, associated decrease of infection spread at school/workplace, and the increase of infection spread at home.
Severity $\alpha_j = 1$ if the individual suffers from a severe infection and $\alpha_j = 0$ otherwise; this is sampled at 50\% probability independently of all other events. 
The functions $\psi_s(\cdot)$ and $\psi_w(\cdot)$ account for absenteeism in case of a severe infection. It can be time-varying and can depend on school or workplace. We take $\psi_s(t) = 0.8$ and $\psi_w(t) = 0.5$ while infective and after one day since infectiousness. School-goers with severe infection contribute lesser to the infection spread, due to higher absenteeism, than those that go to workplaces; moreover, the absenteeism results in an increased spreading rate at home. 

\end{enumerate}

Let $\beta_h$, $\beta_s$, $\beta_w$,  $\beta_c$
denote the transmission coefficients at home, school, workplace,  community spaces, 
respectively. Then the infection rates can be expressed as follows, 

\begin{equation*}
\begin{aligned}
    \lambda_j^h(t)
 & =  \sum_{\tilde{j} : h(\tilde{j}) = h(j)} \frac{\beta^{h} \cdot \left( \gamma_{\tilde{j}} ~  \kappa(t - \tau_{\tilde{j}}) ~ (1 + \alpha_{\tilde{j}})\right)}{n^h_j} \cdot {\mathbf 1}_{\mathrm{inf},\tilde{j}}(t)
\\
 \lambda^s_j(t) &= \sum_{\tilde{j} : s(\tilde{j}) = s(j)} \frac{\beta^{s} \cdot  \left( \gamma_{\tilde{j}} ~  \kappa(t - \tau_{\tilde{j}}) ~ (1 -  \psi_s(t - \tau_{\tilde{j}}) \alpha_{\tilde{j}} )\right)}{n^s_j} \cdot {\mathbf 1}_{\mathrm{inf},\tilde{j}}(t)
 \\
 \lambda^w_j(t) &= \sum_{\tilde{j} : w(\tilde{j}) = w(j)} \frac{\beta^{w} \cdot \left( \gamma_{\tilde{j}} ~  \kappa(t - \tau_{\tilde{j}}) ~ (1 - \psi_w(t - \tau_{\tilde{j}}) \alpha_{\tilde{j}})\right) }{n^w_j} \cdot {\mathbf 1}_{\mathrm{inf},\tilde{j}}(t)
 \\
 \lambda^c_j(t) &= \sum_{\tilde{j}=\{1\ldots N\}} \psi_{\tilde{j},j}  \cdot
  \frac{\beta^{c} \cdot \left(    \gamma_{\tilde{j}} \kappa(t - \tau_{\tilde{j}})  (1 + \alpha_{\tilde{j}})  \right)}{N} \cdot {\mathbf 1}_{\mathrm{inf},\tilde{j}}(t)   
\end{aligned}
\end{equation*}

where
\begin{equation*}
\psi_{\tilde{j},j} =  \frac{r_{{c}(\tilde{j})} \zeta(a_j) \zeta(a_{\tilde{j}})  f(d_{j,c(j)})f(d_{\tilde{j},c(\tilde{j})}) f(d_{c(j),c(\tilde{j})})}
  {\sum_{\tilde{c}}f(d_{c,\tilde{c}})\left(\frac{1}{N}\sum_{i:c(i)=c(\tilde{j})} f(d_{i,c(i)})\right)}
\end{equation*}

The parameter $\psi_{\tilde{j},j}$ models various spatial and social factors, including locality-based community structure, individual mobility patterns within communities, proximity to community centers, and population density variations.  It represents how much individual $\tilde{j}$ in community/locality $\tilde{c}$ impacts individual $j$ in community/locality $c$. 
The factor $r_c$ stands for a high-density interaction multiplying factor. For Mumbai, $r_c = 2$ for some high density areas and $r_c = 1$ for the other areas.
The function $\zeta(a)$ is the relative travel-related contact rate of an individual aged $a$. We take this to be 0.1, 0.25, 0.5, 0.75, 1, 1, 1, 1, 1, 1, 1, 1, 0.75, 0.5, 0.25, 0.1 for the various age groups in steps of 5 years, with the last one being the 80+ category.

 Further each individual contributes in a distance-weighted way in how an individual $\tilde{j}$ in a locality $c(\tilde{j})$ affects another individual $j$ in another locality $c$. To account for the distance between $j$ and $\tilde{j}$, we divide into three component  $d_{j,c(j)}$ is the distance between individual $j$ lives and its locality center $c(j)$,  distance between community centers $d_{c,\tilde{c}}$ and   distance $d_{\tilde{j},c(\tilde{j})}$. 
The function $f(\cdot)$ is a distance kernel that can be matched to the mobility patterns in the city. We set the distance kernel to $f(d) = 1/(1+(d/a)^b)$ and $d \ll a$,
We take $a = 10.751$\,km and $b = 5.384$, based on a data fitted to the city under consideration.

\subsection{Disease progression}

\begin{figure*}[h!]
      \centering
 \includegraphics[width=0.8\linewidth]{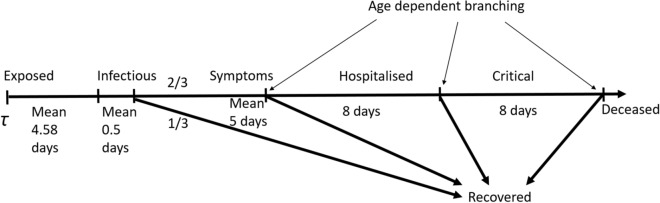}
\caption{ A simplified model of Covid-19 progression.}
\label{fig:diseaseProgression}
\end{figure*}

We have used a disease progression model of Covid-19 disease from \cite{verity2020estimates}.  An individual may have one of the following states: susceptible, exposed, infective (pre-symptomatic or asymptomatic), recovered, symptomatic, hospitalised, critical, or deceased (see  Figure~\ref{fig:diseaseProgression}).

 Let $\tau$ denote the time at which an individual is exposed to the virus, see Figure~\ref{fig:diseaseProgression}. The incubation period is random with the Gamma distribution of shape 2 and scale 2.29; the mean incubation period is then 4.6 days. Individuals are infectious for an exponentially distributed period of mean duration 0.5 of a day. 
 We assume that a third of the patients recover, these are the asymptomatic patients; the remaining two-thirds develop symptoms. 
 Symptomatic patients are assumed to be 1.5 times more infectious during the symptomatic period than during the pre-symptomatic but infective stage. Individuals either recover or move to the hospital after a random duration that is exponentially distributed with a mean of 5 days. See Table \ref{Disease_Progression_parameters} for a summary. The probability that an individual recovers depends on the individual's age.  It is also assumed that recovered individuals are no longer infective nor susceptible to a second infection. While hospitalised individuals may continue to be infectious, they are assumed to be sufficiently isolated, and hence do not further contribute to the spread of the infection. Further progression of hospitalised individuals to critical care and fatality is given in Table \ref{tab:demand}.

\begin{table*}[h!]
  \centering
    \small
    \caption{Disease progression parameters \label{Disease_Progression_parameters}}
    \begin{tabular}{|c|c|}
      \hline
      {\bf Parameter description} & {\bf Values}\\
      \hline
      Incubation Period & Gamma distributed with shape 2 and scale 2.29\\
      Asymptomatic Period & Exponentially distributed with mean duration 0.5 of a day \\ Symptomatic Period &  Exponentially distributed with mean duration of 5 days \\ hospitalization Period & 8 days \\ Critical Period & 8 days \\
      \hline
    \end{tabular}
\end{table*}

\begin{table}[ht]
\caption{Age-wise hospitalization transition probabilities}
\label{tab:demand}
\centering
\begin{tabular}{|l|c|c|c|}
 \hline
 Age-group & \% symptomatic cases       & \% hospitalised cases & \% of critical cases  \\
 (years)   & requiring hospitalization  & requiring critical care & deceased \\
 [0.5ex]
 \hline
 0 to 9   & 0.1\% & 5.0\% & 40\% \\
 10 to 19 & 0.3\% & 5.0\% & 40\% \\
 20 to 29 & 1.2\% & 5.0\% & 50\% \\
 30 to 39 & 3.2\% & 5.0\% & 50\% \\
 40 to 49 & 4.9\% & 6.3\% & 50\% \\
 50 to 59 & 10.2\% & 12.2\% & 50\% \\
 60 to 69 & 16.6\% & 27.4\% & 50\% \\
 70 to 79 & 24.3\% & 43.2\% & 50\% \\
 80+      & 27.3\% & 70.9\% & 50\% \\
 \hline
\end{tabular}
\end{table}

\subsection{Public health safety measures (PHSMs)}
     
 Our simulator has the capability to accommodate PHSMs/interventions employed to control the disease spread Table~\ref{tab:interventions} describes some of the interventions that are implemented in our simulator. It is easy to include any other implementation of interest as well.

\begin{table*}[h!]
  \centering
    \small
    \caption{Interventions as implemented in the simulator \label{tab:interventions}}
    \begin{tabular}{|c|c|}
      \hline
      {\bf Intervention} & {\bf Description}\\
      \hline
      No intervention & Business as usual\\
      Lockdown & For compliant households, household rates are doubled, no workplace \\ & interactions except for 25\% leakage (for essential services), community \\ & interactions reduce by 75\%.  For non-compliant households, workplace \\ & interactions only have a leakage of 25\%,  community interactions are \\ & unchanged, and household interactions increase by 50\%\\
	  Case Isolation & Compliant symptomatic individuals stay at home for 7 days, reducing  non- \\ & household contacts by 75\%. Household contacts remain unchanged.\\
	  Home Quarantine & Following identification of a symptomatic case in a compliant household, \\ & all household members remain at home for 14 days. Household  contact \\ & rates double,  contacts in the community reduce by 75\%\\
	  Social distancing of the elderly & All compliant individuals over 65 years of age reduce their community \\ & interactions by 75\%\\
	  Schools and colleges closed & Self explanatory\\
	  Masks & Reduce community transmission by 20\%\\
      \hline
    \end{tabular}
\end{table*}

\subsection{Seeding of infection}
A small number of individuals can be set to either exposed, presymptomatic/asymptomatic, or symptomatic states, at time $t = 0$, to seed the infection. This can be done randomly based either on locality-level probabilities, which could be input to the simulator, or it can be done uniformly at random across all localities in the city.

\subsection{Calibration}
We calibrate our model by tuning the transmission coefficients at various interaction spaces under the no-intervention scenario in order to match the cumulative fatalities to a target curve observed in the city. 
Calibrated model parameters are given in Table \ref{tab:params}.

\begin{table}[h!]
\caption{Calibrated model parameters}
\label{tab:params}
\centering
\begin{tabular}{|l|c|c|c|}
 \hline
 Parameter & Symbol & Value \\
 \hline
 Transmission coefficient at home   & $\beta_h$ & 1.227 \\
 Transmission coefficient at school & $\beta_s$ & 1.82 \\
 Transmission coefficient at workplace & $\beta_w$ & 0.919 \\
 Transmission coefficient at community & $\beta_c$ & 0.233 \\
 \hline
\end{tabular}
\end{table}

\vspace{3mm}

\subsection{Modelling multiple strains} 
Recall that in the current methodology, a susceptible individual $j$ sees an incoming infection rate of  $\lambda_j(t)$ from all the infected individuals in the city at time $t$, based on which it gets exposed with probability $1 - \exp\{ - \lambda_j(t) \cdot \Delta t\}$. However, with the introduction of a new infectious strain, it is important to identify whether the person was exposed from an infected person with a new strain or an infected person with an original strain. To estimate this, we compute infection rates coming in from individuals infected with original strain ($\lambda^{original}_j(t)$) and those with new infectious strain ($\lambda^{infectious}_j(t)$) separately.  To account for the increased infectiousness of the strain (let's say $k$ times more infectious than the original strain), if an individual is infected with the infectious strain, its infectiousness is increased by the factor $k$. Then, $\lambda_j(t)$ can be written as
$\lambda_j(t)=\lambda^{original}_j(t)+\lambda^{infectious}_j(t)$.  In our model, initially a fixed percentage (2.5\%) of individuals chosen randomly from infected population (active) are assumed to be from a more infectious strain on a fixed day (after 350 days from start of the simulation in this case). After this initial seeding of infectious strain, it spreads in the following manner: every individual $j$ who gets exposed to the disease at time $t$ is deemed to be infected with the infectious strain with probability $\frac{\lambda^{infectious}_j(t)}{\lambda_j(t)}$ and with the original strain with probability $1-\frac{\lambda^{infectious}_j(t)}{\lambda_j(t)}$. The above methodology can be easily extended to more than two strains.

\vspace{3mm}

\section{Additional Experiments}

\subsection{Sparser city scenario 2 results}
\label{sec:sparser_city_scenario_2_results}

Figure \ref{fig:1m-12.m-all-beta-no-inte} (in Appendix \ref{sec:sparser_city_scenario_2_results}) demonstrates that the SSR algorithm's results align with those of the larger model in the no-intervention scenario. Figure \ref{fig:12.8m-all-beta-low-intervention} demonstrates the results when  home quarantine is implemented after 40 days.

  \begin{figure}[h!]
      \centering
      \begin{minipage}[b]{0.49\textwidth}
  \includegraphics[width=\linewidth, height=5cm]{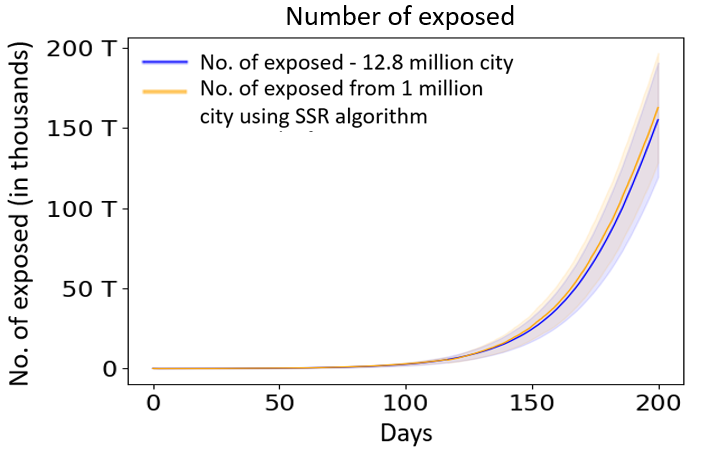}   
    \caption{ \textbf{Sparser city (Scenario 2, No intervention):} Shift-and-scale smaller model (number of exposed) matches the larger model under the no-intervention scenario ($\beta^c_{new} = \beta^c/20$, $\beta^h_{new} = \beta^h/4$, $\beta^w_{new} = \beta^w/4$, $\beta^s_{new} = \beta^s/4$).} 
    \label{fig:1m-12.m-all-beta-no-inte}
  \end{minipage}
\hfill
\begin{minipage}[b]{0.49\textwidth}
 \includegraphics[width=\linewidth, height=5cm]{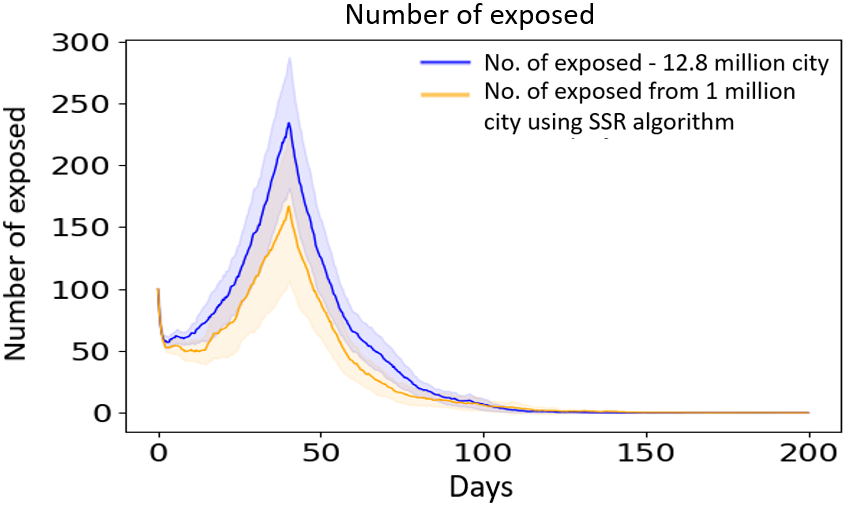}   
    \caption{ \textbf{Sparser city (Scenario 2, With intervention):} Shift-scale-restart smaller model  results (number of exposed) are within 20\% of the larger model under intervention ($\beta_c^{new} = \beta_c/20$, $\beta_h^{new} = \beta_h/4$, $\beta_w^{new} = \beta_w/4$, $\beta_s^{new} = \beta_s/4$).} 
    \label{fig:12.8m-all-beta-low-intervention}
\end{minipage}
  \end{figure}

\subsection{Smaller city results}

\label{sec:smaller_city_under_intervention_results}

\begin{figure}[h!]
      \centering
\begin{minipage}[b]{0.49\textwidth}
\includegraphics[width=\linewidth, height=5cm]{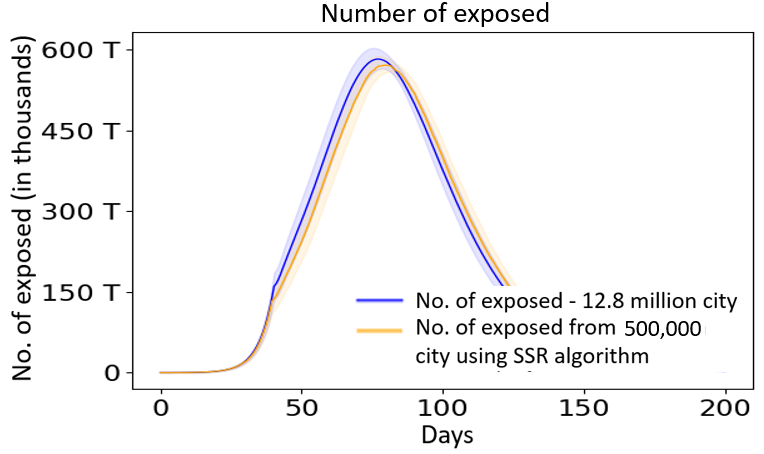}   
    \caption{{\bf 500,000 smaller model (With intervention):} Shift-scale-restart smaller model  (no. of exposed) matches the larger model under intervention (Larger model - 12.8 million population).} 
    \label{fig:12.8m-5l-inte}
\end{minipage}
\hfill
\begin{minipage}[b]{0.49\textwidth}
 \includegraphics[width=\linewidth, height=5cm]{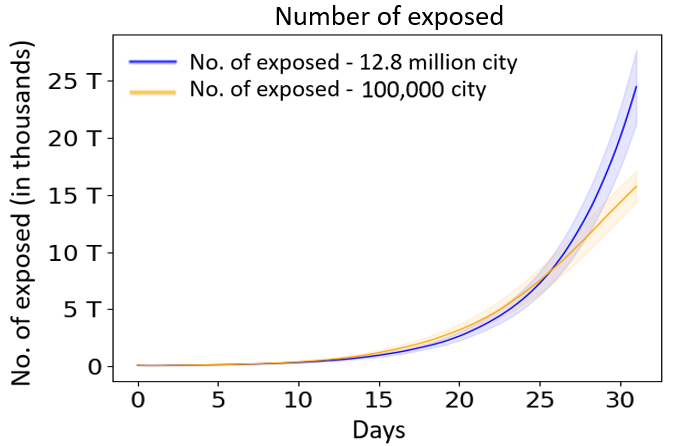}
    \caption{{\bf 100,000 smaller model (Initial phase):} Smaller and larger model are identical initially for about 24 days, when we start with same no. of few, 100 infections, in both cities.
    \label{fig:12.8m-1l-initial-same}}
    \end{minipage}
  \end{figure}

 Figure \ref{fig:12.8m-5l-inte}  demonstrates that the SSR algorithm accurately matches the larger model's outcomes in a scenario with a single intervention—home quarantine implemented from day 40 onward.  Figure \ref{fig:12.8m-1l-initial-same} demonstrates that 12.8-million-person and 100,000-person city models exhibit similar dynamics for only the first 24 days.

\subsection{Experiments with multiple strains}

\label{sec:experiments_with_multiple_strains}

\begin{figure}[h!]
      \centering
\includegraphics[width=0.6\linewidth, height=5cm]{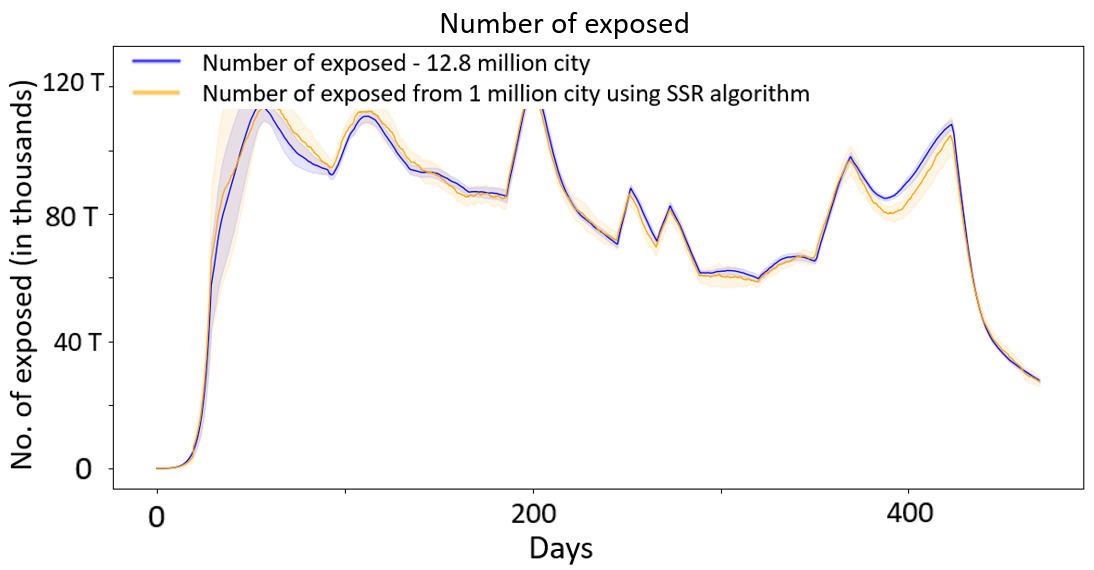}   
   \caption{Shift-scale-restart smaller model matches the larger one under real world interventions over 470 days with a new Delta variant after 350 days. }
    \label{ssr_real_int_with_new_strain}
  \end{figure}

We now evaluate the SSR methodology in the presence of multiple epidemic strains. Figure \ref{ssr_real_int_with_new_strain} compares the two models over an extended 470-day period. This includes the introduction of a new variant after 350 days, alongside the phase preceding its emergence (the methodology for introducing new variants is detailed in Appendix \ref{sec:detailed_abs_model}). Realistic interventions, such as lockdowns, case isolation, home quarantine, and masking, are implemented in the model.
 In this experiment, the smaller and larger cities exhibit similar dynamics until day 22. An intervention is introduced on day 29; consequently, the smaller city is restarted with the intervention on day 18.5, and scaled estimates from day 11.5 of the restarted simulation are appended to day 22 of the initial run (i.e., shifted by 10.5 days).

\section{Experimental details}

\label{sec:additional_numerics}

\subsection{City statistics} 
\label{sec:Numerical_Parameters_Section}

In this section for completeness, we summarize the city statistics  used in our numerical experiments.

Recall that for the numerical experiments, we first create a synthetic city that closely models the actual city. A synthetic model is set to match the numbers employed, numbers in schools, commute distances, etc in the city.  Tables \ref{Household_size} show the household size distribution and school size distribution in the model.  Fraction of working population is set to 40.33\%. 
Table \ref{non_slum_age} summarizes the low-density areas population age distribution  and the high density areas population age distribution.

\begin{table*}[h!]
  \centering
    \small
    \caption{Household and school size distribution \label{Household_size}}
    \begin{tabular}{|c|c|}
      \hline
      {\bf Household size} & {\bf Fraction of households}\\
      \hline
      1 & 0.0485\\
      2 & 0.1030 \\ 
      3 & 0.1715\\
      4 & 0.2589 \\ 
      5 & 0.1819\\
      6 & 0.1035 \\ 
      7-10 & 0.1165\\
      11-14 & 0.0126 \\ 
      15+ & 0.0035\\
      \hline
    \end{tabular}
    \quad
     \begin{tabular}{|c|c|}
      \hline
      {\bf School size} & {\bf Fraction of schools}\\
      \hline
      0-100 & 0.0185\\
      100-200 & 0.1284 \\ 
      200-300 & 0.2315\\
      300-400 & 0.2315 \\ 
      400-500 & 0.1574\\
      500-600 & 0.0889 \\ 
      600-700 & 0.063\\
      700-800 & 0.0481 \\ 
      800-900 & 0.0408\\
      \hline
    \end{tabular}
\end{table*}

   \begin{table*}[h!]
  \centering
  \small
    \caption{Population age distribution \label{non_slum_age}}
    \begin{tabular}{|c|c|}
      \hline
      {\bf Age} & {\bf Fraction of population}\\
      {\bf (yrs)}& {\bf (low density areas)}\\
      \hline
      0-4 & 0.0757\\
      5-9 & 0.0825 \\ 
      10-14 & 0.0608\\
      15-19 & 0.0669 \\ 
      20-24 & 0.0705\\
      25-29 & 0.0692 \\ 
      30-34 & 0.0777\\
      35-39 & 0.0716 \\ 
      40-44 & 0.0762\\
      45-49 & 0.0664 \\ 
      50-54 & 0.0795\\
      55-59 & 0.0632 \\ 
      60-64 & 0.0560\\
      65-69 & 0.0380 \\ 
      70-74 & 0.0227\\
      75-79 & 0.0136 \\
      80+ & 0.0094\\
      \hline
    \end{tabular}
    \quad
     \begin{tabular}{|c|c|}
      \hline
      {\bf Age} & {\bf Fraction of population}\\
       {\bf (yrs)} & {\bf (high density areas)}\\
      \hline
       0-4 & 0.0757\\
      5-9 & 0.0825 \\ 
      10-14 & 0.0875\\
      15-19 & 0.0963 \\ 
      20-24 & 0.0934\\
      25-29 & 0.0917 \\ 
      30-34 & 0.0921\\
      35-39 & 0.0849 \\ 
      40-44 & 0.0606\\
      45-49 & 0.0529 \\ 
      50-54 & 0.0632\\
      55-59 & 0.0503 \\ 
      60-64 & 0.0327\\
      65-69 & 0.0221 \\ 
      70-74 & 0.0073\\
      75-79 & 0.0044 \\
      80+ & 0.0021\\
      \hline
    \end{tabular}
\end{table*}

  \subsection{Other details for experiments}

\noindent\textbf{Experiment setting for Figures \ref{fig:12800_infections}-\ref{fig:shift_scale_100_infections} (Section \ref{sec:speeding_abs}):} We consider one community space with the entire population residing in low-density regions.

\noindent\textbf{Experiment setting for Figure \ref{fig:shift_scale_restart_real_intervention} (Section \ref{sec:speeding_abs}):} We consider 48 community spaces to model the 24 administrative localities in the city, further subdivided into high-density and low-density regions. High-density regions are densely populated, leading to difficulties in maintaining social distancing and increased transmission rates between infected and uninfected individuals. We account for this by implementing a higher community transmission rate in these regions (twice that of low-density regions).
Interventions were implemented based on actual measures enacted in the city, specifically:
\begin{itemize}
    \item No interventions for the first 33 days (simulation day 1–33).
    \item Lockdown from day 34 to day 91.
    \item Mask mandates active from day 53.
    \item Enhanced social distancing rules for elderly populations enforced from day 75.
    \item Schools remain closed throughout the pandemic period except during the initial no-intervention phase.
    \item Additional interventions including home quarantine and case isolation implemented post-lockdown.
\end{itemize}

Workplace attendance schedules after day 91 were set as follows: 5\% from day 92–105, 20\% throughout days 106–135, 33\% throughout days 136–166, and 50\% thereafter. These percentages were selected considering prevailing restrictions and observations from transportation and Google mobility data.
Compliance levels were set at 60\% in low-density areas and 40\% in high-density areas, chosen to fit the fatality data from early in the pandemic.

   In this experiment, both city models (smaller and larger) evolve similarly until day 22. With intervention occurring at day 33, the smaller city model is restarted with intervention measures at day 22.5, and scaled estimates from day 11.5 of the restarted simulation are appended at day 22 of the initial run (representing a 10.5-day shift).

\noindent\textbf{Experiment setting for Figures \ref{fig:1m-12.8m-beta-c-10-no-inte} to \ref{fig:12.8m-all-beta-low-intervention} (Section \ref{sec:sparse_cities}):} We consider one community space with the entire population residing in low-density regions. Interventions are specified in each respective figure.

\noindent\textbf{Experiment setting for Figures \ref{fig:12.8m-5l-nointe} to \ref{fig:12.8m-1l-initial-same} (Section \ref{sec:limits_smaller_city}):} We consider one community space with the entire population residing in low-density regions. Interventions are specified in each respective figure.

\noindent\textbf{Experiment setting for Figures \ref{fig:param-uncertain-no-inte} to \ref{fig:param-uncertain-with-inte} (Section \ref{sec:parameter_uncertainty}):}  We consider 48 community spaces to model the 24 administrative localities in the city, further subdivided into high-density and low-density regions. High-density regions are densely populated, leading to difficulties in maintaining social distancing and increased transmission rates between infected and uninfected individuals. We account for this by implementing a higher community transmission rate in these regions (twice that of low-density regions). Interventions in  Figure \ref{fig:param-uncertain-with-inte}  are identical to those in Figure \ref{fig:shift_scale_restart_real_intervention}.

\noindent\textbf{Experiment setting for Figure \ref{ssr_real_int_with_new_strain}  (Section \ref{sec:multiple_strains}):}  We consider 48 community spaces to model the 24 administrative localities in the city, further subdivided into high-density and low-density regions. High-density regions are densely populated, leading to difficulties in maintaining social distancing and increased transmission rates between infected and uninfected individuals. We account for this by implementing a higher community transmission rate in these regions (twice that of low-density regions).
Interventions were implemented based on actual measures enacted in the city, specifically:
\begin{itemize}
    \item No interventions for the first 33 days (simulation day 1–33).
    \item 
Lockdown from day 34 to day 91.
\item Mask mandates active from day 53.
\item 
Enhanced social distancing rules for elderly populations enforced from day 75.
\item Schools remain closed throughout the pandemic period except during the initial no-intervention phase.
\item Additional interventions including home quarantine and case isolation implemented post-lockdown.
\end{itemize}

Workplace attendance schedules after day 91 were set as follows: 5\% attendance from day 92–105; 15\% from day 106–135; 25\% from day 136–166; 33\% from day 167–198; 50\% from day 199–349; 65\% from day 350; and 20\% from day 419–470.
To account for increased social interaction during festival periods (day 184–196), we increased the community $\beta$ parameter by 2/3 and reduced compliance from 60\% in low-density areas and 40\% in high-density areas to 40\% and 20\%, respectively. Similar adjustments were made for subsequent festival periods.
Compliance levels were initially set at 60\% in low-density areas (NS) and 40\% in high-density areas (S) before day 286 and outside festival periods (during festivals: 40\% NS and 20\% S). These rates subsequently changed to (50\%, 30\%) from day 287–318, (40\%, 20\%) from day 319–349, (20\%, 10\%) from day 350–367, and (40\%, 20\%) from day 368–418. During the lockdown period of day 419–470, compliance was set at (60\%, 40\%).
We assumed a single variant accounted for 2.5\% of the infected population on day 350 in our model, randomly selected from all infected individuals on that date. This variant was assumed to be 2.25 times more infectious than the original strain.

Results presented in all the figures are average of 20 simulation runs.

\subsection{Algorithms for parameter uncertainty}
\label{sec:algorithms_parameter_uncertainty}

The simulation dynamics and SSR algorithm under parameter uncertainty are presented in Algorithms \ref{alg:base_parameter_uncertainty} and \ref{alg:SSR_parameter_uncertainty}, respectively.
 
\begin{algorithm}[h!]
\caption{Simulation dynamics (under parameter uncertainty)}\label{alg:base_parameter_uncertainty}
\begin{algorithmic}[1]
\While{$i<N$} \State Draw a sample of parameters from its distribution. For each $i$, at $t=0$, start the simulation with $I_0$ infections distributed as per $\mu_0(N)$. 
 
 \While {$t < T$}
 \State For each susceptible individual
 $n$, calculate $\lambda_n(t)$.  Its status then changes to exposed 
 with probability $1-\exp({-\lambda_n(t)})$.
 \State All individuals in some state other than susceptible, independently transition to another state as per the disease progression 
 dynamics. 
 \State $t \gets t+1$.
 \EndWhile
 \State $i \gets i+1$.
 \EndWhile
 \State The above simulation is independently repeated many times and average (as well as associated uncertainty) of the performance measures  are reported.
\end{algorithmic}
\end{algorithm}

\begin{algorithm}[h!]
\caption{Shift, scale and restart algorithm (under parameter uncertainty)}\label{alg:SSR_parameter_uncertainty}
\begin{algorithmic}[1]
\While{$i<N$}
\State  Draw a sample of parameters from its distribution. For each $i$, at $t=0$, start the simulation with $I_0$ infections distributed as per $\mu_0(N)$.
 Generate the simulation sample path  
 $[y_1,y_2,...,y_{t_{min}}]$ where $y_t$ denotes the statistics of the affected population (e.g., number exposed, number hospitalised, 
 number of fatalities) at time $t$. 

 \State Restart 
 a new simulation of the city using common random numbers, but with the intervention introduced 
 at time $t_{\frac{x}{k}}+t_{I}-t_{min}$. Simulate it upto time $T-(t_{min}-t_{x/k})$. 
Denote the time series of statistics of the affected population in the restart simulation
by $z_1,z_2,...,z_{t_{x/k}},...,z_{T-(t_{min}-t_{x/k})}.$

 \State  The approximate statistics of the affected population for the larger city is then obtained as   
 $
 [y_1,y_2,...,y_{t_{min}},k \times z_{t_{x/k}+1},...,k \times z_{T+t_{x/k}-t_{min}}]
 .$

 \State $i \gets i+1$.
 \EndWhile
\State 
Same as Step 10, Algorithm 4.
\end{algorithmic}
\end{algorithm}

\subsection{Additional experiment statistics}

\begin{figure}[h!]
      \centering
      \begin{minipage}[b]{0.49\textwidth}
  \includegraphics[width=\linewidth, height =5cm]{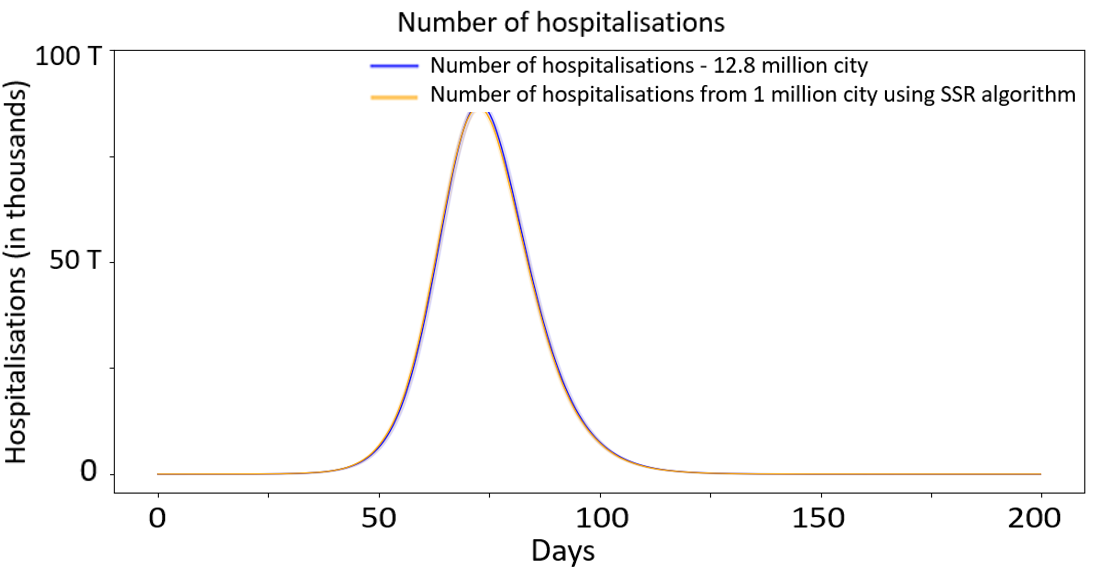}
  \caption{Shift and scale smaller model (no. of hospitalised) matches the larger model under no intervention scenario. }
\label{hosp_no_int}
  \end{minipage}
\hfill
\begin{minipage}[b]{0.49\textwidth}
\includegraphics[width=\linewidth, height =5cm]{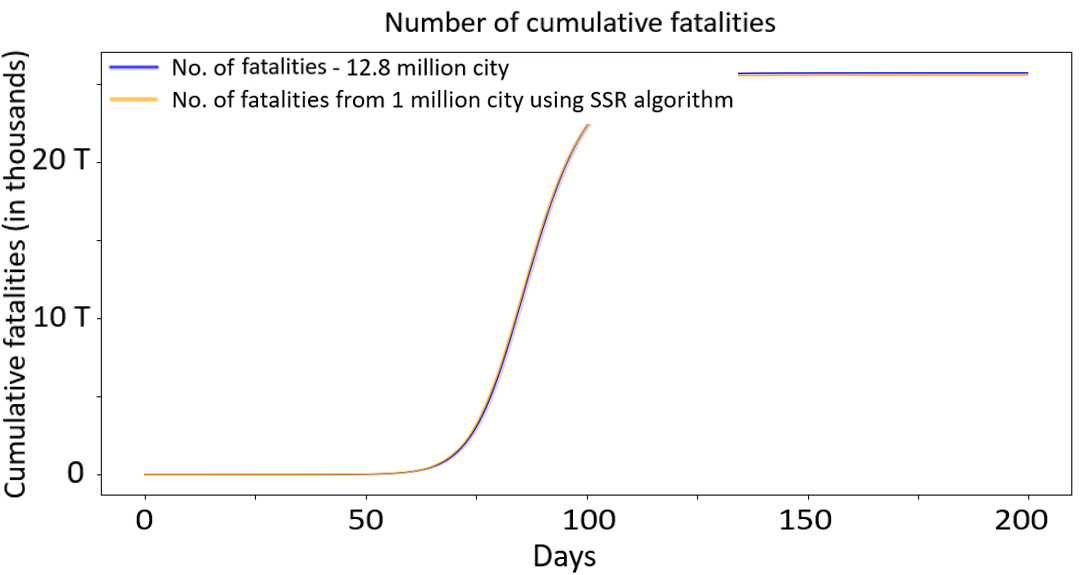}
  \caption{Shift and scale smaller model (no. of cumulative fatalities) matches the larger model under no intervention scenario. }
\label{fatal_no_int}
\end{minipage}
  \end{figure}

Figures \ref{hosp_no_int} and \ref{fatal_no_int} compare the hospitalization counts and cumulative fatalities obtained from a 12.8-million-inhabitant simulation without intervention against the estimates produced by the shift–scale–restart algorithm applied to a 1-million-inhabitant simulation. The city is same as the one considered in Figures \ref{fig:12800_infections}-\ref{fig:shift_scale_100_infections} with one community space and the entire population residing in low-density regions. In this scenario, the epidemic trajectories in the two city sizes match closely up to day 35. Because no control measures are introduced, the scaled estimates from day 21.5 of the base simulation are appended after day 35—corresponding to a 13.5-day temporal shift.

\begin{figure}[h!]
      \centering
\begin{minipage}[b]{0.49\textwidth}
 \includegraphics[width=\linewidth, height =5cm]{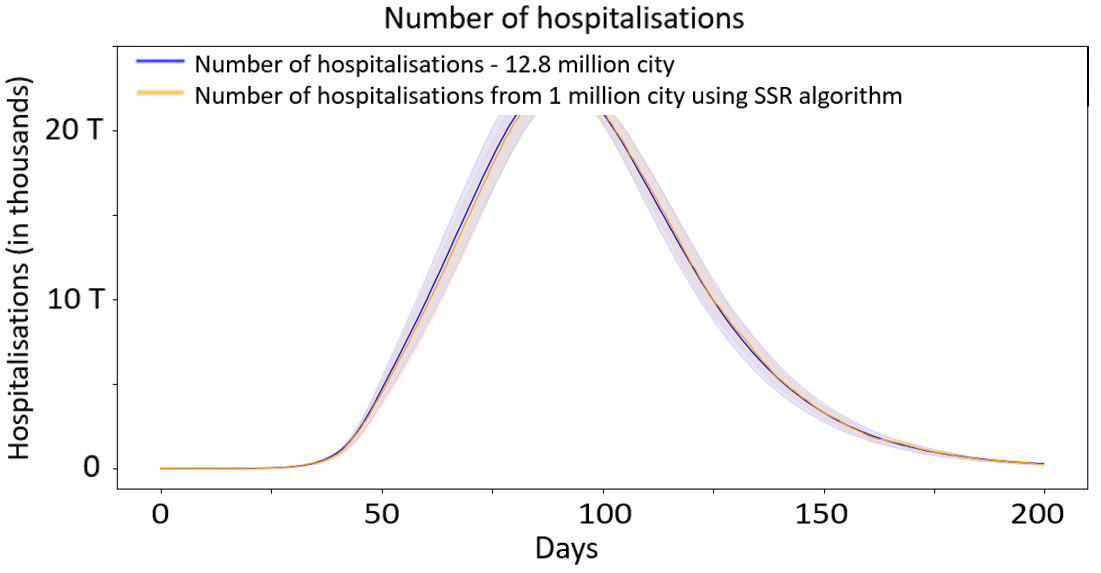}
  \caption{Shift-scale-restart smaller model (no. of hospitalised) matches the larger model (home quarantine from day 40). }
  \label{hosp_hq_40}
\end{minipage}
\hfill
\begin{minipage}[b]{0.49\textwidth}
  \includegraphics[width=\linewidth, height =5cm]{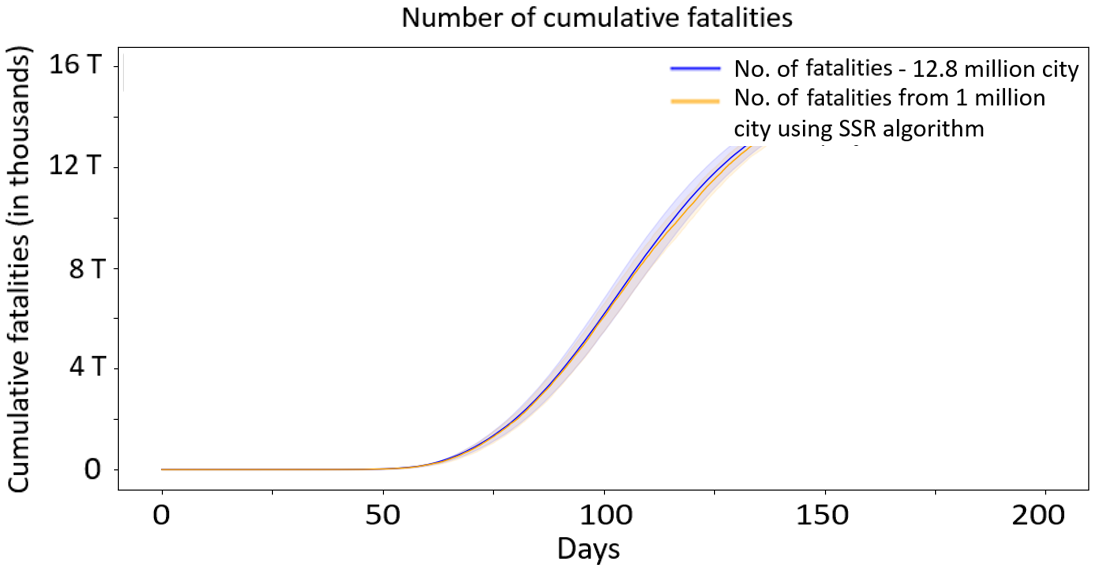}
  \caption{Shift-scale-restart smaller model (no. of cumulative fatalities) matches the larger model (home quarantine from day 40).}
  \label{fatal_hq_40} 
\end{minipage}
  \end{figure}

Figures \ref{hosp_hq_40} and \ref{fatal_hq_40} extend the comparison to a setting in which home quarantine for compliant infectious individuals begins on day 40. They plot the numbers of exposed individuals, hospitalizations, and cumulative fatalities for the 12.8-million-inhabitant simulation alongside the shift–scale–restart estimates derived from the 1-million-inhabitant model. The two city sizes again evolve similarly until day 35. Because the intervention starts on day 40, the smaller-city simulation is restarted with quarantine enabled at day 26.5, and its scaled estimates from day 21.5 of the restarted run are appended to the larger-city trajectory at day 35—again representing a 13.5-day shift.

\end{document}